\documentclass[10pt,a4paper]{article} 
\newcommand{\sect}{\setcounter{equation}{0}\section}

\usepackage{t1enc}
\usepackage{amsmath}
\usepackage{amssymb}
\usepackage{enumerate}
\begin{document} 

\title{Bondi-type systems near space-like infinity and the calculation of the
NP-constants}

\author{Helmut Friedrich\\
Albert-Einstein-Institut\\
Max-Planck-Institut f\"ur Gravitationsphysik\\
Am M\"uhlenberg 5\\
14476 Golm, Germany\\ \\
J\'anos K\'ann\'ar\\
MTA KFKI \\
R\'eszecske \'es Magfizikai Kutat\'oint\'ezet \\
Budapest, Pf. 49, \\
1525, Hungary}

\date{}

\maketitle

\begin{abstract}

\noindent
We relate Bondi systems near space-like infinity to another type of gauge
conditions. While the former are based on null infinity, the latter are
defined in terms of Einstein propagation, the conformal structure, and
data on some Cauchy hypersurface. For a certain class of time symmetric
space-times we study an expansion which allows us to determine the behavior
of various fields arising in Bondi systems in the region of space-time where
null infinity touches space-like infinity. The coefficients of these
expansions can be read off from the initial data. We obtain in particular
expressions for the constants discovered by Newman and Penrose (NP-constants)
in terms of the initial data. For this purpose we calculate a certain
expansion introduced in
\cite{Fr:space-like-infinity} up to 3rd order. 

\end{abstract}

\newpage

\sect{Introduction}

Most studies of gravitational fields near null infinity are based on
the use of ``Bondi-type'' coordinates. In the first investigations of the
behavior of the field near null infinity (cf. \cite{BBM},
\cite{NP:spin-coefficients}, \cite{S}) Bondi-type coordinates played a crucial
role in the specification of the fall-off behavior of the field. 
The characterization of the asymptotic behavior of gravitational
fields near null infinity in terms of the conformal geometry subsequently
suggested by Penrose (\cite{P:asymptotic-properties-of-fields},
\cite{P:zero-rest-mass-fields}) does not require the use of such a
specific class of coordinates. Nevertheless, Bondi-type coordinates
are usually also employed in this context because they allow us to exploit 
in a convenient way certain features of the null cone structure.
If the gravitational field is, however, to be analyzed in detail in the
region where future and past null infinity ${\cal J}^{\pm}$ ``touch''
space-like infinity, and if this is to be done such that 
${\cal J}^{-}$ and ${\cal J}^{+}$ are treated on an equal footing, 
Bondi-type coordinates are not particularly helpful. Already in the
simplest non-trivial case, that of the Schwarzschild solution, the use 
of double null coordinates leads to difficulties.

In \cite{Fr:space-like-infinity} an initial value problem for the conformal
vacuum field equations has been formulated which is designed to analyze near
space-like and null infinity the Einstein propagation of asymptotically flat
data on a Cauchy hypersurface $\tilde{S}$ in a finite picture.  In this
setting, which is based on certain conformally invariant structures,
space-like infinity is represented by a cylinder 
$I \simeq ]-1, 1[ \times S^2$ such that the sets 
${\cal J}^{\pm} \simeq \mathbb{R} \times S^2$, representing future resp. past
null infinity, ``touch'' the cylinder at its two boundary component 
$I^{\pm} = \{\pm1\} \times S^2$. Though the underlying facts about the
evolution equations which have been used here hold for much more general
situations, the picture has been analyzed so far under certain
simplifying assumptions on the initial data. The data are assumed to be
time-symmetric and the conformal structure, which then represents the free
datum, is assumed to extend smoothly through space-like infinity such that
the latter is represented by a point $i$ in an extended manifold 
$S = \tilde{S} \cup \{i\}$. The cylinder $I$ is obtained by blowing up
the point $i$ to a sphere $I^0 \simeq \{0\} \times S^2$ and by smoothly
extending the solution in a particular geometric gauge.  

It can be seen already under these assumptions on the data that the new
picture allows us to relate near $I^{\pm}$ properties of the data on
$\tilde{S}$, which touches $I$ at $I^0$, to properties of the field on
null infinity by solving a hierarchy of differential equations on $I$. These
equations have been used in \cite{Fr:space-like-infinity} to derive certain
``asymptotic regularity conditions'' for the initial data whose imposition
prevents a certain class of logarithmic singularities of the field  at the
sets $I^{\pm}$ from arising. However, it still has to be shown that the
asymptotic regularity conditions ensure a time evolution of the data which
extends near space-like infinity smoothly to null infinity. 

In the present article we analyze the consistency of the early investigations
of fields near null infinity with the picture developed in
\cite{Fr:space-like-infinity} and we demonstrate to some extent the efficiency
of the latter in calculating near space-like infinity quantities on null
infinity from the given data. For this purpose we make two different types of
assumptions. On the one hand we shall consider space-times arising from time
symmetric  vacuum data as described above which satisfy the asymptotic
regularity conditions. Our calculations of fields on the cylinder $I$ rely
only on these assumptions. On the other hand we shall assume that
these data develop into solutions which admit a smooth conformal structure at
null infinity and that the gauge conditions proposed in
\cite{Fr:space-like-infinity} extend in a smooth and regular way to 
${\cal J}^{\pm}$. We expect that our analysis will contribute information on
the solution process which in the end will allow us to remove the second type
of assumptions and to show that the existence of the smooth evolution can be
derived solely from assumptions on the initial data.

The present article can be divided into three different, though related, 
parts.  

\noindent$-$ In \cite{Fr:space-like-infinity} an expansion of the field
near space-like infinity in terms of a ``radial'' coordinate $\rho$, which
vanishes on the cylinder $I$ representing space-like infinity, has been
introduced. We calculate the coefficients of this expansion to third order.
This calculation is not only of interest because it allows us to study the 
NP-constants, which will be discussed below, but also because it provides some
information on the smoothness of the evolution near null infinity for fields
arising from data subject only to our first type of assumptions. Though the
asymptotic regularity conditions referred to above exclude certain types of
logarithmic singularities in the evolution near $I$, there exists another
potential source of singularities. To show that in fact no further
singularities can arise at any order, it is clearly of interest to understand
the situation for the first few orders of the expansion. The potential
singularities should show up for the first time at the order of our
calculation. Our calculations show that at this order they are in fact
excluded by the asymptotic regularity conditions. 

We note that our expansion of the field near space-like and null
infinity, which we carry out in terms of the conformally rescaled
fields and associated gauge conditions, can be translated into an
expansion of the field near space-like infinity in terms of the
``physical'' field and suitable coordinates. We shall not carry
out such a translation because the main point of our consideration 
is the fact that we can relate quantities on null infinity to the 
data on $\tilde{S}$.

\noindent$-$ Bondi-type coordinates and certain related frame fields
(cf. the definition of the ``NP-gauge'' below) are based on the structure
of null infinity. The gauge conditions in \cite{Fr:space-like-infinity}
(cf. the definition of the ``F-gauge'' below) are based on Cauchy data, the
Einstein equations, and certain properties of conformal structures.    
We discuss in general terms how to construct near null infinity the
transformation from the  F-gauge into the NP-gauge. Using the expansion
referred to above we then obtain expansions near $I^+$ of various quantities
given in the NP-gauge in terms of the coordinates arising in the F-gauge
and coefficients which are given directly in terms of the initial data on
$S$. We note that these expansions imply expansions of quantities of physical
interest on null infinity such as the  Bondi-energy-momentum, the angular
momentum  (cf. \cite{P:angular-mom.} for various suggestions), the radiation
field, etc. in terms of the coordinate $\rho$ on null infinity, which
vanishes at $I^+$, and coefficients derived from the initial data.  

Since we need for our considerations quite detailed information on the
structure of the initial data near space-like infinity, our explicit
calculation are done only for time-symmetric data. However, many of our
considerations apply also to more general situations and as soon as
sufficient information on data with non-vanishing extrinsic curvature becomes
available (cf. \cite{dain}), we shall be able to derive by similar
calculations relations between fields on ${\cal J}^{-}$ and 
${\cal J}^{+}$. These relations will contain non-trivial information on
the evolution process. 

\noindent$-$ As a specific application of this discussion we reconsider the
constants which have been associated by Newman and Penrose with asymptotically
simple space-times (cf. \cite{NP:NPQ-letter}, \cite{NP:NPQ}). The
NP-constants are given by certain integrals over spherical cuts of null
infinity and have been shown to be absolutely conserved in the sense of being
independent of the choice of cut. We derive for them expressions in terms of
the initial data on $\tilde{S}$. Such expressions have been given already in
the static case in \cite{NP:NPQ}. We derive analogous expressions for a much
more general class of space-times arising from time-symmetric initial data.
For these data the time evolution of the field is in general not known
explicitly as it is the case in the presence of a time-like Killing vector
field. The fact that we can nevertheless obtain expressions in terms of the
data illustrates to some extent the efficiency of the new picture. Though
various authors (cf. \cite{G:invariant-transformations}, \cite{PB:scepticism},
\cite{R:NPQ}) discuss these constants from different points of view, no
consensus has been found concerning their geometrical/physical significance.
Whether our discussion will help clarify the meaning of the NP-constants
remains to be seen. One of our main reasons for looking at them is the
expectation that they may play a role in the construction of space-times. In
numerical calculations they may certainly provide a check on the numerical
accuracy.

\sect{Relating different gauge conditions near null infinity}

We begin by giving an outline of the {\it finite, regular initial
value problem near space-like infinity}. This has been introduced in 
the article \cite{Fr:space-like-infinity}, to which we refer for more
details. It involves a gauge which we refer to as the {\it F-gauge}.   
We then recall the {\it NP-gauge}, employed in \cite{NP:NPQ}, to discuss
the gravitational field near null infinity. Finally, we discuss how the
NP-gauge is related to the F-gauge.

\subsection{The regular finite initial value problem near space-like
infinity}\label{F-gauge}

We want to discuss asymptotically flat solutions $(\tilde{M}, \tilde{g})$ to
Einstein's field equations $\tilde{R}_{\mu\nu}=0$ in a neighborhood
$\tilde{M}_a$ of space-like infinity which covers parts of future and past
null infinity. The solutions arise from asymptotically flat data on a smooth
space-like Cauchy hypersurface $\tilde{S} \subset \tilde{M}$ which are such
that the intrinsic conformal structure on $\tilde{S}$ admits an extension with
a certain smoothness to a smooth compact manifold $S$ obtained from
$\tilde{S}$ by adjoining a point $i$ which represents space-like
infinity, $S = \tilde{S} \cup \{i\}$. We assume that the solution, i.e. the
evolution in time of these data, possesses a smooth conformal extension 
$(M, g, \Theta)$ such that we can write $M = \tilde{M} \cup {\cal J}^-
\cup{\cal J}^+$, where
${\cal{J}}^{\pm} \simeq \mathbb{R} \times S^2$ represent
future respectively past null infinity and $\Theta$ denotes a smooth
``conformal factor'' on $M$ such that $\Theta > 0$ and $g =
\Theta^2\,\tilde{g}$ on $\tilde{M}$ while $\Theta = 0$, $d\,\Theta
\neq 0$ on ${\cal J}^{\pm}$.

To analyze in detail the consequences of the field equations in a
neighborhood of space-like infinity which covers parts of ${\cal{J}}^{\pm}$,
the situation above has been discussed in the \cite{Fr:space-like-infinity} in
terms of a certain principal fiber bundle $M'_a \rightarrow M_a$ with
projection $\pi$, $4-$dimensional base space $M_a$, and bundle space $M'_a$
which is a $5-$dimensional manifold with boundary and edges. To describe this
setting further we need to introduce some notation. 

We employ the two-components spinor and space-spinor formalisms as used in
\cite{Fr:space-like-infinity} where
$\epsilon_{ab}$, $\epsilon^{ab}$ are the antisymmetric spinors with
$\epsilon_{01} = 1$, $\epsilon^{01} = 1$. We set    
$\tau^{aa'} = \epsilon_0\,^a\,\bar{\epsilon}_{0'}\,^{a'} +
\epsilon_1\,^a\,\bar{\epsilon}_{1'}\,^{a'}$. By $SU(2)$ will be denoted the
group of $2 \times 2$ matrices $t = (t^a\,_b)$ satisfying 
\[
\epsilon_{ac}\,t^a\,_b\,t^c\,_d = \epsilon_{bd},\,\,\,\,\,\,\,
\tau_{ac}\,t^a\,_b\,t^c\,_d = \tau_{bd},
\]
and by $U(1)$ its subgroup of diagonal matrices. A basis of the Lie-algebra
of $SU(2)$ is then given by the $2 \times 2$ matrices  
\begin{equation}
u_1 = \frac{1}{2} \left( \begin{array}{cc}
0 & i \\
i & 0
\end{array} \right), \, \, 
u_2 = \frac{1}{2} \left( \begin{array}{cc}
0 & - 1 \\
1 & 0
\end{array} \right),\,\,
u_3 = \frac{1}{2} \left( \begin{array}{cc}
i & 0 \\
0 & - i 
\label{SU2-Lie-alg}
\end{array} \right), 
\end{equation} 
of which $u_3$ generates $U(1)$.

In the following will be described in detail the regular finite initial value
problem at space-like infinity formulated in \cite{Fr:space-like-infinity}.  
Though we shall remark in passing on the construction of the manifold $M'_a$
and the underlying gauge conditions, we refer for the full details to the
original article. The manifold $M'_a$ is given by
\begin{displaymath}
M'_a = \{(\tau, \rho, t) \in \mathbb{R} \times \mathbb{R} \times SU(2)|\,\,
0 \le \rho < a,\, - \frac{\omega}{\rho} \le \tau \le \frac{\omega}{\rho}\},
\end{displaymath} 
where $a$ is a positive real number and $\omega = \omega(\rho, t)$ a
smooth non-negative function, given below, such that
$\frac{\omega}{\rho}$ extends to a smooth positive function with
$\frac{\omega}{\rho} \rightarrow 1$ as $\rho \rightarrow 0$.
By $\rho$ and $\tau$ will also be denoted the projections of $M'_a$ 
onto the first respectively second component of
$\mathbb{R} \times \mathbb{R} \times SU(2)$. Then any coordinate system on
$SU(2)$ will define together with the functions $\rho$ and $\tau$ a coordinate
system on $M'_a$. There will, however, arise no need for us to
introduce coordinates on $SU(2)$. We denote the projection onto the third
component of $\mathbb{R} \times \mathbb{R} \times SU(2)$ by $t$ and 
regard the $SU(2)$-valued function $t$ as a ``coordinate'' on $M'_a$.

The natural action on the right of $U(1)$ on $SU(2)$ induces a smooth action
of $U(1)$ on $M'_a$. The quotient $M'_a/U(1)$ under this action will be
denoted by $M_a$ and the induced projection of $M'_a$ onto $M_a$ by $\pi$.
We shall write $N = \pi(N')$ for any subset $N'$ of $M'_a$. The
following subsets of $M'_a$ will be important for us:        
\[
{\cal J}^{'\pm} = \{ \tau
= \pm\,\frac{\omega}{\rho}, \rho > 0\} \simeq \mathbb{R} \times S^3,
\]
\[
I' = \{ |\tau| < 1, \rho = 0\} \simeq \mathbb{R}\times S^3,
\,\,\,\,\,\,\,\,\,\,\,
I^{'\pm} = \{\tau = \pm\,1,\, \rho = 0\} \simeq S^3,
\]
\[
C' = \{\tau = 0\},\,\,\,\,\,\,\,\,\,\,\,
I^{'0} = \{\tau = 0,\, \rho = 0\} = C' \cap I' \simeq S^3.
\]
Because they cover only a part of null infinity close to space-like infinity,
we should have denoted the first sets more precisely by ${\cal J}^{'\pm}_a$
but we dropped the subscript $a$ for convenience. By definition the part of
the physical manifold $\tilde{M}$ which is covered by $M_a$ is given by 
$\tilde{M}_a = M_a \setminus ({\cal J}^{-} \cup {\cal J}^{+}  
\cup I \cup I^- \cup I^+)$, the sets ${\cal J}^{\pm}$ represent 
{\it future} resp. {\it past null infinity} while the set $I$ represents
{\it space-like infinity} for $\tilde{M}_a$ and the metric induced on it by
$\tilde{g}$. Thus $\tilde{M}_a$ covers a neighborhood of space-like and
null infinity in $\tilde{M}$. The edges $I^{\pm} \simeq S^2$ of $M_a$ at which
future resp. past null infinity touches space-like infinity will play an
important role in the following. We shall refer to the set $C$ as the 
{\it initial hypersurface} since by definition 
$C \cap \tilde{M}_a = C \setminus I^0 = \tilde{S} \cap \tilde{M}_a$. 
There exists a neighborhood $B_a$ of $i$ in $S$ and smooth surjective map 
$\pi':C \rightarrow B_a$ which is injective on $C \setminus I^0$ and which
maps $I^0$ onto $i$. 

As described in \cite{Fr:space-like-infinity}, the manifold $M'_a$ is
obtained essentially by lifting $M_a$ into the bundle of normalized (with
respect to $\epsilon_{ab}$) spin frames. The set $I^{'0} \simeq SU(2)$
corresponds to the set of normalized (with respect to $\epsilon_{ab}$ and
$\tau_{ab}$) spin frames at the point $i$. With each such spin frame we
associate a unit tangent vector of $S$ at $i$. With this vector we associate
in turn a curve through $i$ in $B_a$ and extend the spin frame along this
curve by a certain transport process. Thus we obtain spin frames at each
point of $B_a \setminus \{i\}$. These frames are transported off 
$B_a \setminus \{i\} \simeq C \setminus I^0$ into the space-time $M_a$ by a
certain propagation law along conformal geodesics orthogonal to $C$. The
latter are given in our description of $M'_a$ by the curves $\rho = const.$,
$t = const.$ with $\tau$ a natural parameter along them. Since for given unit
tangent vector at $i$ the spin frame defining it is determined up to a phase
factor, the spin frames at points of $M_a \setminus (I \cup I^- \cup I^+)$ are
also given up to multiplications by phase factors, which corresponds to the
action of the group $U(1)$. The transport laws as well as further details of
the gauge conditions are encoded in the form of the data and certain
properties of the unknowns for the reduced equations.    

Since it turns out to be most convenient, we will carry out all our
calculations on the manifold $M'_a$ and use for the subsets of  $M'_a$
introduced above the same names as for their images under $\pi$.  

We denote by $Z_{u_i}$ the vector field generated by $u_i$ and the obvious
action of $SU(2)$ on $M'_a$ and define complex vector fields 
$X_{+} = - (Z_{u_2} + i Z_{u_1})$, $X_{-} = - (Z_{u_2} - i Z_{u_1})$, 
$X = - 2\,i\,Z_{u_3}$ which satisfy
the commutation relations
\begin{equation}
[X,\, X_{+}] = 2\,X_{+},\,\,\,\,[X,\, X_{-}] = - 2\,X_{-},\,\,\,\,
[X_{+},\,X_{-}] = -X. 
\label{X-commutators}
\end{equation}

The conformal field equations, in the form used in
\cite{Fr:space-like-infinity}, are given in a particular (conformal,
coordinate, and frame) gauge which is explained, together with the equations,
most naturally in the context of {\it normal conformal Cartan connections}
(cf. \cite{Fr:AdS}). Again, we shall not go through the complete argument but
just describe the unknowns and equations. To obtain the equations on $M'_a$,
we use the fact that the solder and the connection forms on the bundle of
spin frames induce corresponding forms $\sigma^{aa'}$, $\omega^a\,_b$ on 
$M'_a \setminus I'$ which extend smoothly to $M'_a$. The metric
$\epsilon_{ab}\,\bar{\epsilon}_{a'b'}\,\sigma^{aa'}\,\sigma^{bb'}$ on
$M'_a$ is degenerate because $<\sigma^{aa'}, X>\, = 0$ (the angle brackets
denoting the dual pairing), but it descends to
the Lorentz metric $g$ on $\pi(M'_a \setminus I')$.

The equations are written as equations for the ``vector''-valued unknown
\begin{equation*}
u = (c^0\,_{ab},\,c^1\,_{ab},\,c^{\pm}\,_{ab},\,
\chi_{(ab)cd},\,\xi_{abcd},\,f_{ab},\,\Theta_{(ab)cd},\,
\Theta_{g\;ab}^{\phantom{g}g},\,\phi_{abcd}),
\end{equation*}
whose components have the following meaning. We consider the smooth vector
fields
\begin{equation*}
c_{aa'} = c^0\,_{aa'}\,\partial_{\tau} + c^1\,_{aa'}
\partial_{\rho}  + c^{+}\,_{aa'}\,X_{+} + c^{-}\,_{aa'}\,X_{-} 
\label{c-frame1}
\end{equation*} 
which satisfy $<\sigma^{aa'}, c_{bb'}>\, =
\epsilon_b\,^a\,\bar{\epsilon}_{b'}\,^{a'}$ on $M'_a \setminus I'$.
All fields are written in space spinor notation based on the vector
field $\sqrt{2}\,\partial_{\tau} = \tau^{aa'}\,c_{aa'}$. Since
$\tau^{aa'}c_{aa'}$ is invariant under the action of $U(1)$ it descends to a
vector field on $\pi(M'_a \setminus I')$ which is time-like, has norm
$\tau_{aa'}\,\tau^{aa'} = 2$, and is orthogonal to $\tilde{S}$. We
have
\begin{equation}
\label{c-frame2}
c_{aa'} = \frac{1}{\sqrt{2}}\,\tau_{aa'}\,\partial_{\tau} 
- \tau^b\,_{a'}\,c_{ab}
\end{equation} 
with
$c_{ab} \equiv \tau_{(a}\,^{b'}\,c_{b)b'} 
= c^0\,_{ab}\,\partial_{\tau} + c^1\,_{ab}
\partial_{\rho} + c^{+}\,_{ab} X_{+} + c^{-}\,_{ab} X_{-}$.
The connection defines connection coefficients 
$\Gamma_{abcd} = \tau_b\,^{a'}\Gamma_{aa'cd} = 
\tau_b\,^{a'}<\omega_{cd}, c_{aa'}>$   
which can be decomposed in the form   
\begin{displaymath}
\label{Gamma-F-gauge}
\Gamma_{abcd} = \frac{1}{\sqrt{2}}\,(\xi_{abcd} - \chi_{abcd})   
= \frac{1}{\sqrt{2}}\,(\xi_{abcd} - \chi_{(ab)cd}) - 
\frac{1}{2}\,\epsilon_{ab}\,f_{cd},  
\end{displaymath}
with fields satisfying $\chi_{abcd} = \chi_{ab(cd)}$,
$\xi_{abcd} = \xi_{(ab)(cd)}$, $f_{ab} = f_{(ab)}$. 
The curvature is represented by the rescaled conformal Weyl spinor field  
$\phi_{abcd} = \phi_{(abcd)}$ and by a spinor field 
$\Theta_{abcd} = \Theta_{ab(cd)}$ which is the Ricci spinor field of a
certain Weyl connection for $\tilde{g}$.

The pull back $\pi^*\,\Theta$, again referred to as the 
conformal factor and denoted by $\Theta$, extends smoothly to $M'_a$ and is
known in our gauge explicitly. It is given by 
\begin{equation}
\label{conformal-factor}
\Theta = \frac{\Omega}{\rho}\left(1 - \tau^2\,\frac{\rho^2}{\omega^2}\right),
\end{equation}
and appears, together with the 1-form \begin{displaymath}
d_{ab} = 2\,\rho\,\frac{U\,x_{ab} - \rho\,D_{ab}U - \rho^2\,D_{ab}W}
{(U + \rho\,W)^3},
\end{displaymath}
(with $x_{ab}$ as given in appendix [\ref{spinorial-identities}])
which characterizes in a certain way the difference between the Levi-Civita
connection of $g$ and the Weyl connection referred to above,
as coefficient in the conformal field equations. We have set here  
\begin{equation}
\begin{array}{r@{}l}
\multicolumn{2}{c}{\Omega = \frac{\rho^2}{(U + \rho\,W)^2},} \\[6pt]
\omega \equiv & 
\,2\,\Omega\,(-D_{ab}\,\Omega\,D^{ab}\,\Omega)^{- \frac{1}{2}} =
\rho\,(U + \rho\,W)\, \bigl\{ U^2 + 2\,\rho\,U\,x^{ab}D_{ab}U 
- \rho^2\,D^{ab}U\,D_{ab}U \\[3pt]
& + 2\,\rho^2\,U\,x^{ab}D_{ab}W 
- 2\,\rho^3\,D^{ab}U\,D_{ab}W 
- \rho^4\,D^{ab}W\,D_{ab}W \bigr\}^{- \frac{1}{2}}, \label{Omega}
\end{array}
\end{equation}
where the smooth functions $U = U(\rho, t)$, $W = W(\rho, t)$,
which satisfy $U = 1$ and $W = \frac{1}{2}m_{ADM}$ on $I^0$, are given
as part of the initial data on the initial hypersurface $C'$, on which
$D_{ab}$ is the intrinsic covariant derivative. Note that
the fields $\Omega$, $\omega$, $d_{ab}$ do not depend on $\tau$.
The conformal factor satisfies the relations (cf. \cite{Fr:AdS})
\begin{equation}
\begin{array}{l}
\Theta > 0
\quad\mbox{on}\quad M'_a,\,\,\,\,\,
\{\Theta = 0\} = {\cal J}^{'-} \cup I^{'-} \cup I' \cup I^{'+} \cup 
{\cal{J}}^{'+},\\[3pt] 
c_{aa'}(\Theta) \neq 0 \quad\mbox{and}\quad
\epsilon^{ab}\,\bar{\epsilon}^{a'b'}\,c_{aa'}(\Theta)
\,c_{bb'}(\Theta) = 0
\quad\mbox{on}\quad {\cal J}^{'\pm}. 
\label{conf.factor}
\end{array}
\end{equation}
In the following we shall refer to the coordinates $\tau$, $\rho$, $t$, the 
frame $\{c_{aa'}\}$, and the conformal gauge defined by
(\ref{conformal-factor}) as the {\it F-gauge}. 

\subsubsection{The conformal evolution equations}

We recall here a few general features of the conformal field
equations and refer again to \cite{Fr:space-like-infinity} for more details.
The conformal field equations imply on $M'_a$ evolution equations of
the form
\begin{equation}
\label{conformal-field-equations}
\{ A^{0}\,\partial_{\tau} + A^{1}\,\partial_{\rho}
+ A^{+}\,X_{+} + A^{-}\,X_{-} \}\,u = C\,u,
\end{equation} 
where $A^{0}$, $A^{1}$, $A^{\pm}$, $C$ denote matrix-valued functions which
depend on $u$ and the coordinates. The system is, for $u$
close to the data given below and for the coordinates taking values on $M'_a$
near $C'$, symmetric hyperbolic. Writing $u = (v, \phi)$ with   
\begin{equation}
v = (c^0\,_{ab},\,c^1\,_{ab},\,c^{\pm}\,_{ab},\,\chi_{(ab)cd},\,\xi_{abcd},\, 
f_{ab},\,\Theta_{(ab)cd},\,\Theta_{g\;ab}^{\phantom{g}g}),
\hskip .5cm \phi = (\phi_{abcd}), \label{v-variable}
\end{equation}
the evolution equations for $v$ are obtained, with our assumptions on the
gauge, from the structural equations of the normal conformal Cartan
connection associated with $g$. They read explicitly
\begin{equation}
\label{IIc0pe}
\partial_{\tau} c^0\,_{ab} = - \chi_{(ab)}\,^{ef}\,c^0\,_{ef} - f_{ab}, 
\end{equation}

\begin{equation}
\label{IIcape}
\partial_{\tau} c^{\alpha}\,_{ab} = - \chi_{(ab)}\,^{ef}\,c^{\alpha}\,_{ef}, 
\,\,\,\,\,\, \alpha = 1, +, -,
\end{equation}

\begin{equation}
\label{IIksipe}
\partial_{\tau} \xi_{abcd} = - \chi_{(ab)}\,^{ef}\,\xi_{efcd}
+ \frac{1}{\sqrt{2}}\,
(\epsilon_{ac}\,\chi_{(bd)ef} + \epsilon_{bd}\,\chi_{(ac)ef})\,f^{ef} 
\end{equation}
\[
- \sqrt{2}\,\chi_{(ab)(c}\,^e \,f_{d)e}
- \frac{1}{2}\,(\epsilon_{ac}\,\Theta_f\,^f\,_{bd} +
\epsilon_{bd}\,\Theta_f\,^f\,_{ac}) - i\,\Theta\,\mu_{abcd},
\]

\begin{equation}
\label{IIfpe}
\partial_{\tau} f_{ab} = - \chi_{(ab)}\,^{ef}\,f_{ef} 
+ \frac{1}{\sqrt{2}}\,\Theta_f\,^f\,_{ab},
\end{equation}

\begin{equation}
\label{IIchipe}
\partial_{\tau} \chi_{(ab)cd} = - \chi_{(ab)}\,^{ef}\,\chi_{efcd}
- \Theta_{(cd)ab} + \Theta\,\eta_{abcd},
\end{equation}

\begin{equation}
\label{IIPpesy}
\partial_{\tau}\Theta_{(ab)cd} = -  \chi_{(cd)}\,^{ef}\,\Theta_{(ab)ef}
- \partial_{\tau}\Theta\,\eta_{abcd} 
+ i\,\sqrt{2}\,d^e\,_{(a}\mu_{b)cde},
\end{equation}

\begin{equation}
\label{IIPpe}
\partial_{\tau}\Theta_g\,^g\,_{ab} = -  \chi_{(ab)}\,^{ef}\,\Theta_g\,^g\,_{ef}
+ \sqrt{2}\,d^{ef}\,\eta_{abef},
\end{equation}
where $\eta_{abcd} = \frac{1}{2}\,(\phi_{abcd} + \phi^{+}_{abcd})$ and
$\mu_{abcd} = - \frac{i}{2}\,(\phi_{abcd} - \phi^{+}_{abcd})$, with
$\tau_a\,^{a'}\tau_b\,^{b'}\tau_c\,^{c'}\tau_d\,^{d'}
\bar{\phi}_{a'b'c'd'}$ $= \phi^{+}_{abcd}$, denote the electric and the
magnetic part of
$\phi_{abcd}$ respectively. These equations are of the form
\begin{equation}
\partial_{\tau}\,v = K(v) + Q(v,v) + L(\phi), \label{v-equations}
\end{equation}
with a linear function $K$ and a quadratic function $Q$ of $v$, both with
constant coefficients, and a linear function $L$ of $\phi$ with
coefficients which depend on the coordinates. We have $L = 0$ on $I'$.
The evolution equations for $\phi$, derived from the Bianchi identities, are
genuine partial differential equations. They will be considered in more detail
below. 

\subsubsection{The initial data}

Consequences of the finite regular initial value problem have been
worked out so far for Cauchy data which are time symmetric and admit a
smooth extension through space-like infinity. In fact, it has been
assumed in \cite{Fr:space-like-infinity}, as will be done in the
following, that the conformal structure is analytic near space-like
infinity. We note that this condition is imposed only for convenience
and could be relaxed. The free Cauchy data on $\tilde{S}$ are then
given by the conformal structure of a smooth metric $h$ on $S$ which
is analytic in some $h-$normal coordinates near $i$.

We assume $h$ to be given near $i$ in a certain conformal gauge, the
{\it cn-gauge} (cf. \cite{Fr:space-like-infinity}). This reduces the
freedom of performing conformal rescalings $h \rightarrow \theta^2\,h$
to the choice of the $4$ real parameters $\theta(i)$,
$\theta_{,a}(i)$, the value of $\theta$ in a neighborhood of $i$ then
being determined by the conformal gauge.  We assume that $B_a$ is a
convex $h-$normal neighborhood of $i$ and that $\rho$ descends to a
radial normal coordinate on $B_a$.

The metric $\tilde{h}$ induced by $\tilde{g}$ on $\tilde{S}$ is
related to $h$ by a rescaling $\tilde{h} = \Omega^{- 2}\,h$, where the
conformal factor $\Omega$ satisfies $\rho\,\Omega^{-\frac{1}{2}}
\rightarrow 1$ as $\rho \rightarrow 0$ and the Lichnerowicz (Yamabe)
equation
\begin{equation}
(D_{\alpha}D^{\alpha}-\frac{1}{8}\,r)(\Omega^{-\frac{1}{2}})=0. 
\label{Yamabe-equation}  
\end{equation}
Here $D$ denotes the covariant derivative and $r$ the Ricci scalar of
$h$. The form (\ref{Omega}) of $\Omega$ in terms of the functions
$U$ and $W$ is a consequence of this equation and the
required asymptotic behavior of $\Omega$, which ensures that
$\tilde{h}$ is asymptotically flat.

The initial data on $C'$ for the conformal field equations are derived
from $h$ and $\Omega$. They are given by  
\begin{equation}
\begin{array}{llll}
c^0\,_{ab} = 0, & c^1\,_{ab} = \rho\,x_{ab}, & 
c^{+}\,_{ab} = z_{ab} + \rho\,\check{c}^{+}\,_{ab}, &
c^{-}\,_{ab} = y_{ab} + \rho\,\check{c}^{-}\,_{ab}, \\[6pt]
\multicolumn{4}{c}{\chi_{(ab)cd}= 0,\;\;
\xi_{abcd} =\sqrt{2}\,\rho\,\check{\gamma}_{abcd},\;\;
f_{ab} = x_{ab},} \\[6pt]
\multicolumn{4}{c}{\Theta_{abcd} = 
-\frac{\rho^2}{\Omega}\,D_{(ab}\,D_{cd)}\Omega 
+\frac{1}{12}\,\rho^2\,r\,h_{abcd},\;\;
\phi_{abcd} = \frac{\rho^3}{\Omega^2}\,(D_{(ab}\,D_{cd)}\Omega 
+\Omega\,s_{abcd}),}  \label{initial-data}
\end{array}
\end{equation}
with $x_{ab}$, $y_{ab}$, $z_{ab}$, and the expression $h_{abcd}$ of the metric
$h$ in space spinor notation as given in appendix
[\ref{spinorial-identities}], and $s_{abcd} = s_{(abcd)}$ the trace free part
of the Ricci tensor of $h$. 

In chapter [\ref{3rd-order-solution}] we shall discuss how
the coefficients $\check{c}^{\pm}\,_{ab}$, $\check{\gamma}_{abcd}$
defining the frame and the connection coefficients are determined on
$C'$ by the ($3-$dimensional) structure equations from $r$ and
$s_{abcd}$. The observation (cf. \cite{Fr:space-like-infinity}) that
the data above extend smoothly to $I^{'0} \subset C'$ is most
important for our construction.

\subsubsection{The transport equations on $I$}\label{transp+asreg}

At first sight it may appear that the initial data on $\tilde{S}$,
thus in particular on $C'$, should be complemented by boundary data on
$I'$ for the solutions of equations (\ref{conformal-field-equations})
to be uniquely determined. However, it turns out that for any smooth
solution to the evolution equations on $M'_a$ which coincides on $C'$
with the initial data above, we have the important relation
\begin{equation}
\label{A1deg}
A^1 = 0 \quad\mbox{on}\quad I'.
\end{equation}
As a consequence, equations (\ref{conformal-field-equations}) reduce
to a symmetric hyperbolic system of the form $\{
A^{0}\,\partial_{\tau} + A^{+}\,X_{+} + A^{-}\,X_{-} \}\,u = C\,u$ on
$I'$ which allows us to determine the unknown $u$ on $I'$ uniquely in
terms of the value of $u$ on $I^{'0}$. Thus we find, as was to be
expected, that any smooth solution of
(\ref{conformal-field-equations}) on $M'_a$ taking on $C'$ our initial
data is determined uniquely by its data on $\tilde{S}$.

More generally, by applying repeatedly the derivative operator
$\partial_{\rho}$ to the evolution equations, restricting to $I'$, and
observing (\ref{A1deg}), we obtain symmetric hyperbolic transport
equations
\begin{equation}
\{A^{0}\,\partial_{\tau} + A^{+}\,X_{+} + A^{-}\,X_{-} \}\,u^p = 
C_p\,u^p + g_p 
\quad\mbox{on}\quad I',\,\,\,\,\,p = 0, 1, 2, \ldots, 
\label{transport-equations}
\end{equation} 
for the quantities $u^p = (\partial^p_{\rho}\,u)|_{I'}$. Here the
matrix-valued function $C_p$ and the vector valued function $g_p$
depend on $p$ and the quantities $u^0, \ldots, u^{p - 1}$, but the
matrices $A^{0}$, $A^{\pm}$ are universal in the sense that they
depend neither on $p$ nor on the initial data. We shall
employ the notation above more generally, such that applying it to the 
fields $s_{abcd}$ and $r$ on the Cauchy hypersurface we have 
$s^p_{abcd}=(\partial^p_{\rho}\,s_{abcd})|_{I^{'0}}$
and $r^p=(\partial^p_{\rho}\,r)|_{I^{'0}}$, respectively. 

To integrate the transport equations (\ref{transport-equations}) on
$I'$, we expand all fields in terms of the matrix elements of unitary
representations of $SU(2)$ which are given, in terms of the matrix
elements $(t^a\,_b)_{a, b = 0, 1}$ of the $2-$dimensional standard
representation of $t \in SU(2)$, by the complex-valued functions
\begin{equation}
\begin{array}{c}
SU(2, C) \ni t \rightarrow T_m\,^j\,_k(t) 
= {\binom{m}{j}}^{\frac{1}{2}}\,{\binom{m}{k}}^{\frac{1}{2}}\,
t^{(b_1}\,_{(a_1} \ldots t^{b_m)_j}\,_{a_m)_k},\,\,\,\,\,T_0\,^0\,_0(t) =
1, \\[6pt]
j,k = 0, \ldots ,m,\,\,\,\,\,m = 1,2,3, \ldots \label{T-functions}
\end{array}
\end{equation}
Here, as in the following, setting a string of indices into brackets with 
a lower index $k$ is meant to indicate that the indices are symmetrized and
then $k$ of them are set equal to 1 while the remaining ones are set equal
to 0. The functions $\sqrt{m + 1}\,T_m\,^j\,_k(t)$ form a complete
orthonormal set in the Hilbert space  $L^2(\mu,SU(2))$ where $\mu$
denotes the normalized Haar-measure on $SU(2)$. Under complex conjugation 
we have  
\begin{displaymath}
\overline{T_m\,^j\,_k(t)} = (-1)^{j + k}\,T_m\,^{m - j}\,_{m -
k}(t),\,\,\,\,\,
t \in SU(2),
\end{displaymath}
and, for $0 \le k, j \le m$, $m = 0, 1, 2, \ldots$, we have with  
$\beta_{m,j} = \{ j\,(m - j + 1) \}^{\frac{1}{2}}$ 
\begin{equation}
X\,T_m\,^k\,_j = (m - 2j)\,T_m\,^k\,_j,\,\,
X_{+}\,T_m\,^k\,_j = \beta_{m,j}\,T_m\,^k\,_{j - 1},\,\,
X_{-}\,T_m\,^k\,_j = - \beta_{m,j + 1}\,T_m\,^k\,_{j + 1}.
\label{Xact}
\end{equation}
A function $f$ satisfying a relation $X f= 2 s f$ with an integer or
half integer number $s$, is said to have spin weight $s$. We note the
spin raising (lowering) property of the action of $X_{\pm}$ on such
functions implied by (\ref{X-commutators}), i.e. $X\,X_{\pm}\,f= 2\,(s
\pm 1)\,X_{\pm}\,f$. By construction of the manifold $M'_a$ any function
occuring in our formalism has a well defined spin weight. This leads to a
simplification of the expansion in terms of the functions $T_{m\;j}^{\;\;k}$.
The general form of these expansions has been discussed in detail in
\cite{Fr:space-like-infinity} and will be assumed here without further
explanation.

The quantities $u^0$, $u^1$, $u^2$ have been determined
in \cite{Fr:space-like-infinity}. They are given here (with a
correction and a useful change of notation) at the beginning of
chapter [\ref{3rd-order-solution}]. The functions $u^3$
will be calculated in chapter [\ref{3rd-order-solution}]. The
quantities $u^p$, $p=2,3,\ldots$ have been shown (cf.
\cite{Fr:space-like-infinity}) to develop a certain type of
logarithmic singularity on the sets $I^{'\pm}$ unless the free datum
$h$ on $S$ satisfies the {\it asymptotic regularity condition}
\begin{equation}
D_{(a_q b_q} \ldots D_{a_1 b_1}\,b_{abcd)}(i) = 0,
\label{asregcond}
\end{equation}
for $q = 0,1,2,\ldots$, where the spinor field $b_{abcd} = b_{(abcd)}$
represents the Cotton tensor of $h$. The values of the functions $u^p$, 
$p \le 3$, which will be given below, have been calculated on $I'$ under the
assumption that (\ref{asregcond}) is satisfied for  $q \le 1$. The
analysis of the quantities $u^p$, to the extent to which it has been
carried out in \cite{Fr:space-like-infinity}, indicates another
potential source for a singular behavior of the fields $u^p$, $p \ge
3$, at $I^{'\pm}$. This will be discussed further in
chapter [\ref{3rd-order-solution}].

\subsection{The NP-gauge}\label{NP-gauge-conditions}

For simplicity we restrict our discussions now to the future of
$\tilde{S}$ in $M$, we refer to future null infinity simply as to null
infinity and we denote it by ${\cal J}$. In the following we shall
describe a certain class of gauge conditions on $(M,g)$ near null
infinity, referred to as the {\it{NP-gauge}}, which comprise certain
requirements on the conformal gauge, certain coordinates, and a
certain orthonormal frame field. Though this gauge is known, our
description will be quite detailed, because we will have to refer to
it later. The Levi-Civita connection induced by the conformal metric
$g$ will be denoted by $\nabla$.

Suppose $\{E^{\circ}_{aa'}\}$ is a smooth frame field, satisfying
$g(E^{\circ}_{aa'},E^{\circ}_{bb'})=\epsilon_{ab}\,\bar{\epsilon}_{a'b'}$,
which is defined in a neighborhood of null infinity. We call it an ``adapted
frame'', if it satisfies the following conditions. The vector field
$E^{\circ}_{11'}$ is tangent to and parallel propagated along null
infinity. On the neighborhood on which the frame is given there is exists a
smooth function $u^{\circ}$ which induces an affine parameter on the null
generators of ${\cal J}$ such that $E^{\circ}_{11'}\,(u^{\circ}) = 1$,
which is constant on null hypersurfaces transverse to ${\cal J}$, and
which satisfies $E^{\circ\,\alpha}_{00'} =
g^{\alpha\beta}\nabla_{\beta}u^{\circ}$.  Thus $E^{\circ}_{00'}$ is
tangent to the hypersurfaces $\{u^{\circ} = const.\}$ and geodesic.
The fields $E^{\circ}_{11'}$, $E^{\circ}_{00'}$ as well as the fields
$E^{\circ}_{01'}$, $E^{\circ}_{10'}$, which are necessarily tangent to
the slices $\{u^{\circ} = const.\} \cap {\cal J}$, are parallelly
propagated in the direction of $E^{\circ}_{00'}$.

In terms of its NP-spin-coefficients (note the slight difference of
our notation with that of \cite{NP:spin-coefficients})
\begin{equation}
\Gamma^{\circ}_{aa'bc}=\frac{1}{2}\left\{E^{\circ\,\alpha}_{aa'} 
E^{\circ\,\beta}_{b1'}\nabla_{\alpha}E^{\circ}_{c0'\beta}
+E^{\circ\,\alpha}_{aa'} 
E^{\circ\,\beta}_{c1'}\nabla_{\alpha}E^{\circ}_{b0'\beta} \right\},
\label{spin-coefficients} 
\end{equation}
an adapted frame is characterized by the properties 
\begin{equation}
\begin{array}{lllll}
\Gamma^{\circ}_{10'11} =0, & \Gamma^{\circ}_{11'11} =0 
\quad\mbox{on}\quad {\cal J}, & & & \\[6pt]
\Gamma^{\circ}_{10'00 } = \bar{\Gamma}^{\circ}_{01'0'0'}, &
\Gamma^{\circ}_{11'00} = 
\bar{\Gamma}^{\circ}_{01'0'1'}+\Gamma^{\circ}_{01'01}, &
\Gamma^{\circ}_{00'ab} =0, & a,b = 0,1, & 
\quad\mbox{near}\quad {\cal J}. \label{Adapted-frame}
\end{array}
\end{equation}
The first of these conditions tells us that ${\cal J}$ is shear free.
This well known fact follows from the equation for the trace free part
$s_{\alpha\beta}$ of the Ricci tensor of the conformal vacuum metric
$g$,
\begin{equation}
\Theta\,s_{\alpha\beta}=
\frac{1}{2}g_{\alpha\beta}\nabla_{\gamma}\nabla^{\gamma} 
\Theta -2\nabla_{\alpha}\nabla_{\beta}\Theta.
\label{eq:trf-Ricci-spinor1} 
\end{equation}
Transvection with $E^{\circ\;\alpha}_{10'}E^{\circ\;\beta}_{10'}$ and
restriction to ${\cal J}$ gives
$\Gamma^{\circ}_{10'11}\,E^{\circ}_{00'}(\Theta) = 0$, while
$E^{\circ}_{00'}(\Theta) \neq 0$ on ${\cal J}$.
We shall combine now the construction of an adapted frame with the
freedom to perform rescalings
\begin{equation}
g \rightarrow g^{\star} = \theta^2g,\,\,\,\,\,\, 
\Theta \rightarrow \Theta^{\star} = \theta\,\Theta \label{conf-scale}
\end{equation}
with some positive function $\theta$, to obtain another adapted frame
$\{E^{\bullet}_{aa'}\}$ for which we get further simplifications besides
(\ref{Adapted-frame}).  We start with an adapted frame
$\{E^{\circ}_{aa'}\}$ as described above. For arbitrary $\theta > 0$ and
for arbitrary function $p > 0$ which is constant on the generators of
${\cal J}$ we set
\begin{equation}
E^{\bullet}_{11'} = \theta^{-2}\,p\,E^{\circ}_{11'}
\quad\mbox{and}\quad
u^{\bullet}(u^{\circ})=\int^{u^{\circ}}_{u^{\circ}_*}
\theta^2(u')\,p^{-1}(u')\,du' +u^{\bullet}_*  
\quad\mbox{on}\quad {\cal J}, \label{affine-parameters}
\end{equation}
where the integration is performed along the generators of ${\cal J}$.
Then $E^{\bullet}_{11'}$ will be parallelly propagated and 
$E^{\bullet}_{11'}(u^{\bullet}) = 1$ will hold. We assume that 
$u^{\circ} = u^{\circ}_*$ and $u^{\bullet} = u^{\bullet}_*$ on ${\cal C}$
and set
\begin{equation}
\begin{array}{ccccc}
E^{\bullet}_{00'}= p^{-1}\,E^{\circ}_{00'}, & 
E^{\bullet}_{11'}=\theta^{-2}\,p\,E^{\circ}_{11'}, &
E^{\bullet}_{01'}=\theta^{-1}E^{\circ}_{01'} 
\quad\mbox{on}\quad {\cal C}.
\label{frame-on-C}
\end{array}
\end{equation}

Since ${\cal C}$ is diffeomorphic to $S^2$ and thus carries (up to 
diffeomorphisms) precisely one Riemannian conformal structure, 
we can fix coordinates $x^3 = \vartheta$, $x^4 = \varphi$ as well as 
the function $\theta$ on ${\cal C}$ 
such that the metric $h^{\star}$ induced by $g^{\star}$ on
$\cal{C}$ is given by the standard $S^2-$metric
$h^{\star} = d\,\vartheta^2 + \sin^2\vartheta\,d\,\varphi^2$. 
Using the transformation laws 
$\Gamma^{\bullet}_{10'00}= p^{-1}\,\bigl[\Gamma^{\circ}_{10'00}
-E^{\circ}_{00'}(\rm{log}\,\theta)\bigr]$ and
$\Gamma^{\bullet}_{01'11}= p\,\theta^{-2}\,\bigl[\Gamma^{\circ}_{01'11}
+E^{\circ}_{11'}({\rm log}\,\theta)\bigr]$ on $\cal{C}$, we
can achieve, by suitable choice of $d\,\theta$ and $p$,
\begin{equation}
\Gamma^{\bullet}_{10'00} = 0,\,\,\,\,\, 
\Gamma^{\bullet}_{01'11} = 0,\,\,\,\,\, 
E^{\bullet}_{00'}(\Theta^{\star}) = const. \neq 0
\quad\mbox{on}\quad {\cal C}. \label{romu=0onC}
\end{equation} 
The transformation $s^{\star}_{\alpha\beta}=
-\frac{2}{\theta}\left\{(\nabla_{\alpha}\nabla_{\beta}\theta
  -\frac{2}{\theta}\nabla_{\alpha}\theta\nabla_{\beta}\theta)
  -\frac{1}{4}g_{\alpha\beta} (\nabla_{\gamma}\nabla^{\gamma}\theta
  -\frac{2}{\theta}\nabla_{\gamma}\theta\nabla^{\gamma}\theta)
\right\} +s_{\alpha\beta}$ of the trace free part $s_{\alpha \beta}$
of the Ricci tensor under the rescaling (\ref{conf-scale}) implies a
transformation of $\Phi_{22} = \frac{1}{2}s_{\alpha\beta}
E^{\circ\,\alpha}_{11'}\,E^{\circ\,\beta}_{11'}$ into
$\Phi^{\star}_{22} = \frac{1}{2}s^{\star}_{\alpha\beta}
E^{\bullet\,\alpha}_{11'}\,E^{\bullet\,\beta}_{11'}$, which yields,
with the assumption that $\Phi^{\star}_{22} = 0$ on ${\cal J}$, on the
generators of ${\cal J}$ the ODE
\begin{equation}
E^{\circ}_{11'}\bigl(E^{\circ}_{11'}(\theta)\bigr)
-\frac{2}{\theta}\bigl(E^{\circ}_{11'}(\theta)\bigr)^2-\theta\,\Phi_{22} = 0. 
\label{eq:conf.factor}
\end{equation}
This equation can be rewritten as a linear ODE for $\theta^{-1}$ which
can be solved on the generators of ${\cal J}$ with $\theta > 0$. Using
the initial data $\theta$, $E^{\circ}_{11'}(\theta)$ on ${\cal C}$
determined above, we solve for $\theta$ to obtain
\begin{equation} 
\Phi^{\star}_{22} = 0,\,\,\,\,\, 
\Gamma^{\bullet}_{01'11} = 0
\quad\mbox{on}\quad {\cal J}.   \label{phi22=0}
\end{equation}
Here the second equation is a consequence of the first, the field
equations, and (\ref{romu=0onC}). We assume in the following
(\ref{affine-parameters}). We observe that the induced metric on the
sections $\{u^{\bullet} = const.\}$ is given as a consequence 
everywhere on ${\cal J}$ by the $S^2$-standard metric.

Once $\theta$ and $E^{\bullet}_{11'}$ have been fixed on ${\cal J}$, the
vector field $E^{\bullet}_{01'}$ (whence $E^{\bullet}_{10'}$) tangent to
$\{u^{\bullet} = const.\}$ is determined up to rotations. We choose
some smooth field $E^{\bullet}_{01'}$ on ${\cal J}$, solve the equation 
\begin{equation}
E^{\bullet}_{11'}(c) = -i\,E^{\bullet\,\alpha}_{10'}\, 
E^{\bullet\,\beta}_{11'}\,\nabla^{\star}_{\beta}
\,E^{\bullet}_{01'\alpha}
\label{ode:phase}
\end{equation}
for the function $c$ with initial value $c = 0$ on ${\cal C}$, and replace
$E^{\bullet}_{01'}$ by $e^{i\,c}\,E^{\bullet}_{01'}$ to achieve
\begin{equation}
\label{gamma=0}
\Gamma^{\bullet}_{11'01} = 0
\quad\mbox{on}\quad {\cal J}.
\end{equation}  
Observing the simplifications above, we contract the analogue of
(\ref{eq:trf-Ricci-spinor1}) for $g^{\star}$ with
$E^{\bullet\,\alpha}_{01'}E^{\bullet\,\beta}_{10'}$ to conclude that
$\nabla^{\star}_{\,\alpha}\nabla^{\star\,\alpha}\Theta^{\star}=0$ on
${\cal J}$.  A further contraction with
$E^{\bullet\,\alpha}_{00'}E^{\bullet\,\beta}_{11'}$ gives
\begin{equation}
E^{\bullet}_{11'}\bigl(E^{\bullet}_{00'}(\Theta^{\star})\bigr)=0,
\quad\mbox{i.e.}\quad E^{\bullet}_{00'}(\Theta^{\star}) = const.
\quad\mbox{on}\quad {\cal J}, 
\label{nl(theta)}
\end{equation}
while a contraction with
$E^{\bullet\,\alpha}_{00'}E^{\bullet\,\beta}_{01'}$ yields now
$E^{\bullet}_{01'}\bigl(E^{\bullet}_{00'}(\Theta^{\star})\bigr) =
\Gamma^{\bullet}_{11'00}E^{\bullet}_{00'}(\Theta^{\star})$,
which implies
\begin{equation}
\Gamma^{\bullet}_{11'00} = 0
\quad\mbox{on}\quad {\cal J}. \label{tau=0}
\end{equation}
To fix also $d\,\theta$ on ${\cal J}$, we use the conformal
transformation law for the Ricci scalar, i.e.
\begin{equation}
R[g^{\star}]= \frac{1}{\theta^2}R[g]+
\frac{12}{\theta^2}\nabla^{\star}_{\,\alpha}\,\theta
\nabla^{\star\,\alpha}\,\theta
-\frac{6}{\theta}\nabla^{\star}_{\,\alpha}\nabla^{\star\,\alpha}\,\theta.
\label{eq:curv.scalar}
\end{equation}
If we require that $R[g^{\star}] = 0$ along ${\cal J}$, this equation takes 
on the generators of the null hypersurface ${\cal J}$ the form    
\begin{equation}
E^{\bullet}_{11'}\bigl(E^{\bullet}_{00'}(\theta)\bigr) 
-\frac{2}{\theta}E^{\bullet}_{11'}(\theta)E^{\bullet}_{00'}(\theta)= 
F^{\star},
\label{ode:tr.derivative}
\end{equation}
of a linear ODE for the unknown $E^{\bullet}_{00'}(\theta)$, where the
right hand side 
\begin{equation*}
F^{\star}= {\mathfrak{Re}\,} 
\left\{ E^{\bullet}_{01'}(E^{\bullet}_{10'}(\theta))
-2\,\Gamma^{\bullet}_{01'01}E^{\bullet}_{10'}(\theta)
-\frac{2}{\theta}E^{\bullet}_{01'}(\theta)E^{\bullet}_{10'}(\theta) 
+\frac{1}{12\,\theta}R[g] \right\}
\end{equation*}
is given in terms of quantities which have been determined already on
${\cal J}$.  Using the initial value 
$E^{\bullet}_{00'}(\theta)= p^{-1}\theta\,\Gamma^{\circ}_{10'00}|_{\cal{C}}$,
fixed on ${\cal C}$ by (\ref{romu=0onC}), we can integrate
the equation to achieve
\begin{equation}
R[g^{\star}] = 0,\,\,\,\,\, \Gamma^{\bullet}_{10'00} = 0 
\quad\mbox{on}\quad {\cal J}, \label{Rstar=0}
\end{equation}
where the second equation follows again from our previous results and
the field equations.

We do not require conditions of higher order on the conformal gauge.
Assuming a conformal gauge as described here, we shall refer to an
adapted frame $\{E^{\bullet}_{aa'}\}$ satisfying the conditions above
as to an {\it NP-frame}, and to a normalized spin frame
$\varepsilon_a^{\bullet\,A}
\equiv\{o^{\bullet\,A},\iota^{\bullet\,A}\}$ which implies a NP-frame
as to a {\it NP-spin-frame}.

We extend the coordinates $x^3$, $x^4$ to ${\cal J}$ such that they
are constant on the null generators of ${\cal J}$. As described above,
we define null hypersurfaces $\{u^{\bullet} = const.\}$ transverse to
${\cal J}$ and we denote by $r^{\bullet}$ the affine parameter on the
null generators of these hypersurfaces which satisfies
$E^{\bullet}_{00'}(r^{\bullet}) = 1$ and, on ${\cal J}$, $r^{\bullet}
= 0$. The coordinates $x^3$, $x^4$ are extended such that they are
constant on the null generators of $\{u^{\bullet} = const.\}$. Thus we
get a {\it{Bondi-type system}} $(u^{\bullet},r^{\bullet},x^3,x^4)$ in
some neighborhood of null infinity. Occasionally 
we shall change from the coordinates $\vartheta$, $\varphi$, to a
complex stereo-graphical coordinate given by
$\zeta= e^{i\varphi}ctg{\frac{\vartheta}{2}}$. 
We write the volume element and the volume form alternatively
\begin{equation*}
ds^2= -(d\vartheta^2+sin^2\vartheta\,d\varphi^2)= 
-P(\zeta)^{-2}d\zeta\,d\bar\zeta,\,\,\,\,\,\,\,\, 
\epsilon= {\rm{sin}}\vartheta\,d\vartheta\wedge d\varphi
=[2P(\zeta)]^{-2}d\zeta \wedge d\bar\zeta,  
\end{equation*}
where we set $P(\zeta)=\frac{1}{2}(1+\zeta\bar\zeta)$. We shall refer
to the conditions on the conformal scaling, the frame field, and the
coordinates as to the {\it{NP-gauge}}.

\subsection{Relating the NP-gauge to the F-gauge}\label{FNP-gauge-tr.}

While the NP-gauge is hinged on null infinity, the F-gauge is based on
a Cauchy hypersurface and these gauge conditions are in general
completely different. In the following we will study the
transformation which relates one to the other.  It is important for
this that the conformal factor $\Theta$, whence ${\cal J}$, is known
explicitly in the F-gauge.

The vector fields $\{c_{aa'}\}$ tangent to the $5$-dimensional bundle
space $M'_a$ are not directly related to the NP-gauge
on the subset $M_a \setminus I$ of $M$. Let $S^2
\supset U \ni p \stackrel{s}{\rightarrow} s(p) \in SU(2)$ be a smooth
local section, defined on some open subset $U$ of $S^2$, of the Hopf
fibration $SU(2) \rightarrow SU(2)/U(1) \simeq S^2$. It induces a smooth
section $U \times \mathbb{R} \times \mathbb{R} \ni (p, \tau, \rho)
\stackrel{S}{\rightarrow} (s(p), \tau, \rho) \in M'_a$. We denote the
image of $S$ by $M^*_a$.  The vector fields tangent to $s(U)$ which
have projection identical to that of $X_{\pm}$ are of the form
$X_{\pm} + a_{\pm}\,X$ with some smooth functions $a_{\pm}$ on $s(U)$,
satisfying $a_- = - \bar{a}_+$.  Because of (\ref{X-commutators})
$a_{\pm}$ cannot vanish on open subsets of $s(U)$.  Consequently, the
tangent vector fields $c^*_{aa'}$ of $M^*_a$ satisfying
$\pi_{*}(c^*_{aa'}) = \pi_{*}(c_{aa'})$ are given on $M^*_a$ by
\begin{equation*}
c^*_{aa'} = c_{aa'} + (a_+\,c^+\,_{aa'} + a_-\,c^-\,_{aa'})\,X, 
\end{equation*}
with functions $a_{\pm}$ which are independent of $\tau$ and $\rho$. 
The connection coefficients defined on $M^*_a$ by the connection form
$\omega^b\,_c$ and the vector fields $c^*_{aa'}$ are given by 
\begin{equation*}
\Gamma^*_{aa'}\,^b\,_c = \Gamma_{aa'}\,^b\,_c +
(a_+\,c^+\,_{aa'} + a_-\,c^-\,_{aa'})\,
(\epsilon_0\,^b\,\epsilon_c\,^0 - \epsilon_1\,^b\,\epsilon_c\,^1).
\end{equation*}  
In the remaining part of this section we shall work on $\pi(M'_a)$ and
denote the projection of the vector fields $c^*_{aa'}$, which define a
smooth orthonormal frame field on $\pi(M^*_a \setminus I')$,
and the pull-back of $\Gamma^*_{aa'}\,^b\,_c$ by $S$ again by $c^*_{aa'}$ and
$\Gamma^*_{aa'}\,^b\,_c$. Similarly, the projections of ${\cal J}' \cap
M^*_a$ and $I^{'+} \cap M^*_a$ will be denoted by ${\cal J}$ and $I^+$.  

The frame field $\{c^*_{aa'}\}$, which is in general not adapted to null
infinity, will now be related close to $I^+$ to an adapted frame
$\{E^{\circ}_{aa'}\}$.  On ${\cal J}$ the vector field $E^{\circ}_{11'}$ must
be of the form
\begin{equation}
E^{\circ\,\alpha}_{11'}=f\nabla^{\alpha}\Theta, 
\label{n-vector}
\end{equation}
where $\nabla$ and $\Theta$ denote the Levi-Civita connection and the
conformal factor associated with the F-gauge. The requirement $0 =
E^{\circ\,\beta}_{11'}\nabla_{\beta}E^{\circ\,\alpha}_{11'} =
f\,\nabla^{\beta}\,\Theta\,\nabla_{\beta}\,f\,\nabla^{\alpha}\Theta +
f^2\nabla_{\beta}\,(\frac{1}{2}\,
\nabla_{\alpha}\Theta\,\nabla^{\alpha}\Theta)$ that
$E^{\circ\,\alpha}_{11'}$ be parallelly propagated, gives after
contraction with a vector field $Z$ transverse to ${\cal J}$ the ODE
\begin{equation}
\nabla^{\alpha}\Theta\nabla_{\alpha}({\rm{log}}\,f)=
-\frac{Z(\frac{1}{2}\nabla_{\beta}\Theta 
\nabla^{\beta}\Theta)}{Z(\Theta)}
\label{f-equation}
\end{equation}
for $f$ on the generators of ${\cal J}$. To fix $f$, we set 
$f = f_0 = const. > 0$ on some section ${\cal C}$ of ${\cal J}$. The function
$u^{\circ}$ satisfying  $E^{\circ\,\alpha}_{11'}(u^{\circ}) = 1$ on ${\cal J}$ 
and $u^{\circ} = u^{\circ}_*$ on ${\cal C}$ can be now be determined.
   
Let $\lambda^a_{\phantom{a}b}\,\in\,SL(2,C)$ satisfy
\begin{equation}
E^{\circ}_{aa'} = \lambda^b_{\phantom{b}a} 
\bar{\lambda}{}^{b'}_{\phantom{b'}a'}\,c^*_{bb'}
\label{Lorentz-transf.-1}.
\end{equation}
Rewriting (\ref{n-vector}) in the form $E^{\circ}_{11'}=
f\,c^*_{bb'}(\Theta)\epsilon^{ab}\bar{\epsilon}^{a'b'}c^*_{aa'}$,
we find the relations
\begin{equation}
\begin{array}{ccc}
\lambda^0_{\phantom{0}1}\bar{\lambda}{}^{0'}_{\phantom{0'}1'}= 
f\,c^*_{11'}(\Theta), & 
\lambda^0_{\phantom{0}1}\bar{\lambda}{}^{1'}_{\phantom{1'}1'}= 
-f\,c^*_{10'}(\Theta), & 
\lambda^1_{\phantom{1}1}\bar{\lambda}{}^{1'}_{\phantom{1'}1'}= 
f\,c^*_{00'}(\Theta).
\label{matrix-elements1} 
\end{array}
\end{equation}
From (\ref{Lorentz-transf.-1}) we obtain 
$\lambda^0_{\phantom{0}1}E^{\circ}_{01'}=
\lambda^0_{\phantom{0}0}E^{\circ}_{11'}
-\bar{\lambda}{}^{0'}_{\phantom{0'}1'}c^*_{10'}
-\bar{\lambda}{}^{1'}_{\phantom{1'}1'}c^*_{11'}$. Applying this to the
function $u^{\circ}$, we get
\begin{equation}
\lambda^0_{\phantom{0}0}= 
\bar{\lambda}{}^{0'}_{\phantom{0'}1'}c^*_{10'}(u^{\circ})
+\bar{\lambda}{}^{1'}_{\phantom{1'}1'}c^*_{11'}(u^{\circ}). 
\label{lambda00}
\end{equation}
Together with the condition ${\rm{det}}(\lambda^a_{\phantom{a}b})=1$
the relations (\ref{matrix-elements1}), (\ref{lambda00}) allow us to
determine the matrix elements $\lambda^a_{\phantom{a}b}$
on ${\cal J}$ up to replacements $\lambda^a_{\phantom{a}b}
\rightarrow \lambda^a_{\phantom{a}b}\,\eta^b_{\phantom{b}c}$ with
$(\eta^a\,_b) = diag(e^{i\,\alpha},e^{- i\,\alpha}) \in U(1)$.
After making here an arbitrary choice, the adapted frame
$\{E^{\circ}_{aa'}\}$ is determined uniquely near ${\cal J}$.

To determine an NP-frame $\{E^{\bullet}_{aa'}\}$ near ${\cal J}$, we
need to find an appropriate rescaling (\ref{conf-scale}) and a
scaling factor $p$. We set 
\begin{equation}
c^{\star}_{aa'} =\theta^{-1}\,c^*_{aa'},\,\,\,\,\,\,\,\,
E^{\bullet}_{aa'}= \Lambda^b_{\phantom{b}a}
\bar{\Lambda}{}^{b'}_{\phantom{b'}a'}\,c^{\star}_{bb'} 
\label{Lorentz-transf.-2}
\end{equation}
with $\Lambda^a_{\phantom{a}b}\,\in\,SL(2,C)$.
Assuming (\ref{affine-parameters}), we have   
$E^{\bullet\,\alpha}_{11'}=f^{\star}\,\nabla^{\star\,\alpha}\Theta^{\star}$ 
with 
\begin{equation}
f^{\star}=\frac{f\,p}{\theta} 
\quad\mbox{and}\quad 
E^{\bullet}_{00'}(\Theta^{\star})=\frac{1}{f^{\star}}
\quad\mbox{on}\quad {\cal J}. 
\label{f-conf.tr.}
\end{equation}
We choose now $\theta$, $d\,\theta$, and coordinates $x^3$, $x^4$ 
such that the induced metric on ${\cal C}$ is given by the 
$S^2$-standard metric and, with $p$ chosen such that $p = \theta$ on 
${\cal C}$, conditions (\ref{romu=0onC}) are satisfied with
$E^{\bullet}_{00'}(\Theta^{\star})=f_0^{-1}$. 

Following the procedure of the previous section, we can determine the
conformal factor $\theta$ on ${\cal J}$ such that (\ref{phi22=0}) is
satisfied. The transformation $\Lambda^a_{\phantom{a}b}$ can be
determined in the same way as $\lambda^a_{\phantom{a}b}$. Imposing
condition (\ref{gamma=0}), we determine $\Lambda^a_{\phantom{a}b}$ up
to $U(1)-$transformations on ${\cal C}$. Conditions (\ref{nl(theta)}),
(\ref{tau=0}) will now be satisfied as well and we can determine
$d\,\theta$ on ${\cal J}$ such that (\ref{Rstar=0}) holds. Extending
the tetrad to a neighborhood of ${\cal J}$ such that it is parallelly
propagated in the direction of $E^{\bullet}_{00'}$, we get the desired
NP-frame.

In our later calculations we will need the quantities 
$E^{\bullet}_{00'}(\Lambda^a_{\phantom{a}b})$.
Using our gauge condition $\Gamma^{\bullet}_{00'ab} = 0$ 
and the transformation laws for the connection coefficients,
\begin{equation*}
\Gamma^{\star}_{aa'bc}=\frac{1}{\theta}\left\{\Gamma^*_{aa'bc}
+\epsilon_{a(b}c^*_{c)a'}({\rm{log}}\,\theta)\right\},
\end{equation*}
\begin{equation*}
E^{\bullet}_{aa'}(\Lambda^b\,_c) = - \Lambda^f\,_a\,\bar{\Lambda}^{f'}\,_{a'}
\,\Lambda^h\,_c\,\Gamma^{\star}_{ff'}\,^b\,_h
+ \Lambda^b\,_d\,\Gamma^{\bullet}_{aa'}\,^d\,_c, 
\end{equation*}

where $\Gamma^{\star}_{aa'bc}$ denotes the connection
coefficients with respect to $\nabla^{\star}$ and
$\{c^{\star}_{aa'}\}$, we find
\begin{equation}
\label{transversal-derivatives}
E^{\bullet}_{00'}(\Lambda^b\,_c) = - \Lambda^f\,_0\,\bar{\Lambda}^{f'}\,_{0'}
\,\Lambda^h\,_c\,\Gamma^{\star}_{ff'}\,^b\,_h.
\end{equation}

In the considerations above we had to fix various quantities by
prescribing data on the section ${\cal C}$. When we shall determine
later the expansion of a NP-frame near $I^+$, it will be natural to
try pushing ${\cal C}$ to $I^+$. A priori it is not clear, however,
whether this can be done in a continuous way. We shall see, that for
certain quantities the limits to $I^+$ do exist, while others
quantities can only described in terms of their growth behavior near
$I^+$.

\sect{The NP-constants}

In 1965 Newman and Penrose discovered certain non-trivial quantities,
defined by certain integrals over a 2-dimensional cross-section of
${\cal J}^+$, which are absolutely conserved in the sense that their
values do not depend on the choice of the section
(cf. \cite{NP:NPQ-letter,NP:NPQ}).  The interpretation of these ten real
{\it NP-constants} is still open. In the case where the
space-time admits a smooth conformal extension containing a point
$i^+$ (``future time-like infinity'') whose past light cone represents
${\cal J}^+$, these constants are essentially given by the five
complex components of the rescaled conformal Weyl spinor
(cf. \cite{NP:NPQ,FrS}). However, these quantities do not
allow us a simple interpretation either. More interesting is the case
of stationary vacuum space-times. In this case the constants have been
calculated and have been given in the form
$(mass)\times(quadrupole\;moment)-(dipole\;moment)^2$ 
(cf. \cite{NP:NPQ,P:conserved-quantities-survay}).

If the evolution of the field in time is not given explicitly as in the
presence of a time-like Killing vector field, there appears to be no obvious
way to calculate the NP-constants. It turns out, however, that under
suitable assumptions on the asymptotic behavior of the field near
space-like infinity the constants can be calculated by integrating the
transport equations on $I'$ to a sufficiently high order. In the following
we shall derive a formula for the constants in terms of quantities which
can be determined by solving the transport equations. 

To explain the original formula (cf. \cite{NP:NPQ}),
which is given in the Bondi-Sachs-Newman-Penrose framework, let
$(u,r,\vartheta,\varphi)$ denote Bondi-coordinates on the physical
space-time, where $r$ denotes an affine parameter along the generators
of the null hypersurfaces $\{u=const.\}$ and the generators are
labeled by the standard coordinates $(\vartheta,\varphi)$ on the
two-sphere. The null frame $\{\tilde{E}_{aa'}\}$ as well as a
corresponding spinor dyad 
$\{\tilde{o}^A,\tilde{\iota}^A\}$, both defined on the
physical space-time, are normalized with respect to the physical
metric $\tilde{g}$. They are adapted to the Bondi-coordinates such
that $\tilde{E}_{00'} = \partial_r$.

We assume that the conformal space-time with metric
$g^{\star}:=r^{-2}\tilde{g}$ admits a smooth extension as 
$r \rightarrow \infty$ to a smooth Lorentz space with boundary 
${\cal J}^+ = \{r^{\bullet} = 0\}$ and that the functions 
$u^{\bullet}:=u$, $r^{\bullet}:=r^{-1}$, $\vartheta$, and
$\varphi$ extend such as to define a smooth system of
Bondi-type coordinates near ${\cal J}^+$. Furthermore, we assume that
the frame $\{E^{\bullet}_{aa'}\}$ and the spinor dyad 
$\{o^{\bullet\,A},\iota^{\bullet\,A}\}$, defined by 
\begin{equation}
\begin{array}{cc}
\multicolumn{2}{c}{E^{\bullet}_{aa'}=r^{2 - a - a'}\tilde{E}_{aa'}} \\[6pt]
o^{\bullet\,A}=r\,\tilde{o}^A, & \iota^{\bullet\,A}=\tilde{\iota}^A, 
\label{frame-dyad-conformal-tr.}
\end{array}
\end{equation}
such that they are normalized with respect to $g^{\star}$,
extend to smooth frame resp. dyad near ${\cal J}^+$. The results of
Newman and Unti (cf. \cite{NU}) then imply that $\{E^{\bullet}_{aa'}\}$
defines in fact a NP-frame.

Under our assumptions the component 
$\psi_0 =\psi_{ABCD}\tilde{o}^A\tilde{o}^B\tilde{o}^C\tilde{o}^D$
of the conformal Weyl spinor has an expansion 
$\psi_0=\psi^0_0r^{-5}+\psi^1_0r^{-6}+O(r^{-7})$ with coefficients 
$\psi^p_0$ which are independent of $r$. In terms of the physical
space-time the NP-constants are given with this notation by the integrals
\begin{equation}
G_m=
\oint{_2\bar{Y}_{2,m}}\psi^1_0\,{\rm{sin}}\vartheta\,d\vartheta\,d\varphi,
\label{traditional-definition}
\end{equation}
which are calculated for fixed value of $u$. The functions
${_2Y_{2,m}}$, $m = -2, -1, 0, 1, 2 $, denote spin-2 spherical
harmonics (cf. \cite{G:edth-operator}) which are obtained from the
standard spherical harmonics by
\begin{equation}
{_2Y_{2,m}}= \frac{1}{2\sqrt{6}}E^{\bullet\,\alpha}_{01'}
E^{\bullet\,\beta}_{01'}\delta_{\alpha}\delta_{\beta}Y_{2,m}= 
\frac{1}{2\sqrt{6}}{\textrm{\dh}}^2Y_{2,m}.
\label{spin-2-harmonics}
\end{equation}
Here $\delta$ and \dh\; denote the standard covariant differential
operator on the unit $2$-sphere and the ``edth''-operator,
respectively. In evaluating (\ref{traditional-definition}), it will 
be important that the operator \dh\; is defined with
respect to the complex null vector field $E^{\bullet}_{01'}$ (cf.  
\cite{Ko-Newman}).

We reexpress the constants in terms of the fields
$g^{\star}$, $E^{\bullet}_{aa'}$, $o^{\bullet\,A}$,
$\iota^{\bullet\,A}$ satisfying the NP-gauge, in particular
(\ref{Rstar=0}). Using the component $\phi_0 = r\,\psi_{ABCD}
o^{\bullet\,A}o^{\bullet\,B}o^{\bullet\,C}o^{\bullet\,D}$ of the
rescaled conformal Weyl spinor, and performing the obvious lift to
$M'$, we obtain for the NP-constants the formula
\begin{equation}
G_m= -\frac{1}{2\,\pi} \oint {_2\bar{Y}_{2,m}}\,E^{\bullet}_{00'}(\phi_0)\,
d{\cal{S}}\,d\,\alpha.
\label{NPQ-semifinal-definition}
\end{equation}
Here $d{\cal{S}} = {\rm{sin}}\vartheta\,d\vartheta\,d\varphi$ denotes
the surface element on the cross-section
$\{r^{\bullet},\,u^{\bullet}=const.\} \subset {\cal J}^+$
and $\alpha$ denotes a parameter on the fibers of the
principal fiber bundle $M'\rightarrow{M}$. The second integration can
be performed without changing the result because the integrand is
independent of the variable $\alpha$. 

The values of these integrals are independent of the value of the
constant defining the cross-section as well as of the choice of the   
Bondi-coordinate $u^{\bullet}$ itself. Thus they are invariant under
supertranslations (cf. \cite{NP:NPQ}).

We shall determine the NP-constants by integrating the transport
equations on $I'$. Since these equations and their unknowns are given
in the F-gauge, we express (\ref{NPQ-semifinal-definition}) in this
gauge. Using (\ref{Lorentz-transf.-2}), we obtain in the notation of
the previous chapter
\begin{equation}
\begin{split}
G_m= -\frac{1}{2\,\pi} \oint {_2\bar{Y}_{2,m}} \frac{1}{\theta^4} \Bigl\{ 
&\Lambda^b_{\phantom{A}0}\Lambda^c_{\phantom{A}0}\Lambda^d_{\phantom{A}0} 
\Lambda^e_{\phantom{A}0}
\bigl[\Lambda^a_{\phantom{A}0}\bar{\Lambda}^{a'}_{\phantom{A}0'} 
\,c^*_{aa'}(\phi_{bcde}) -3\,\phi_{bcde}E^{\bullet}_{00'}(\theta)\bigr] \\
& +4\,\theta\,\Lambda^b_{\phantom{A}0}\Lambda^c_{\phantom{A}0}
\Lambda^d_{\phantom{A}0}
E^{\bullet}_{00'}(\Lambda^e_{\phantom{A}0})\,\phi_{bcde} \Bigr\} 
\,d{\cal{S}}\,d\,\alpha.
\label{NPQ-final-definition}
\end{split}
\end{equation}
This is the expression for the NP-constants which will be used in the
calculations of section [\ref{explicit-NP-constants}].

\sect{Time symmetric space-times}

In this section we will use the assumptions of the regular finite
initial value problem near space-like infinity and thus restrict our
considerations to time symmetric space-times. We
begin by solving the third order transport equations on $I'$. This
calculation is of interest for two quite different reasons. First of
all, it will give us a first insight into the potential source of
singular behavior of the quantities $u^p$ pointed out in section
[\ref{transp+asreg}]. Further, besides giving information on this
question of principle, the calculation will allow us to analyze the
relation between the NP-constants and the initial data for
asymptotically flat solutions. Under our assumptions, we will be able
to evaluate the integral (\ref{NPQ-final-definition}) in terms of
quantities derived from the initial data.

\subsection{Solving the third-order transport equation} 
\label{3rd-order-solution}

The solutions $u^p$ of equations (\ref{transport-equations}) have been given
in \cite{Fr:space-like-infinity} for $p \le 2$. Since they will be used in
the following calculations we reproduce them here, in a notation, though,
which is more convenient for a systematic discussion of the higher order
expansion coefficients. We also take the opportunity to correct a misprint in
\cite{Fr:space-like-infinity}.    

The solution $u^0$ of the transport equations
(\ref{transport-equations}) has the form
\begin{equation}
\begin{array}{r@{}lr@{}lr@{}lr@{}lr@{}l}
(c^0_{\;ab})^0= & \;-\tau x_{ab}, & (c^1_{\;ab})^0= & \;0 , & 
(c^+_{\;ab})^0= & \;z_{ab}, & (c^-_{\;ab})^0= & \;y_{ab}, &
\xi^0_{abcd}= & \;0, \\[6pt]
\chi^0_{(ab)cd}= & \;0, 
& f^0_{ab}= & \;x_{ab}, & (\Theta^{\;g}_{g\;ab})^0= & \;0, &
\Theta^0_{(ab)cd}= & \;0, & 
\phi^0_{abcd}= & \;-6\,m\,\varepsilon^2_{abcd}, 
\label{0-order-solution}
\end{array}
\end{equation}
where $m = m_{ADM}$ denotes the ADM-mass of the initial data set. The
spinors appearing on the right hand side of these and the following
formulae are listed in (\ref{primary-spinors}) of appendix
[\ref{tau-polynomials}]. The solution $u^1$ is given by
\begin{equation}
\begin{array}{r@{}lr@{}lr@{}l}
(c^0_{\;ab})^1= & \;c^{01}(\tau)\,x_{ab}, & 
(c^1_{\;ab})^1= & \;x_{ab}, &
(c^+_{\;ab})^1= & \;c^{\pm1}(\tau)z_{ab}, \\[6pt]
(c^-_{\;ab})^1= & \;c^{\pm1}(\tau)y_{ab}, & 
\xi^1_{abcd}= & 
\;S^1(\tau)(\epsilon_{ac}x_{bd}+\epsilon_{bd}x_{ac}), &
\chi^1_{(ab)cd}= & \;K^1(\tau)\varepsilon^2_{abcd}, \\[6pt]
f^1_{ab}= & \;F^1(\tau)x_{ab}, &
(\Theta^{\;g}_{g\;ab})^1= & \;t^1(\tau)x_{ab}, &
\Theta^1_{\ (ab)cd}= & \;T^1(\tau)\varepsilon^2_{abcd}, \\[6pt]
\phi^1_{abcd}= & \multicolumn{5}{l}{ 
\phi^1_1(\tau)X_+W_1 \varepsilon^1_{abcd} 
+[\phi^1_2(\tau)+\phi^1_3(\tau)W_1] \varepsilon^2_{abcd}
-\phi^1_1(-\tau)X_-W_1\varepsilon^3_{abcd},}
\label{1-order-solution}
\end{array}
\end{equation}
while $u^2$ takes the form 
\begin{equation}
\begin{split}
(c^0_{\;ab})^2= &\;[c^{02}_1(\tau)+c^{02}_2(\tau)W_1]x_{ab}
+c^{02}_3(\tau)[X_-W_1y_{ab}+X_+W_1z_{ab}], \\
(c^1_{\;ab})^2= &\;c^{12}(\tau)x_{ab}, \\
(c^+_{\;ab})^2= &\;[c^{\pm2}_1(\tau)+c^{\pm2}_2(\tau)W_1]z_{ab}
+c^{\pm2}_3(\tau)X_-W_1x_{ab}, \\
(c^-_{\;ab})^2= &\;[c^{\pm2}_1(\tau)+c^{\pm2}_2(\tau)W_1]y_{ab}
+c^{\pm2}_3(\tau)X_+W_1x_{ab}, \\
\xi^2_{abcd}= 
&\;[S^2_1(\tau)+S^2_2(\tau)W_1](\epsilon_{ac}x_{bd}+\epsilon_{bd}x_{ac}) 
+S^2_3(\tau)(\epsilon_{ac}y_{bd}+\epsilon_{bd}y_{ac})X_-W_1 \\ 
&\;+S^2_3(\tau)(\epsilon_{ac}z_{bd}+\epsilon_{bd}z_{ac})X_+W_1 
+S^2_4(\tau)(\varepsilon^1_{abcd}X_+W_1+\varepsilon^3_{abcd}X_-W_1), \\
\chi^2_{(ab)cd}= 
&\;[K^2_1(\tau)+K^2_2(\tau)W_1]\varepsilon^2_{abcd}+K^2_3(\tau)h_{abcd} 
+K^2_4(\tau)(\epsilon_{ac}y_{bd}+\epsilon_{bd}y_{ac})X_-W_1 \\
&\;-K^2_4(\tau)(\epsilon_{ac}z_{bd}+\epsilon_{bd}z_{ac})X_+W_1 
+K^2_5(\tau)(\varepsilon^1_{abcd}X_+W_1-\varepsilon^3_{abcd}X_-W_1), \\  
f^2_{ab}= &\;[F^2_1(\tau)+F^2_2(\tau)W_1]x_{ab} 
+F^2_3(\tau)(X_-W_1y_{ab}+X_+W_1z_{ab}), \\
(\Theta^{\;g}_{g\;ab})^2= &\;[t^2_1(\tau)+t^2_2(\tau)W_1]x_{ab} 
+t^2_3(\tau)(X_-W_1y_{ab}+X_+W_1z_{ab}), \\
\Theta^2_{(ab)cd}= &
\;[T^2_1(\tau)+T^2_2(\tau)W_1]\varepsilon^2_{abcd}+T^2_3(\tau)h_{abcd}
+T^2_4(\tau)(\epsilon_{ac}y_{bd}+\epsilon_{bd}y_{ac})X_-W_1 \\
&\;-T^2_4(\tau)(\epsilon_{ac}z_{bd}+\epsilon_{bd}z_{ac})X_+W_1
+T^2_5(\tau)(\varepsilon^1_{abcd}X_+W_1-\varepsilon^3_{abcd}X_-W_1), \\
\phi^2_{abcd}= &\;\phi^2_1(\tau)X_+X_+W_2 \varepsilon^0_{abcd}
+[\phi^2_2(\tau)X_+W_1+\phi^2_3(\tau)X_+W_2]\varepsilon^1_{abcd} \\
&\;+[\phi^2_4(\tau)+\phi^2_5(\tau)W_1 
+\phi^2_6(\tau)W_2]\varepsilon^2_{abcd}
-[\phi^2_2(-\tau)X_-W_1+\phi^2_3(-\tau)X_-W_2]\varepsilon^3_{abcd} \\
&\;+\phi^2_1(-\tau)X_-X_-W_2 \varepsilon^4_{abcd}.
\label{2-order-solution}
\end{split}
\end{equation}
The $\tau$-dependent functions in these expressions are polynomials 
which are given in appendix [\ref{tau-polynomials}].

The calculation of $u^3$ is facilitated by the following properties 
of the transport equations (\ref{transport-equations}). For $p \ge 1$ they are
of the form 
\begin{equation}
\begin{array}{cc}
\partial_{\tau}v^p = L_p\,v^p + l_p, &
B^{\alpha}\partial_{\alpha}\phi^p= M_p \phi^p, 
\label{separated-transport-equations} 
\end{array}
\end{equation}
where, using the notation (\ref{v-variable}), we set $v^p =
(\partial^p_{\rho}v)|_{I'}$, $\phi^p = (\partial^p_{\rho}\phi)|_{I'}$
and denote by $L_p$ and $l_p$ a matrix- resp. vector-valued function
of the quantities $u^0,\dots,u^{p-1}$, while $M_p$ denotes a
matrix-valued function which depends on the variables
$u^0,\dots,u^{p-1}$, $v^p$. The matrices $B^{\alpha}$ neither depend
on $p$ nor on the initial data.  Thus, given the quantities $u^q$, $q
\le p -1$, we can integrate the first of equations
(\ref{separated-transport-equations}), which is an ODE. To integrate
the second equation, we expand the quantities $u^p$ in terms of the
functions $T_{m\;j}^{\;\;k}$ given in (\ref{T-functions}) and use
(\ref{Xact}) to reduce the integration to that of a system of ODE's.

To determine the initial data for $u^3$ on $I^{'0}$, we have to expand the
unknowns  (\ref{initial-data}) in terms of $\rho$. Instead of prescribing the
conformal metric $h$ on the initial slice, which represents the free datum, we
shall prescribe, in a fashion consistent with the $3$-dimensional Bianchi
identities, certain curvature quantities and use the $3$-dimensional
structure equations and the Yamabe equation to determine the remaining
quantities.  

The conformal factor, which appears in the expressions (\ref{initial-data}), 
is given in (\ref{Omega}) in terms of the functions $U$ and $W$. 
The function $U$, which is determined locally by $h$ near space-like
infinity, is given, by a procedure explained in \cite{Fr:space-like-infinity},
in the form 
\begin{equation}
U=\sum_{p=0}^{\infty}U_p\,\rho^{2p}, \label{U:rho-expansion}
\end{equation}
with $\rho$-dependent coefficients $U_p$. As shown in
\cite{Fr:space-like-infinity}, the Taylor expansion of $U$ in terms of $\rho$
has in our gauge the form \begin{equation}
U=1 +\sum_{k=4}^{\infty}\frac{1}{k!}\hat{U}_k\rho^k. 
\label{U:Taylor-expansion}
\end{equation}
For our calculations we shall need the coefficient $\hat{U}_4$, which  
will be determined later in this chapter.  

The function $W$, which contains global information on the free initial data,
is determined by solving the Yamabe equation on the initial hypersurface. We
shall consider here a larger class of functions which are subject to the
Yamabe equation only in a small neighborhood of space-like infinity.
The coefficients in the Taylor expansion
$W=W_0 +W_1\,\rho +\frac{1}{2}W_2\,\rho^2 +\frac{1}{3!}W_3\,\rho^3
+O(\rho^4)$ have expansion (cf. \cite{Fr:space-like-infinity})
\begin{equation*}
W_i=\sum_{m=0}^{2i}\sum_{k=0}^m\,W_{i;m,k}\,T_{m\;{\frac{m}{2}}}^{\;\;k}.
\end{equation*}
They are restricted by the requirement that the Yamabe equation
$(h^{\alpha\beta}D_{\alpha}D_{\beta}-\frac{1}{8}r_h)[W]=0$ holds
near $\{\rho = 0\}$, which implies the simplification 
\begin{equation}
W_i=\displaystyle{\sum_{k=0}^2}\,W_{i;2i,k}\,T_{2i\;i}^{\;\;k},
\,\,\,\,\,\,\,\,i \le 3.
\label{W:T-expansion-2}
\end{equation}
We get for the conformal factor and the trace-free part of
its second covariant derivative 
\begin{equation}
\begin{array}{l@{}l}
\Omega=&\rho^2 -m\,\rho^3 +\bigl[\frac{3}{4}\,m^2-2\,W_1\bigr]\,\rho^4
+\bigl[-\frac{1}{2}m^3+3\,m\,W_1-W_2\bigr]\,\rho^5 \\[6pt]
&+\bigl[\frac{5}{16}\,m^4-3\,m^2W_1+3W_1^2+\frac{3}{2}\,m\,W_2 
-\frac{1}{3}W_3-\frac{1}{12}\hat{U}_4\bigr]\,\rho^6+O(\rho^7), \\[12pt]
D_{(ab}&D_{cd)}\Omega= \bigl[-6\,m\,\varepsilon^2_{abcd}\bigr]\rho 
+\bigl[(12\,m^2-36\,W_1)\varepsilon^2_{abcd}
-12\,(\varepsilon^1_{abcd}X_+-\varepsilon^3_{abcd}X_-)W_1\bigr]\rho^2 \\[6pt]
&+\bigl[(-15\,m^3+96\,m\,W_1-36\,W_2)\varepsilon^2_{abcd} 
+(\varepsilon^1_{abcd}X_+-\varepsilon^3_{abcd}X_-)(24\,m\,W_1-8\,W_2) \\[6pt]
&-\frac{1}{2}(\varepsilon^0_{abcd}X_+X_+
+\varepsilon^4_{abcd}X_-X_-)W_2 \bigr]\rho^3 
+\bigl[(156\,W_1^2-150\,m^2W_1+15\,m^4+81\,m\,W_2 \\[6pt]
&-20\,W_3-4\,\hat{U}_4
+\frac{1}{12}X_+X_-\hat{U}_4-6\,X_+W_1X_-W_1)\varepsilon^2_{abcd} \\[6pt]
&+(\varepsilon^1_{abcd}X_+-\varepsilon^3_{abcd}X_-)(30\,W_1^2
-30\,m^2W_1+15\,m\,W_2-\frac{10}{3}W_3-\frac{5}{6}\hat{U}_4) \\[6pt]
&+\frac{1}{2}(\varepsilon^0_{abcd}X_+X_++\varepsilon^4_{abcd}X_-X_-)(3\,W_1^2 
+\frac{3}{2}m\,W_2-\frac{1}{3}W_3-\frac{1}{12}\hat{U}_4) 
-\frac{2}{3}x_{e(a}\gamma^{3\;\;e}_{\;bc\;d)}\bigr]\rho^4 \\[6pt]
&+O(\rho^5).
\label{Omega:Taylor-expansion} 
\end{array}
\end{equation}
From this we obtain as initial data for $u^3$ on $I^{'0}$ 
\begin{equation}
\begin{array}{r@{}l}
(c^0_{\;ab})^3\,&=\,0,\;\;(c^1_{\;ab})^3=0,\;\;
(c^+_{\;ab})^3=0,\;\;(c^-_{\;ab})^3=0, \\[6pt]
\xi^3_{abcd}\,&=\,0,\;\;\chi^3_{(ab)cd}=0,\;\;
f^3_{ab}=0,\;\;(\Theta^{\;g}_{g\;ab})^3=0, \\[6pt]
\Theta^3_{(ab)cd}\,&=\, 3\,X_+X_+W_2\varepsilon^0_{abcd} 
+(-72\,m\,X_+W_1+48\,X_+W_2)\varepsilon^1_{abcd} \\[6pt]
&+(27\,m^3-288\,m\,W_1+216\,W_2)\varepsilon^2_{abcd} \\[6pt]
&+(72\,m\,X_-W_1-48\,X_-W_2)\varepsilon^3_{abcd}
+3\,X_-X_-W_2\varepsilon^4_{abcd}, \\[6pt]
\phi^3_{abcd}\,&=\, (\varepsilon^0_{abcd}X_+X_+ 
+\varepsilon^4_{abcd}X_-X_-)(9W_1^2
-\frac{3}{2}mW_2-W_3-\frac{1}{4}\hat{U}_4) \\[6pt]
&+4(\varepsilon^1_{abcd}X_+-\varepsilon^3_{abcd}X_-) 
(9W_1^2 -\frac{3}{2}mW_2 -5W_3 -\frac{5}{4}\hat{U}_4) \\[6pt]
&+6\,\varepsilon^2_{abcd}(12\,W_1^2-3m\,W_2-20W_3-4\hat{U}_4
+\frac{1}{12}X_+X_-\hat{U}_4-6X_+W_1X_-W_1) \\[6pt]
&-4x_{e(a}\gamma^{3\;\;e}_{\;bc\;d)}+3s^2_{abcd},
\label{detailed-initial-values}
\end{array}
\end{equation}
where $\gamma_{abcd} = (2\,\rho)^{-1}(\epsilon_{ac}x_{bd}+\epsilon_{bd}x_{ac})
+ \check{\gamma}_{abcd}$ denote the connection coefficients on $C'$.

We determine now how the functions $\hat{U}_4$, $\gamma^3_{abcd}$, and
$s^2_{abcd}$ are related to the free data on the initial hypersurface $C'$. 
As shown in \cite{Fr:space-like-infinity}, the structure equations on $C'$,
which relate the connection coefficients to the curvature, read 
\begin{equation*}  
\begin{array}{@{}l}
\frac{1}{\sqrt2}\bigl\{\partial_{\rho}\check{\gamma}_{00ab} 
+\frac{\sqrt2}{\rho}\bigl[\check{\gamma}_{0000}z_{ab} 
-\check{\gamma}_{0011}y_{ab} 
+\frac{1}{\sqrt2}\check{\gamma}_{00ab}\bigr]\bigr\}= 
\check{\gamma}_{0000}\check{\gamma}_{11ab} 
-\check{\gamma}_{0011}\check{\gamma}_{00ab} 
-\frac{1}{2}s_{ab00}-\frac{1}{6\sqrt2}\,r\,y_{ab}, \\[6pt]
\frac{1}{\sqrt2}\bigl\{\partial_{\rho}\check{\gamma}_{11ab} 
+\frac{\sqrt2}{\rho}\bigl[\check{\gamma}_{1100}y_{ab} 
-\check{\gamma}_{1111}y_{ab} 
+\frac{1}{\sqrt2}\check{\gamma}_{11ab}\bigr]\bigr\}= 
\check{\gamma}_{1100}\check{\gamma}_{11ab} 
-\check{\gamma}_{1111}\check{\gamma}_{00ab}+\frac{1}{2}s_{ab11} 
-\frac{1}{6\sqrt2}\,r\,z_{ab}, 
\end{array}
\end{equation*}
and the components of
$\check{\gamma}_{abcd}$ have Taylor expansions
\begin{equation*}
\check{\gamma}_{01ab}=0, \;\;\;
\check{\gamma}_{00ab}=\frac{1}{3!}\check{\gamma}^3_{00ab}\,\rho^3
+ O(\rho^4), \;\;\;
\check{\gamma}_{11ab}=\frac{1}{3!}\check{\gamma}^3_{11ab}\,\rho^3
+ O(\rho^4).
\end{equation*}
From this we get
\begin{equation*}
\begin{array}{l@{}lc@{}lr@{}l}
\check{\gamma}^3_{0001}= &\,-\frac{3}{4\sqrt2}s^2_{0001}, &
\check{\gamma}^3_{1101}= &\,\frac{3}{4\sqrt2}s^2_{0111}, &
\check{\gamma}^3_{0000}= &\,-\frac{3}{5\sqrt2}s^2_{0000}, \\[6pt]
\check{\gamma}^3_{1100}= &\,\frac{3}{5\sqrt2}s^2_{0011} 
-\frac{1}{10\sqrt2}r^2, &
\check{\gamma}^3_{0011}= &\,-\frac{3}{5\sqrt2}s^2_{0011} 
+\frac{1}{10\sqrt2}r^2, & 
\check{\gamma}^3_{1111}= &\,\frac{3}{5\sqrt2}s^2_{1111}, 
\end{array}
\end{equation*}
and obtain thus for the quantity 
$F_{abcd} = -4\,x_{e(a}\gamma^{3\;\;e}_{\;bc\;d)}+3\,s^2_{abcd}$ 
the concise expressions
\begin{equation}
\begin{array}{lllll}
F_0=\frac{9}{5}\,s^2_0, & F_1=3\,s^2_1, &
F_2=\frac{17}{5}\,s^2_2-\frac{1}{15}r^2, & F_3=3\,s^2_3, &
F_4=\frac{9}{5}\,s^2_4,  
\label{F-components}
\end{array}
\end{equation}
where we set $F_i = F_{(abcd)_i}$, $s_i = s_{{(abcd)}_i}$, using the notation
introduced in (\ref{T-functions}).

In the cn-gauge the curvature vanishes at zeroth and first order at
space-like infinity. At second order this is in general not true and the
prescription of the free data on $S$ in terms of curvature quantities has to
be consistent with the cn-gauge, the Bianchi identity, and the regularity
condition (\ref{asregcond}) for $q = 1$. The content of the
cn-gauge is expressed in second order in the curvature by the conditions
\begin{equation*}
D_{ab}D^{ab}\,r = 0,\,\,\,\,\,\,
D_{ab}D^{ab}\,s_{cdef} = - \frac{5}{4}\,D_{cd}\,D_{ef}\,r,\,\,\,\,\,\,
D_{(ab}\,D_{cd}\,s_{efgh)} = 0
\quad\mbox{at}\quad i.
\end{equation*} 
It follows that the spinor 
\begin{equation*}
t_{abcd}\,_{efgh} = D_{ab}\,D_{cd}\,s_{efgh} 
- \frac{1}{3}\,h_{abcd}\,\Delta_h\,s_{efgh},
\end{equation*}
where $\Delta_h$ denotes the Laplacian corresponding to the
metric $h$, is symmetric in the first and the last four indices 
separately. Using the Bianchi identity
\begin{equation*}
D^{ab}\,s_{abcd} = \frac{1}{6}\,D_{cd}\,r,
\end{equation*} 
we thus get
\[
\frac{1}{6}\,D_{ab}\,D_{cd}\,r - \frac{1}{3}\,\Delta_h\,s_{abcd}
= t^{ef}\,_{ab\,cdef} = t_a\,^e\,_b\,^f\,_{\,cdef}
= D_a\,^e\,D_b\,^f\,s_{cdef} + \frac{1}{6}\,\Delta_h\,s_{abcd},
\]
whence
\begin{equation*}
D_a\,^e\,D_b\,^f\,s_{cdef} = \frac{19}{24}\,D_{ab}\,D_{cd}\,r.
\end{equation*}
No further conditions are implied at $i$ on the Ricci scalar
$r$ at this order. Finally, we get from (\ref{asregcond}) for $q = 1$ 
\[
D^h\,_{(a}\,D_{bc}\,s_{def)h} = 0
\quad\mbox{at}\quad i.
\]
The relations above imply that the expansion of $t_{abcd}\,_{efgh}$ in
terms of symmetric spinors and $\epsilon_{ab}$'s can be expressed completely
in terms of symmetrized twofold contractions of this spinor, which in turn can
all be expressed in terms of the symmetric spinor $D_{ab}\,D_{cd}\,r$. Working
out this expansion we get
\begin{equation}
\label{ddsexp}
D^{ab}\,D^{cd}\,s_{efgh} = h^{(ab}\,_{(ef}\,D^{cd)}\,_{gh)}\,r
- \frac{5}{15}\,h^{abcd}\,D_{ef}\,D_{gh}\,r
\quad\mbox{at}\quad i,
\end{equation}
in our gauge. Going through the procedure described in section (3.5) of
\cite{Fr:space-like-infinity} we get 
$s_{(abcd)_j} = s^2_j\,\rho^2 + O(\rho^3)$ and 
$r = r^{2}\,\rho^2 + O(\rho^3)$ with
\begin{equation}
\begin{array}{lcr}
s^2_j=\frac{3^{|2-j|}}{12}\displaystyle{\sum_{k=0}^4}
R_k^*{\tbinom{4}{j}}^{-\frac{1}{2}} T_{4\;j}^{\;k}, &\;\;\;&
r^2=\frac{2}{\sqrt{6}}\displaystyle{\sum_{k=0}^4}
R_k^*\,T_{4\;2}^{\;k}, 
\label{SR:T-expansion}
\end{array}
\end{equation}
where we set $R^{*}_k = \frac{1}{2}\,
{\binom{4}{k}}^{\frac{1}{2}}\,D_{(ab}\,D_{cd)_k}\,r^{*}$, with the star
indicating that the quantities are given in our gauge at $i$. 
The $5$ real numbers $R^{*}_k$ contain precisely the information on the
metric $h$ which can at this order be freely specified in the cn-gauge.

We note that the Cotton spinor is then given at $i$ by 
\begin{equation*}
D_{ab}\,b_{cdef} =
- \frac{5}{8}\,\left\{ \epsilon_{a(b}\,D_{cd}\,D_{ef)}\,r +
\epsilon_{b(a}\,D_{cd}\,D_{ef)}\,r \right\},
\end{equation*}
and the deviation of $h$ from conformal flatness at $i$ is encoded at 
this order in the symmetric spinor $D_{ab}\,D_{cd}\,r(i)$.

From (\ref{F-components}), (\ref{SR:T-expansion}) we obtain 
\begin{equation*}
\begin{array}{lll}
F_0= \frac{27}{20}\displaystyle{\sum_{k=0}^4} R_k^*\, T_{4\;0}^{\;k}, &
F_1= \frac{3}{8}\displaystyle{\sum_{k=0}^4} R_k^*\, T_{4\;1}^{\;k}, &
F_2= \frac{3}{20\sqrt{6}}\displaystyle{\sum_{k=0}^4} R_k^*\,T_{4\;2}^{\;k}, 
\\[6pt]
F_3= \frac{3}{8}\displaystyle{\sum_{k=0}^4} R_k^*\, T_{4\;3}^{\;k}, &
F_4= \frac{27}{20}\displaystyle{\sum_{k=0}^4} R_k^*\, T_{4\;4}^{\;k}.  &
\end{array}
\end{equation*}

Finally, we will calculate the coefficient $\hat{U}_4$ in the Taylor
series (\ref{U:Taylor-expansion}). Only the coefficients $U_0$,
$U_1$ and $U_2$ of the expansion (\ref{U:rho-expansion}) contribute to
$\hat{U}_4$.  
These functions have the following expansions (cf.
\cite{Fr:space-like-infinity} for the defining integrals).
\begin{equation}
U_0= exp\left\{{\frac{1}{4}}\int_0^\rho \left(\Delta {\rho'}^2 +6\right) 
{\frac{d\rho'}{\rho'}}\right\}= 
1+\frac{1}{4!}\bigl[\sqrt2\,\gamma^3_{1100}\bigr]\rho^4+O(\rho^5), 
\label{U0:rho-expansion}
\end{equation}
where we used the expansion
\begin{equation*}
\Delta \rho^2= -\,6 + \frac{2\sqrt2}{3}\gamma^3_{1100}\,\rho^4 +O(\rho^5).
\end{equation*}
Further we have, with $L$ denoting the Yamabe operator,
\begin{equation}
U_1=\frac{U_0}{2\,\rho}\int_0^{\rho}\frac{L[U_0]}{U_0}d\rho'= 
\frac{1}{2}\left[-\frac{7\sqrt{2}}{36} \gamma^3_{1100} 
-\frac{1}{48}\,r^2\right]\rho^2 +O(\rho^3).
\label{U1:rho-expansion}
\end{equation}
Finally, observing (\ref{SR:T-expansion}), we obtain 
\begin{equation*}
U_2=-\frac{U_0}{2\,\rho^2}\int_0^{\rho} 
\frac{L[U_1]\,\rho'}{U_0}d\rho'= O(\rho).
\label{U2:rho-expansion}
\end{equation*}
Collecting results, we arrive at the expansion
\begin{equation}
U=1+\frac{1}{4!}\left[-\frac{4\sqrt{2}}{3}\,\gamma^3_{1100} 
-\frac{1}{4}\,r^2\right]\rho^4 +O(\rho^5)= 1 
+\frac{1}{4!}\left[-\frac{3}{10\sqrt{6}}\sum_{k=0}^4 R_k^*\, 
T_{4\;2}^{\;k}\right]\rho^4 +O(\rho^5).
\label{U:detailed-expansion}
\end{equation}

Since the initial datum for the conformal Weyl spinor is a non-linear
function of the basic quantities and the transport equations are quadratic
in the unknowns, we have to make use of the Clebsch-Gordan expansions of
products like $T_{2\;m}^{\;k}T_{2\;n}^{\;l}$. These are readily calculated 
by using the definition (\ref{T-functions}). For the quantities relevant in
our calculation we thus obtain
\begin{equation}
\begin{array}{@{}r@{}lr@{}l}
X_-W_1X_+W_1= &\;-\displaystyle{\sum_{k=0}^4}\,a_kT_{4\;2}^{\;k}+2\,b, &
W_1^2= &\;\displaystyle{\sum_{k=0}^4}\,a_kT_{4\;2}^{\;k}+b, \\[6pt]
W_1X_-W_1= 
&\;-\frac{\sqrt6}{2}\displaystyle{\sum_{k=0}^4}\,a_kT_{4\;3}^{\;k}, &
W_1X_+W_1= 
&\;\frac{\sqrt6}{2}\displaystyle{\sum_{k=0}^4}\,a_kT_{4\;1}^{\;k}, \\[6pt]
(X_-W_1)^2= &\;\sqrt6\displaystyle{\sum_{k=0}^4}\,a_kT_{4\;4}^{\;k}, &
(X_+W_1)^2= &\;\sqrt6\displaystyle{\sum_{k=0}^4}\,a_kT_{4\;0}^{\;k},
\label{quadratic-W-expansions} 
\end{array}
\end{equation}
with coefficients
\begin{equation}
\begin{array}{lll}
a_0 =\;\frac{2}{\sqrt6}W_{1;2,0}^2, &
a_1 =\;\frac{2}{\sqrt3}W_{1;2,0}W_{1;2,1}, &
a_2 =\;\frac{2}{3}(W_{1;2,0}W_{1;2,2}+W_{1;2,1}^2), \\[6pt]
a_3 =\;\frac{2}{\sqrt3}W_{1;2,2}W_{1;2,1}, &
a_4 =\;\frac{2}{\sqrt6}W_{1;2,2}^2, &
b_{\phantom{1}} =\;-\frac{2}{3}(W_{1;2,0}W_{1;2,2}-\frac{1}{2}W_{1;2,1}^2).
\label{def:a}
\end{array}
\end{equation}
It was shown in \cite{Fr:space-like-infinity} that the quantity $\phi^3_i$
has an expansion of the form
\begin{equation}
\phi^3_i=\sum_{m=|4-2i|}^q\sum_{k=0}^m \phi^3_{i;m,k}
T_{m\;\frac{m}{2}-2+i}^{\;k}. 
\label{phi-T-decomposition} 
\end{equation}
Using the results above in the last equation of 
(\ref{detailed-initial-values}), this expansion reduces to 
\begin{equation}
\begin{array}{r@{}lr@{}lr@{}l}
\multicolumn{6}{l}{\phi^3_{i;m,k}=0,{\rm\;\;\;for\;\;\;}i=\{0,\dots,4\} 
{\rm\;\;\;and\;\;} m\geq 8,} \\[6pt]
\phi^3_{0;6,k}=&\,-2\sqrt{30}\,W_{3;6,k}, &
\phi^3_{1;6,k}=&\,-10\sqrt{3}\,W_{3;6,k}, &
\phi^3_{2;6,k}=&\,-20\,W_{3;6,k}, \\[6pt]
\phi^3_{3;6,k}=&\,-10\sqrt{3}\,W_{3;6,k}, &
\phi^3_{4;6,k}=&\,-2\sqrt{30}\,W_{3;6,k}, & & \\[6pt]
\multicolumn{3}{l}{\phi^3_{0;4,k}=\,18\sqrt{6}\,a_k
-3\sqrt{6}\,m\,W_{2;4,k}+\frac{3}{2}R_k^*,} &
\multicolumn{3}{l}{\phi^3_{1;4,k}=\,9\sqrt{6}\,a_k 
-\frac{3}{2}\sqrt{6}\,m\,W_{2;4,k}+\frac{3}{4}R_k^*,} \\[6pt]
\multicolumn{3}{l}{\phi^3_{2;4,k}=\,18\,a_k 
-3\,m\,W_{2;4,k} +\frac{3}{2\sqrt{6}}R_k^*,} &
\multicolumn{3}{l}{\phi^3_{3;4,k}=\,9\sqrt{6}\,a_k 
-\frac{3}{2}\sqrt{6}\,m\,W_{2;4,k}+\frac{3}{4}R_k^*,} \\[6pt]
\multicolumn{3}{l}{\phi^3_{4;4,k}=\,18\sqrt{6}\,a_k 
-3\sqrt{6}\,m\,W_{2;4,k} +\frac{3}{2}R_k^*,} &&& \\[6pt]
\multicolumn{3}{l}{\phi^3_{i;2,k}=0 {\rm\;\;for\;\;\;}i=\{1,2,3\},} &
\multicolumn{3}{l}{\phi^3_{2;0,0}= 0.}
\label{T-decomposed-phi-initial-values} 
\end{array} 
\end{equation}
Given these data on $I^{'0}$, we are in the position to solve the transport
equations on $I'$. The first of the systems
(\ref{separated-transport-equations}) can be integrated step by step with
the result 
\begin{equation} 
\begin{split}
(c^0_{\;ab})^3= &\;[c^{03}_1(\tau)+c^{03}_2(\tau)W_1 
+c^{03}_3(\tau)W_2]x_{ab}+[c^{03}_4(\tau)X_+W_1+c^{03}_5(\tau)X_+W_2]z_{ab} \\
&\; +[c^{03}_4(\tau)X_-W_1+c^{03}_5(\tau)X_-W_2]y_{ab}, \\
(c^1_{\;ab})^3= &\;[c^{13}_1(\tau)+c^{13}_2(\tau)W_1]x_{ab}
+c^{13}_3(\tau)[X_+W_1z_{ab}+X_-W_1y_{ab}], \\
(c^+_{\;ab})^3= &\;[c^{\pm{3}}_1(\tau)X_-W_1 
+c^{\pm{3}}_2(\tau)X_-W_2]x_{ab}+[c^{\pm{3}}_3(\tau)+c^{\pm{3}}_4(\tau)W_1 
+c^{\pm{3}}_5(\tau)W_2]z_{ab} \\
&\; +c^{\pm{3}}_6(\tau)X_-X_-W_2y_{ab}, \\
(c^-_{\;ab})^3= &\;[c^{\pm{3}}_1(\tau)X_+W_1 
+c^{\pm{3}}_2(\tau)X_+W_2]x_{ab}+[c^{\pm{3}}_3(\tau)+c^{\pm{3}}_4(\tau)W_1 
+c^{\pm{3}}_5(\tau)W_2]y_{ab} \\
&\;+c^{\pm{3}}_6(\tau)X_+X_+W_2z_{ab}, \\
\xi^3_{abcd}= &\;S^3_1(\tau)X_+X_+W_2\varepsilon^0_{abcd}
+[S^3_2(\tau)X_+W_1+S^3_3(\tau)X_+W_2]\varepsilon^1_{abcd} \\
&\;+[S^3_2(\tau)X_-W_1+S^3_3(\tau)X_-W_2]\varepsilon^3_{abcd}
-S^3_1(\tau)X_-X_-W_2\varepsilon^4_{abcd} \\
&\;+[S^3_4(\tau)+S^3_5(\tau)W_1+S^3_6(\tau)W_2]
(\epsilon_{ac}x_{bd}+\epsilon_{bd}x_{ac}) \\
&\;+[S^3_{7}(\tau)X_+W_1+S^3_{8}(\tau)X_+W_2] 
(\epsilon_{ac}z_{bd}+\epsilon_{bd}z_{ac}) \\
&\;+[S^3_{7}(\tau)X_-W_1+S^3_{8}(\tau)X_-W_2] 
(\epsilon_{ac}y_{bd}+\epsilon_{bd}y_{ac}), \\
\chi^3_{(ab)cd}= &\;K^3_1(\tau)X_+X_+W_2\varepsilon^0_{abcd}
+[K^3_2(\tau)X_+W_1+K^3_3(\tau)X_+W_2]\varepsilon^1_{abcd} \\
&\;+[K^3_4(\tau)+K^3_5(\tau)W_1 
+K^3_6(\tau)W_2]\varepsilon^2_{abcd}
-[K^3_2(\tau)X_-W_1+K^3_3(\tau)X_-W_2]\varepsilon^3_{abcd} \\
&\;+K^3_1(\tau)X_-X_-W_2\varepsilon^4_{abcd}
+[K^3_7(\tau)+K^3_8(\tau)W_1]h_{abcd} \\
&\;+[K^3_9(\tau)X_-W_1+K^3_{10}(\tau)X_-W_2] 
(\epsilon_{ac}y_{bd}+\epsilon_{bd}y_{ac}) \\
&\;-[K^3_9(\tau)X_+W_1+K^3_{10}(\tau)X_+W_2] 
(\epsilon_{ac}z_{bd}+\epsilon_{bd}z_{ac}), \\
f^3_{ab}= &\;[F^3_1(\tau)+F^3_2(\tau)W_1+F^3_3(\tau)W_2]x_{ab}
+[F^3_4(\tau)X_-W_1+F^3_5(\tau)X_-W_2]y_{ab} \\
&\;+[F^3_4(\tau)X_+W_1+F^3_5(\tau)X_+W_2]z_{ab}, \\
(\Theta^{\;g}_{g\;ab})^3= 
&\;[t^3_1(\tau)+t^3_2(\tau)W_1+t^3_3(\tau)W_2]x_{ab}
+[t^3_4(\tau)X_-W_1+t^3_5(\tau)X_-W_2]y_{ab} \\
&\;+[t^3_4(\tau)X_+W_1+t^3_5(\tau)X_+W_2]z_{ab}, \\ 
\Theta^3_{\;(ab)cd}= &\;T^3_1(\tau)X_+X_+W_2\varepsilon^0_{abcd}
+[T^3_2(\tau)X_+W_1+T^3_3(\tau)X_+W_2]\varepsilon^1_{abcd} \\
&\;+[T^3_4(\tau)+T^3_5(\tau)W_1+T^3_6(\tau)W_2] 
\varepsilon^2_{abcd}
-[T^3_2(\tau)X_-W_1+T^3_3(\tau)X_-W_2]\varepsilon^3_{abcd} \\
&\;+T^3_1(\tau)X_-X_-W_2\varepsilon^4_{abcd}
+[T^3_7(\tau)+T^3_8(\tau)W_1]h_{abcd} \\
&\;+[T^3_9(\tau)X_-W_1+T^3_{10}(\tau)X_-W_2] 
(\epsilon_{ac}y_{bd}+\epsilon_{bd}y_{ac}) \\
&\;-[T^3_9(\tau)X_+W_1+T^3_{10}(\tau)X_+W_2] 
(\epsilon_{ac}z_{bd}+\epsilon_{bd}z_{ac}). \\
\label{3-order-solution-1}
\end{split}
\end{equation}
The $\tau$-dependent functions in these expressions are given in appendix
[\ref{tau-polynomials}].

We now turn to the second of the transport equations
(\ref{separated-transport-equations}), which is a partial differential
equation. The system for the expansion coefficients $\phi^3_i$ of the
rescaled conformal Weyl spinor on $I'$ has the form
\begin{equation}
\begin{array}{r@{}l}
(1+\tau)&\partial_{\tau}\phi^3_0 +X_+\phi^3_1-\phi^3_0= 
\;R_0, \\[6pt]
&\partial_{\tau} \phi^3_1 +\frac{1}{2}X_-\phi^3_0 
+\frac{1}{2}X_+\phi^3_2+\phi^3_1= \;R_1, \\[6pt]
&\partial_{\tau} \phi^3_2 +\frac{1}{2}X_-\phi^3_1 
+\frac{1}{2}X_+\phi^3_3= \;R_2, \\[6pt]
&\partial_{\tau} \phi^3_3 +\frac{1}{2}X_-\phi^3_2 
+\frac{1}{2}X_+\phi^3_4 -\phi^3_3= \;R_3, \\[6pt]
(1-\tau)&\partial_{\tau}\phi^3_4 +X_-\phi^3_3 +\phi^3_4= \;R_4, 
\label{phi-equations}
\end{array}
\end{equation}
where the right hand sides are given by
\begin{equation}
\begin{array}{@{}l}
R_0=\;A_1(\tau)X_+X_+W_2 +A_2(\tau)(X_+W_1)^2, \\[6pt] 
R_1=\;B_1(\tau)X_+W_1 +B_2(\tau)W_1X_+W_1+B_3(\tau)X_+W_2, \\[6pt]
R_2=\;C_1(\tau) +C_2(\tau)W_1 +C_3(\tau)(W_1)^2 
+C_4(\tau)W_2 +C_5(\tau)X_+W_1X_-W_1, \\[6pt]
R_3=\;B_1(-\tau)X_-W_1 +B_2(-\tau)W_1X_-W_1 +B_3(-\tau)X_-W_2, \\[6pt]
R_4=\;-A_1(-\tau)X_-X_-W_2 -A_2(-\tau)(X_-W_1)^2, 
\label{phi-equations-right-side} 
\end{array}
\end{equation}
with $\tau$-dependent functions $A_i(\tau),\;B_j(\tau),\;C_k(\tau)$ which are
listed in appendix [\ref{tau-polynomials}]. These functions have been
calculated from the lower order expansion coefficients
(\ref{0-order-solution})-(\ref{2-order-solution}) and from  
(\ref{3-order-solution-1}). The symmetry inherent in these expressions
reflects the time-symmetry of the underlying space-time.

Using the expansion (\ref{phi-T-decomposition}) and corresponding expansions
of the terms above, we decompose (\ref{phi-equations}) into the following
equations. For $m\geq{6}$ the coefficients $\phi^3_{i;m,k}$, $k = 0,\dots,m$,
satisfy the homogeneous system
\begin{equation}
\begin{array}{r@{}l}
(1+\tau)&\partial_{\tau}\phi^3_{0;m,k} -\phi^3_{0;m,k}
+\sqrt{(\frac{m}{2}-1)(\frac{m}{2}+2)}\phi^3_{1;m,k}=\, 0, \\[6pt]
&\partial_{\tau}\phi^3_{1;m,k} +\phi^3_{1;m,k}
-\frac{1}{2}\sqrt{(\frac{m}{2}-1)(\frac{m}{2}+2)}\phi^3_{0;m,k}
+\frac{1}{2}\sqrt{\frac{m}{2}(\frac{m}{2}+1)}\phi^3_{2;m,k}=\, 0, \\[6pt]
&\partial_{\tau}\phi^3_{2;m,k}
-\frac{1}{2}\sqrt{\frac{m}{2}(\frac{m}{2}+1)}\phi^3_{1;m,k}
+\frac{1}{2}\sqrt{\frac{m}{2}(\frac{m}{2}+1)}\phi^3_{3;m,k}=\, 0, \\[6pt]
&\partial_{\tau}\phi^3_{3;m,k} -\phi^3_{3;m,k}
-\frac{1}{2}\sqrt{(\frac{m}{2}+1)\frac{m}{2}}\phi^3_{2;m,k}
+\frac{1}{2}\sqrt{(\frac{m}{2}+2)(\frac{m}{2}-1)}\phi^3_{4;m,k}=\, 0, \\[6pt]
(1-\tau)&\partial_{\tau}\phi^3_{4;m,k} +\phi^3_{4;m,k}
-\sqrt{(\frac{m}{2}+2)(\frac{m}{2}-1)}\phi^3_{3;m,k}=\, 0. 
\label{phi-m-equation}
\end{array}
\end{equation}
The coefficients $\phi^3_{i;4,k}$, $k = 0,\dots,4$, solve
\begin{equation}
\begin{array}{r@{}l}
(1+\tau)&\partial_{\tau}\phi^3_{0;4,k} -\phi^3_{0;4,k} +2\,\phi^3_{1;4,k} = 
2\sqrt{6}\,A_1(\tau)\,W_{2;4,k} +\sqrt{6}\,A_2(\tau)\,a_k, \\[6pt] 
&\partial_{\tau}\phi^3_{1;4,k} +\phi^3_{1;4,k} -\phi^3_{0;4,k}
+\frac{1}{2}\sqrt{6}\,\phi^3_{2;4,k}= \frac{1}{2}\sqrt{6}\,B_2(\tau)\,a_k 
+\sqrt{6}\,B_3(\tau)\,W_{2;4,k}, \\[6pt]
&\partial_{\tau}\phi^3_{2;4,k} -\frac{1}{2}\sqrt{6}\,\phi^3_{1;4,k} 
+\frac{1}{2}\sqrt{6}\,\phi^3_{3;4,k} =    
[C_3(\tau)-C_5(\tau)]\,a_k +C_4(\tau)W_{2;4,k}, \\[6pt]
&\partial_{\tau}\phi^3_{3;4,k} +\phi^3_{3;4,k} 
+\phi^3_{4;4,k}-\frac{1}{2}\sqrt{6}\,\phi^3_{2;4,k}= 
-\frac{1}{2}\sqrt{6}\,B_2(-\tau)\,a_k-\sqrt{6}\,B_3(-\tau)\,W_{2;4,k}, \\[6pt]
(1-\tau)&\partial_{\tau}\phi^3_{4;4,k} +\phi^3_{4;4,k} -2\,\phi^3_{3;4,k} = 
-2\sqrt{6}\,A_1(-\tau)\,W_{2;4,k} -\sqrt{6}\,A_2(-\tau)\,a_k,
\label{phi-4-equation}
\end{array}
\end{equation}
with the coefficients $a_k$ defined in (\ref{def:a}). The functions
$\phi^3_{i;2,k}$, $k = 0, 1, 2$, satisfy
\begin{equation}
\begin{array}{r@{}l}
&\partial_{\tau}\phi^3_{1;2,k} +\phi^3_{1;2,k} 
+\frac{1}{\sqrt{2}\,}\phi^3_{2;2,k}=\,\sqrt{2}\,B_1(\tau)\,W_{1;2,k}, \\[6pt]
&\partial_{\tau}\phi^3_{2;2,k} -\frac{1}{\sqrt{2}\,}\phi^3_{1;2,k}
+\frac{1}{\sqrt{2}\,}\phi^3_{3;2,k}= \,C_2(\tau)\,W_{1;2,k}, \\[6pt]
&\partial_{\tau}\phi^3_{3;2,k} -\phi^3_{3;2,k} 
-\frac{1}{\sqrt{2}\,}\phi^3_{2;2,k}=\,-\sqrt{2}\,B_1(-\tau)\,W_{1;2,k},
\label{phi-2-equation}
\end{array}
\end{equation}
while $\phi^3_{2;0,0}$ is subject to
\begin{equation}
\partial_{\tau}\phi^3_{2;0,0} =\,C_1(\tau) +[C_3(\tau)+2\,C_5(\tau)]\,b, 
\label{phi-0-equation}
\end{equation}
with $b$ as defined in (\ref{def:a}). 

These ordinary differential systems have to be integrated for the initial data
(\ref{T-decomposed-phi-initial-values}) at $\tau = 0$. Since the equations are
already quite complicated, we used the program MapleV.4 for this purpose.   
Synthesizing the result of these integrations according to
(\ref{phi-T-decomposition}), we obtain the following concise
expressions for $\phi^3_i$ on $I'$.
\begin{equation}
\begin{array}{r@{}l}
\phi^3_0= &\,-(1+\tau)(1-\tau)^5X_+X_+W_3
+\frac{1}{12}f_0(\tau)\,m\,X_+X_+W_2 \\[6pt] 
&\,+\frac{1}{6}g_0(\tau)(X_+W_1)^2 +\frac{1}{4}h_0(\tau)X_+X_+r^2, 
\\[6pt]
\phi^3_1= &\,-5\,(1+\tau)^2(1-\tau)^4X_+W_3
+\frac{1}{6}f_1(\tau)\,m\,X_+W_2 \\[6pt] 
&\,+\frac{1}{3}g_1(\tau)W_1X_+W_1 +\frac{1}{2}h_1(\tau)X_+r^2 
+\frac{1}{2}k_1(\tau)\,m^2X_+W_1, \\[6pt]
\phi^3_2= &\,-20\,(1+\tau)^3(1-\tau)^3W_3 +f_2(\tau)\,m\,W_2 \\[6pt] 
&\,+g_2(\tau)(W_1)^2 +3\,h_2(\tau)r^2 +k_2(\tau)\,m^2W_1 \\[6pt]
&\,+p(\tau)\,m^4 +[q(\tau)-g_2(\tau)]\,b, \\[6pt]
\phi^3_3= &\,5\,(1+\tau)^4(1-\tau)^2X_-W_3
-\frac{1}{6}f_1(-\tau)\,m\,X_-W_2 \\[6pt] 
&\,-\frac{1}{3}g_1(-\tau)W_1X_-W_1 -\frac{1}{2}h_1(-\tau)X_-r^2 
-\frac{1}{2}k_1(-\tau)\,m^2X_-W_1, \\[6pt]
\phi^3_4= &\,-(1+\tau)^5(1-\tau)X_-X_-W_3
+\frac{1}{12}f_0(-\tau)\,m\,X_-X_-W_2 \\[6pt] 
&\,+\frac{1}{6}g_0(-\tau)(X_-W_1)^2 +\frac{1}{4}h_0(-\tau)X_-X_-r^2,
\label{3-order-solution-2}
\end{array}
\end{equation}
with $\tau$-dependent functions which can be found in appendix
[\ref{tau-polynomials}]. All the functions $\phi^3_i$ have polynomial
dependence on $\tau$.

The most interesting feature of this solution is its smoothness at 
$\tau = \pm 1$, which, in view of the singular behavior of equations
(\ref{phi-m-equation}), (\ref{phi-4-equation}) at these points, was not to be
expected from the beginning. To explain its significance we indicate the
argument which led to the asymptotic regularity condition (\ref{asregcond}).
The Bianchi equations, which were used to obtain the evolution equations for
the rescaled conformal Weyl spinor and, consequently, the second of the
transport equations (\ref{separated-transport-equations}), form an
overdetermined system. Thus there are further equations, to which we refer as
to the constraints. In the present case the constraints take the form     
\begin{equation}
\begin{array}{r@{}l}
\tau\partial_{\tau}\phi^3_1
+\frac{1}{2}(X_+\phi^3_2-X_-\phi^3_0) -3\phi^3_1= &\,S_1, \\[6pt]
\tau\partial_{\tau}\phi^3_2
+\frac{1}{2}(X_+\phi^3_3-X_-\phi^3_1) -3\phi^3_2= &\,S_2, \\[6pt]
\tau\partial_{\tau}\phi^3_3
+\frac{1}{2}(X_+\phi^3_4-X_-\phi^3_2) -3\phi^3_3= &\,S_3, 
\label{phi-constraints}
\end{array}
\end{equation}
where
\begin{equation}
\begin{array}{r@{}l}
S_1= &\, F_1(\tau)X_+W_1 +F_2(\tau)W_1X_+W_1 +F_3(\tau)X_+W_2, \\[6pt]
S_2= &\, G_1(\tau) +G_2(\tau)W_1 +G_3(\tau)(W_1)^2 +G_4(\tau)W_2
+G_5(\tau)X_-W_1X_+W_1, \\[6pt]
S_3= &\, -F_1(-\tau)X_-W_1 -F_2(-\tau)W_1X_-W_1 -F_3(-\tau)X_-W_2,
\label{rhs phi-constraints}
\end{array}
\end{equation}
with functions which are given in appendix [\ref{tau-polynomials}].
As before, we obtain equations for the coefficients in the expansion
(\ref{phi-T-decomposition}). Together with (\ref{phi-m-equation}),
(\ref{phi-4-equation}) these equations imply the systems
\begin{equation}
\begin{array}{l}
(1+\tau)(5\,\tau^2+3)\partial_{\tau}\phi^3_{0;6,k}
+(5\,\tau^3-5\,\tau^2+5\,\tau+7)\phi^3_{0;6,k}
-5\,(\tau-1)^3\phi^3_{4;6,k}=0,  \\[6pt]
(1-\tau)(5\,\tau^2+3)\partial_{\tau}\phi^3_{4;6,k}
+(5\,\tau^3+5\,\tau^2+5\,\tau-7)\phi^3_{0;6,k} -5\,(\tau+1)^3\phi^3_{4;6,k}=0,
\label{hom reduced-equations}
\end{array}
\end{equation}
and 
\begin{equation}
\begin{array}{l}
\phantom{-}4(3+\tau^2)(1+\tau)\partial_{\tau}\phi^3_{0;4,k} 
-2(1-\tau)^3\phi^3_{0;4,k} +2(1-\tau)^3\phi^3_{4;4,k}= 
\;T_{1}(\tau)\,a_k +T_{2}(\tau)\,W_{2;4,k}, \\[6pt]
-4(3+\tau^2)(1-\tau)\partial_{\tau}\phi^3_{4;4,k} -2(1+\tau)^3\phi^3_{4;4,k}
+2(1+\tau)^3\phi^3_{0;4,k}= \;T_{1}(-\tau)\,a_k+T_{2}(-\tau)\,W_{2;4,k},
\label{reduced-equations}
\end{array}
\end{equation}
with functions $T_1$ and $T_2$ (given in appendix
[\ref{tau-polynomials}]) derived from the functions $R_i$ and $S_j$.

It turns out that once these equations have been solved, the remaining
expansion coefficients in (\ref{phi-T-decomposition}) can be obtained
either by purely algebraic operations or by solving ODE's which are regular
for $\tau \in [-1, 1]$. This situation is the same for all orders 
$p \ge 3$ in (\ref{separated-transport-equations}). The solutions
$y(\tau)$, with $y$ denoting in the case above the column vector with
entries given by the two unknowns of (\ref{hom reduced-equations}) resp. of 
(\ref{reduced-equations}), can then be given for $p \ge 3$ in the form
(suppressing here all indices)
\begin{equation}
\label{gensol}
y(\tau) = X(\tau)\,X(0)^{-1}\,y_0
+  X(\tau)\,\int_0^{\tau} X(\tau')^{-1}\,b(\tau')\,d\tau',
\end{equation}
with $X(\tau)$ denoting a fundamental matrix of the system of ODE's
under study. The vector-valued function $b(\tau)$ is built from solutions
which are obtained by solving the equations of lower order. In
\cite{Fr:space-like-infinity} the equations (written there in a
slightly different form) have been discussed in general and the
fundamental matrices $X(\tau)$ have been derived. As in the case of
(\ref{hom reduced-equations}), (\ref{reduced-equations}), there occur
homogeneous as well as inhomogeneous systems for general $p \ge 3$. 
Thus for certain values of the indices (i.e. $p$ and the indices which arise
from expanding $u^p$ in terms of the functions $T_m\,^i\,_j$) the functions
$b(\tau)$ vanish and the solutions are of the form 
$y(\tau) = X(\tau)\,X(0)^{-1}\,y_0$. In these cases some of the entries
of $X(\tau)$ have logarithmic singularities. The latter drop out of the final
expression precisely if the asymptotic regularity conditions 
(\ref{asregcond}) are satisfied. In the remaining cases the entries of the
matrices $X(\tau)$ are polynomials in $\tau$ but 
$\det(X) = c\,f(\tau)\,(1 - \tau^2)^{p - 2}$ with some constant
$c \neq 0$ and some polynomial $f(\tau)$ satisfying $|f(\tau)| \ge 1$ for 
$|\tau| \le 1$. Furthermore, the column vector $b(\tau)$ has poles. However,
it has no logarithmic singularities if the solutions of the equations of lower
order have no logarithmic singularities. Assuming condition (\ref{asregcond}),
the remaining potential source of singularities of $u^p$, $p \ge 3$,  
at $|\tau| = \pm 1$ are the integrals on the right hand sides of the
expressions (\ref{gensol}). These have not been analyzed yet. To understand
the general situation, it is clearly of interest to study the problem for
the first few values of $p$. Remarkably, in the present case, $p = 3$,
we find that the integrand in (\ref{gensol}) has poles at 
$|\tau| = \pm 1$ and also outside the interval $[-1, 1]$, that the integral
has poles and no logarithmic terms, but that the final solution is
a polynomial in $\tau$.

\subsection{The detailed transformation formulae}
\label{explicit-gauge-tr.}

In this section we will determine expansions for the conformal scale factor 
$\theta$ and the $SL(2,C)$-valued function $\Lambda^a_{\phantom{a}b}$
which define the transformation from the F-gauge into the NP-gauge as
described in section [\ref{FNP-gauge-tr.}]. To calculate the NP-constants in
terms of the initial data we shall determine the values of the integrals
defining these quantities by taking their limits as $\rho\to{0}$. The gauge
in which these integrals are given is based on a section ${\cal C}$ of the
generators of ${\cal{J}}^{+}$. We shall try to push this section to $I^+$.
The usefulness of this procedure depends, of course, on the resulting form
of the ODE's on ${\cal{J}}^{+}$ which were used in [\ref{FNP-gauge-tr.}]
to fix the F-gauge. 

Near $I^+$ the hypersurface ${\cal{J}}^{\pm}$ can be given as the graph 
$\{\tau = \tau^s, \rho > 0\}$ of the function 
$\tau^s = \tau^s(\rho, t^a\,_b)$ which is given by
\begin{equation}
\tau^s=\frac{2\,\Omega}{\rho}[-D_{ab}\Omega D^{ab}\Omega]^{-\frac{1}{2}}.
\label{tau-scri}
\end{equation}
Substituting the expansions (\ref{Omega:Taylor-expansion}) of $\Omega$ and
those of the frame vectors into the expression above, we get 
the expansion
\begin{equation}
\tau^s= 1 +\frac{1}{2}m\,\rho +2\,W_1\rho^2 +O(\rho^3). 
\label{tau-rho-expansion}
\end{equation}

Setting in (\ref{f-equation}) $Z=\partial_{\tau}$, we obtain for the right
hand side of this equation the expansion 
\begin{equation}
\frac{Z(\frac{1}{2}\nabla_{\beta}\Theta
\nabla^{\beta}\Theta)}{Z(\Theta)}= \frac{5}{3}m\,\rho^2 
-\left(\frac{229}{63}m^2-\frac{24}{5}W_1\right)\rho^3
+O(\rho^4).
\end{equation}
Suppose $T = T^0\,\partial_{\tau} + T^1\,\partial_{\rho}
+ T^+\,X_+ + T^-\,X_-$ is a vector field defined near and tangent to
${\cal{J}}^+$. Denote by $T^*$ the vector field which is induced by it
on ${\cal{J}}^+$. If $\rho$ and $t^a\,_b$ are used as coordinates on 
${\cal{J}}^+$, one finds for $T^*$ the expression
$T^* =  T^1\,\partial_{\rho} + T^+\,X_+ + T^-\,X_-$. Applying this to
the gradient of $\Theta$ on ${\cal{J}}^+$, we find that the left
hand side of (\ref{f-equation}) is given by         
\begin{equation*}
\begin{array}{@{}l}
\Bigl(\bigl\{-2\rho^2+\frac{19}{3}m\,\rho^3+O(\rho^4)\bigr\}\partial_{\rho}
+\bigl\{\frac{36}{5}X_-W_1\,\rho^3+O(\rho^4)\bigr\}X_+ 
+\bigl\{\frac{36}{5}X_+W_1\,\rho^3+O(\rho^4)\bigr\}X_-\Bigr)(\log f).
\end{array}
\end{equation*}
Thus, dividing (\ref{f-equation}) on both sides by $\rho^2$, we get a
differential equation of the form $T^*(\log f) = g$ on ${\cal{J}}^+$  
with a vector field $T^*$ and a function $g$ which extend smoothly to
$I^+$ such that $T^* = - 2\,\partial_{\rho} + O(\rho)$ near $I^+$.  
For given datum $f_0$ on $I^+$ this equation has a unique smooth solution 
which can be expanded in terms of $\rho$. As shown in our general
discussion, the value of $f_0$ has to be  constant on ${\cal C}$ to fulfill
the NP-gauge conditions. We choose $f_0 = -\frac{1}{2\sqrt2}$
on $I^+$ and find for the solution of (\ref{f-equation}) the expansion
\begin{equation}
f= -\frac{1}{2\sqrt{2}}\left\{1 +\frac{5}{6}m\,\rho 
+\left(\frac{191}{252}m^2+\frac{6}{5}W_1\right)\rho^2 +O(\rho^3)\right\}. 
\label{f-rho-eqpansion}
\end{equation}

To obtain the matrix elements $\lambda^a_{\phantom{a}b}$ of
(\ref{Lorentz-transf.-1}) by using (\ref{matrix-elements1}) we have to
calculate the derivatives $c^*_{aa'}(\Theta)$ of the conformal factor. 
Using the expansion coefficients derived in [\ref{3rd-order-solution}],
we get 
\begin{equation}
\begin{array}{ll}
\multicolumn{2}{l}{c^*_{00'}(\Theta)= O(\rho^4),} \\[6pt]
c^*_{01'}(\Theta)= \sqrt{2}\Bigl\{X_+W_1\rho^3+O(\rho^4)\Bigr\}, &
c^*_{10'}(\Theta)= \sqrt{2}\Bigl\{X_-W_1\rho^3+O(\rho^4)\Bigr\}, \\[6pt]
\multicolumn{2}{l}{c^*_{11'}(\Theta)= \sqrt{2}\Bigl\{-2\,\rho+3\,m\rho^2 
+(8\,W_1-3\,m^2)\rho^3 +O(\rho^4)\Bigr\}.}
\end{array}
\end{equation}
Substituting these expressions into the formulae (\ref{matrix-elements1})
the matrix elements $\lambda^0_{\phantom{0}1}$ and
$\lambda^1_{\phantom{1}1}$ can be calculated explicitly up to a $U(1)$
phase transformation. Since the choice of the latter is not important for
the following we choose it suitably to obtain  
\begin{equation}
\begin{array}{@{}ll}
\lambda^0_{\phantom{0}1}= \rho^\frac{1}{2}\Bigl\{1-\frac{1}{3}m\rho
+(-\frac{7}{5}W_1+\frac{113}{252}m^2)\rho^2
+O(\rho^3)\Bigr\}, &
\lambda^1_{\phantom{1}1}= \rho^\frac{5}{2}\Bigl\{\frac{1}{2}X_+W_1 
+O(\rho)\Bigr\}, \label{matrix1}
\end{array}
\end{equation}
which allows us to determine also the expansion 
\begin{equation}
\begin{array}{r@{}l}
E^{\circ}_{11'}= &\sqrt{2}\Bigl\{\frac{1}{4}m\rho^2 
+(-\frac{7}{12}m^2+2W_1)\rho^3+O(\rho^4)\Bigr\}\partial_{\tau} \\[6pt]
&+\sqrt{2}\Bigl\{\frac{1}{2}\rho^2-\frac{7}{6}m\rho^3
+(\frac{577}{252}m^2-\frac{31}{5}W_1)\rho^4 
+O(\rho^5)\Bigr\}\partial_{\rho} \\[6pt]
&+\sqrt{2}\Bigl\{-\frac{9}{5}X_-W_1\rho^3
+O(\rho^4)\Bigr\}X_+ +\sqrt{2}\Bigl\{-\frac{9}{5}X_+W_1\rho^3
+O(\rho^4)\Bigr\}X_-  \label{E^circ_11'}.
\end{array}
\end{equation}
To solve the differential equation for the affine parameter on the
generators of ${\cal{J}}^+$, we observe that already in the case of
Minkowski space-time this parameter is a singular function of $\rho$, given
by $u^{\circ}=-\sqrt{2}\rho^{-1}+u^{\circ}_*$. The inspection of the
expansion (\ref{E^circ_11'}) suggests to search for a solution of the form  
\begin{equation}
u^{\circ} = w + \sqrt{2}\left(-\frac{1}{\rho}
+\frac{7}{3}\,m\,{\rm{log}}\,\rho\right).
\end{equation}
This ansatz does indeed lead to a smooth regular equation for $w$ near
$I^+$. It allows us to calculate the expansion
\begin{equation}
u^{\circ}= \sqrt2\left\{-\frac{1}{\rho} +\frac{7}{3}m\,{\rm{log}}\,\rho
+u^{\circ}_* + \left(\frac{109}{126}m^2+\frac{62}{5}W_1\right)\rho 
+O(\rho^2)\right\}, \label{affine-par.-u}
\end{equation}
where $u^{\circ}_*$ denotes an arbitrary constant initial datum on $I^+$.
As described in chapter [\ref{FNP-gauge-tr.}], the matrix elements
$\lambda^0_{\phantom{0}0}$ and $\lambda^1_{\phantom{1}0}$ can now be
determined. We obtain the expansions 
\begin{equation}
\begin{array}{@{}ll}
\lambda^0_{\phantom{0}0}= \rho^\frac{3}{2}\Bigl\{ 
\frac{77}{10}X_-W_1+O(\rho) \Bigr\}, & 
\lambda^1_{\phantom{1}0}= \rho^{-\frac{1}{2}}\Bigl\{-1-\frac{1}{3}m\,\rho
+O(\rho^2)\Bigr\}.   \label{matrix2}
\end{array}
\end{equation}

Knowing the matrix $\lambda^a_{\phantom{a}b}$ on null infinity, we can
calculate the limits of the NP-spin-coefficients
$\Gamma^{\circ}_{01'11}$ and $\Gamma^{\circ}_{10'00}$ at $I^+$ as 
$\rho \rightarrow 0$. Substituting our expansions into the formula
for the connection coefficients
\begin{equation}
\Gamma^{\circ}_{aa'bc}= \lambda^f_{\phantom{f}a}
\bar{\lambda}{}^{f'}_{\phantom{f'}a'}\lambda^g_{\phantom{g}b}
\lambda^h_{\phantom{h}c}\Gamma^*_{ff'gh}
-\epsilon_{gh}\lambda^g_{\phantom{g}b}
E^{\circ}_{aa'}(\lambda^h_{\phantom{h}c}),
\end{equation}
we arrive at the expressions
\begin{equation}
\left.\Gamma^{\circ}_{01'11}\right|_{I^+}=
\lim_{\rho\rightarrow0}\Gamma^{\circ}_{01'11}=0, \;\;\;
\left.\Gamma^{\circ}_{10'00}\right|_{I^+}=
\lim_{\rho\rightarrow0}\Gamma^{\circ}_{10'00}=
\frac{11}{6\sqrt{2}}\,m.
\end{equation}

The next step is to calculate the conformal scale factor $\theta$
by solving equation (\ref{eq:conf.factor}). To determine the Ricci spinor
component $\Phi_{22}=
\frac{1}{2}R_{\alpha\beta}E^{\circ\,\alpha}_{11'}E^{\circ\,\beta}_{11'}$,
we have to determine the Ricci tensor $R_{\alpha\beta}$ of the metric $g$.
The components of the tensor
\begin{equation}
\Theta_{\alpha\beta}:=
\frac{1}{2}\hat{R}_{(\alpha\beta)} -\frac{1}{12}g_{\alpha\beta}\hat{R} 
+\frac{1}{4}\hat{R}_{[\alpha\beta]}
\end{equation}
in the frame $\{c^*_{aa'}\}$, where $\hat{R}_{\alpha\beta}$ resp. $\hat{R}$
denote the Ricci tensor and the curvature scalar induced by the Weyl
connection $\hat{\nabla}$ with coefficients
$\hat{\Gamma}_{\alpha\;\gamma}^{\;\;\beta}=
\Gamma_{\alpha\;\gamma}^{\;\;\beta}
+\delta^{\beta}_{\alpha}f_{\gamma}+ \delta^{\beta}_{\gamma}f_{\alpha}
-g_{\alpha\gamma}f^{\beta}$ (cf. \cite{Fr:AdS}), are among the variables
of the conformal field equations. Thus they are known to $3^{rd}$-order
in the $\rho$-coordinate. From the general transformation law 
\begin{equation}
\hat{R}_{\alpha\beta}=R_{\alpha\beta}-2\nabla_{(\alpha}f_{\beta)}
+2f_{\alpha}f_{\beta} -g_{\alpha\beta}(\nabla_{\gamma}f^{\gamma}
+2f_{\gamma}f^{\gamma}) +4\nabla_{[\alpha}f_{\beta]},
\end{equation}
we get the relation
\begin{equation}
\Theta_{\alpha\beta}=\frac{1}{2}\left(R_{\alpha\beta}
-\frac{1}{6}g_{\alpha\beta}R\right)-\nabla_{\beta}f_{\alpha}
+f_{\alpha}f_{\beta}-\frac{1}{2}g_{\alpha\beta}f_{\gamma}f^{\gamma}.
\label{Ricci-spinor} 
\end{equation}
From this we derive the expression
\begin{equation}
\Phi_{22}=\Theta_{\alpha\beta}E^{\circ\,\alpha}_{11'}E^{\circ\,\beta}_{11'} 
+E^{\circ}_{11'}(E^{\circ\,\alpha}_{11'}f_{\alpha}) 
-(E^{\circ\,\alpha}_{11'}f_{\alpha})^2. 
\end{equation}
Substituting here (\ref{E^circ_11'}) and the expansion of
the one-form $f$ obtained from the solution of the field
equations we get the expansion
\begin{equation}
\Phi_{22}= \frac{5}{6}m\rho^3
+\left(-\frac{167}{42}m^2+\frac{18}{5}W_1\right)\rho^4 +O(\rho^5)
\end{equation}
on ${\cal J}^+$. 

On $I^+$ is induced in our gauge the standard $S^2$-metric. Therefore
we solve equation (\ref{eq:conf.factor}) with the initial condition
\begin{equation}
\lim_{\rho\rightarrow{0}}\theta=1.
\end{equation}
For the conformal scale factor we obtain then the expansion 
\begin{equation}
\theta= 1 +\frac{5}{6}m\rho 
+\left(\frac{6}{5}W_1+\frac{191}{252}m^2\right)\rho^2+O(\rho^3).
\label{theta}
\end{equation}
By the choice of the initial value for the conformal factor the scale
function $p$ appearing in the gauge transformations is also fixed with 
\begin{equation}
\begin{array}{ccc}
p\equiv 1 & {\rm{on}} & {\cal{J}}^+.
\end{array}
\end{equation}
In the conformal gauge characterized by the conformal
factor $\Theta^{\star}:=\theta\,\Theta$ the generators of null
infinity are expansion free. Proceeding as indicated before, we 
construct the NP-frame $\{E^{\bullet}_{aa'}\}$.  
Observing the expansions (\ref{Lorentz-transf.-1}) and
(\ref{Lorentz-transf.-2}) of the null vectors $E^{\circ}_{11'}$ resp. 
$E^{\bullet}_{11'}$ and taking into account the properties of the conformal
rescaling we get the relations
\begin{equation}
\begin{array}{cc}
\Lambda^0_{\phantom{0}1}= 
\theta^{-\frac{1}{2}}\lambda^0_{\phantom{0}1}e^{ic}, &
\Lambda^1_{\phantom{1}1}= 
\theta^{-\frac{1}{2}}\lambda^1_{\phantom{1}1}e^{ic},
\end{array}
\end{equation}
with function $c$, characterizing the phase freedom, which will be 
fixed later. Using (\ref{matrix1}) and (\ref{theta}) we get the expansions 
\begin{equation}
\begin{array}{@{}ll}
\Lambda^0_{\phantom{0}1}= {\rho}^{\frac{1}{2}} \Bigl\{1 -\frac{3}{4}\,m{\rho} 
+\left({\frac{15}{32}}\,{m}^{2}-2\,W_{1}\right){\rho}^{2} 
+O(\rho^3)\Bigr\}e^{ic},  &
\Lambda^1_{\phantom{1}1}= {\rho}^{\frac{5}{2}}\Bigl\{\frac{1}{2}\,X_+W_1 
+O(\rho)\Bigr\}e^{ic},
\end{array}
\end{equation}
from which we derive in turn the expansion
\begin{equation}
\begin{array}{r@{}l}
E^{\bullet}_{11'}= &\sqrt{2}\Bigl\{\frac{1}{4}m\rho^2 
+(-m^2+2W_1)\rho^3+O(\rho^4)\Bigr\}\partial_{\tau} \\[6pt]
&+\sqrt{2}\Bigl\{\frac{1}{2}\rho^2-2\,m\rho^3
+(\frac{253}{56}m^2-\frac{37}{5}W_1)\rho^4 
+O(\rho^5)\Bigr\}\partial_{\rho} \\[6pt]
&+\sqrt{2}\Bigl\{-\frac{9}{5}X_-W_1\rho^3+O(\rho^4)\Bigr\}X_+
+\sqrt{2}\Bigl\{-\frac{9}{5}X_+W_1\rho^3+O(\rho^4)\Bigr\}X_-, 
\label{tangent-vekt.-2}
\end{array}
\end{equation}
of the vector field $E^{\bullet}_{11'}$ tangent to the null generators
of ${\cal J}^+$.
Furthermore the new affine parameter has the form
\begin{equation}
u^{\bullet}=\sqrt{2}\left\{-\frac{1}{\rho} +4\,m\,{\rm{log}\,}\rho 
+u^{\bullet}_* +\left(\frac {195}{28}\,{m}^{2}+\frac{74}{5}\,W_1\right)\rho
+O(\rho^2) \right\}, \label{affine-par.-v}
\end{equation}
with a free constant $u^{\bullet}_*$. 
Using the formula analogous to (\ref{lambda00}) we derive
\begin{equation}
\begin{array}{@{}ll}
\Lambda^0_{\phantom{0}0}= \rho^{\frac{1}{2}}\Bigl\{ 
-\frac{101}{10}\,X_-W_1\rho +O(\rho^2)\Bigr\}e^{-ic}, &
\Lambda^1_{\phantom{1}0}= \rho^{-\frac{1}{2}}\Bigl\{
-1 -\frac{3}{4}\,m\,\rho +O(\rho^2)\Bigr\}e^{-ic}.
\end{array}
\end{equation}
To determine of the phase factor $e^{\pm{ic}}$ we solve equation
(\ref{ode:phase}) along the generators of null infinity. Expanding the
right hand side, we get
\begin{equation}
E^{\bullet}_{11'}(c)= 2\,{\mathfrak{Im}\,} \Bigl\{
\hat{\Lambda}^f_{\phantom{f}1}\bar{\hat{\Lambda}}^{f'}_{\phantom{f'}1'}
\hat{\Lambda}^g_{\phantom{g}1}\hat{\Lambda}^h_{\phantom{h}0}\Gamma^{\star}_{ff'gh}
-\hat{\Lambda}^{0}_{\phantom{0}0}E^{\bullet}_{11'}(\hat{\Lambda}^{1}_{\phantom{1}1}) 
+\hat{\Lambda}^{1}_{\phantom{1}0}E^{\bullet}_{11'}(\hat{\Lambda}^{0}_{\phantom{0}1})
\Bigr\}, \label{ode:phase-1} 
\end{equation}
where $\hat{\Lambda}^{a}_{\phantom{a}b}$ has been obtained from the matrix
$\Lambda^{a}_{\phantom{a}b}$ above by setting $c = 0$.
Substituting the known data into the equation above, the solution $c$
which is needed to satisfy the gauge condition
$\Gamma^{\bullet}_{11'01}|_{\cal{J}}=0$, is found to have an expansion
\begin{equation} 
c=O(\rho^2), \label{phase-1}
\end{equation}
which entails the expansions
\begin{equation}
\begin{array}{ccc}
e^{ic}= 1+O(\rho^2), & E^{\bullet}_{11'}(e^{ic})= O(\rho^3), &
E^{\bullet}_{01'}(e^{ic})= O(\rho^2). \label{phase-2}
\end{array}
\end{equation}

The matrix elements $\Lambda^a_{\phantom{a}b}$ are now determined
on null infinity to the precision needed in our later calculations, but in
the definition
(\ref{NPQ-final-definition}) of the NP-constants appear some of the
transversal derivatives
${E^{\bullet}_{00'}(\Lambda^a_{\phantom{a}b})}$ of the matrix
elements as well. Using the general formulae
(\ref{transversal-derivatives}) we get the expansions
\begin{equation}
\begin{array}{@{}ll}
E^{\bullet}_{00'}(\Lambda^0_{\;0})= 
\sqrt{2}\rho^{\frac{1}{2}}\Bigl\{\frac{113}{40}X_-W_1 +O(\rho)\Bigr\}, &
E^{\bullet}_{00'}(\Lambda^1_{\;0})= 
\sqrt{2}\rho^{-\frac{3}{2}}\Bigl\{\frac{1}{4}
+\frac{85}{48}m\rho +O(\rho^2)\Bigr\}, \\[6pt]
E^{\bullet}_{00'}(\Lambda^0_{\;1})= \sqrt{2}
\rho^{-\frac{1}{2}}\Bigl\{\frac{1}{4}+\frac{67}{48}\,m\,\rho
+O(\rho^2)\Bigr\}, &
E^{\bullet}_{00'}(\Lambda^1_{\;1})= \sqrt{2}
\rho^{\frac{3}{2}}\Bigl\{-\frac{47}{40}X_+W_1 +O(\rho)\Bigr\}, 
\end{array}
\end{equation}
where we have taken the expressions (\ref{phase-2}) for the phase factor
into account.

The transversal derivative of the conformal scale factor
$E^{\bullet}_{00'}(\theta)$ is fixed on null infinity by the
requirement $R[g^{\star}]|_{{\cal{J}}^+} =0$. Thus it has to satisfy
equation (\ref{ode:tr.derivative}) with initial datum 
\begin{equation}
\left.E^{\bullet}_{00'}(\theta)\right|_{I^+} = 
\displaystyle{\lim_{\rho\rightarrow{0}}\,\theta\,p^{-1}\Gamma^{\circ}_{10'00}=
\lim_{\rho\rightarrow{0}}\Gamma^{\circ}_{10'00}}.
\end{equation}
Given the matrix $\Lambda^a_{\phantom{a}b}$ and the
conformal scale factor $\theta$, all the terms appearing in 
equation (\ref{ode:tr.derivative}) can be calculated in a straightforward
way, with the exception of the curvature scalar $R[g]$, whose calculation
requires some explanation. Contracting equation (\ref{Ricci-spinor}) we get
the identity
\begin{equation}
R[g]=6\,(\Theta_{aa'bb'}+\nabla_{aa'}f_{bb'}+f_{aa'}f_{bb'})
\epsilon^{ab}\bar{\epsilon}^{a'b'}, \label{curvature-scalar}
\end{equation}
where 
\begin{equation*}
\nabla_{aa'}f_{bb'}=c^*_{aa'}(f_{bb'})-(\Gamma^*_{aa'cb}\bar{\epsilon}_{b'c'}
+\bar{\Gamma}^*_{aa'c'b'}\epsilon_{bc})f^{cc'}.
\end{equation*}
Expanding these quantities we get
\begin{equation}
\begin{array}{l}
R[g]=\Bigl(\frac{23}{3}\,m^2-\frac{168}{5}W_1\Bigr)\rho^2 +O(\rho^3), \\[6pt]
F^{\star}=\Bigl(\frac{23}{36}\,m^2-4W_1+
\frac{6}{5}a_-X_+W_1-\frac{6}{5}a_+X_-W_1\Bigr)\rho^2+O(\rho^3),
\end{array}
\end{equation}
which entail with (\ref{ode:tr.derivative}) the expansion
\begin{equation}
E^{\bullet}_{00'}(\theta)= \sqrt{2}\left\{\frac{11}{12}\,m
+\left(\frac{13}{6}\,m^2-4W_1
+\frac{6}{5}a_-X_+W_1-\frac{6}{5}a_+X_-W_1\right)\rho + O(\rho^2)\right\}.
\end{equation}

Given the expansion above, we can calculate expansions of various quantities
of physical interest, such as the Bondi energy momentum, the angular
momentum, and the radiation field on ${\cal J}^+$. Since the
coefficients in these expansions are given directly in terms of the initial
data on the Cauchy hypersurface $S$, the expansions contain information
about the evolution of the field  over an infinite range. As an example we
will calculate below the NP-constants. 

We close this section with a remark on the
BMS group, the group of transformation between different Bondi-type
systems. It was shown in \cite{NP:BMS} that for solutions for which
the the condition
$\displaystyle{\lim_{u^{\bullet}\rightarrow{-\infty}}
\Gamma^{\bullet}_{[e]\;01'00}=0}$ could be realized at space-like
infinity, where the subscript ``e'' is to denote the electric part of
the considered spin-coefficient, one can single out the inhomogeneous
Lorentz group as the group of transformations preserving this
condition. It turns out that under our assumptions, which include in
particular the time-symmetry of the solution, the even stronger
condition $\displaystyle{\lim_{u^{\bullet}\rightarrow{-\infty}}}
\Gamma^{\bullet}_{01'00}=0$ is satisfied. This means that for our
solutions there is a natural way to single out the inhomogeneous
Lorentz group as asymptotic symmetry group.

\subsection{The NP-constants in time symmetric space-times}
\label{explicit-NP-constants} 

Using the formulae of the previous chapters we can express the
NP-constants in terms of the initial data for the corresponding time
symmetric solutions. All the quantities appearing in the integral
(\ref{NPQ-final-definition}) are known in terms of the initial data
to the precision needed to perform the limit $\rho\rightarrow{0}$.

We have to express the spin-2 spherical harmonics $_2\bar{Y}_{2,m}$ in
terms of the functions $T_{m\;\;k}^{\;\;j}$. By
(\ref{spin-2-harmonics}) the definition of the \dh-operator is based
on the choice of the complex null vector field $E^{\bullet}_{01'}$. In
appendix [\ref{edth-operator}] we have applied the standard choice and
derived the relations between the operators $X_+$ and \dh\; and
between the spin-2 spherical harmonics $_2Y_{2,m}$ and the functions
$T_{m\;\;k}^{\;\;j}$. By this choice we should have 
$E^{\bullet}_{01'} = \frac{i}{\sqrt{2}}X_+$ on $I^+$.
However, calculating the vector $E^{\bullet}_{01'}$ in the conventions
used above, we get
\begin{equation}
E^{\bullet}_{01'}|_{I^+}= \frac{1}{\sqrt{2}}X_-.
\label{E01}
\end{equation}
There are two causes of the difference. We fixed the phase factor such
as to simplify the calculations and the conventions used in the
F-gauge and the NP-gauge are such that one has to swap the two spinors
of the dyad to get from one to the other 
convention. The form (\ref{E01}) of $E^{\bullet}_{01'}$ corresponds to
$-i\sqrt{2}\bar{m}$, if $m$ denotes the standard complex null vector
used in appendix [\ref{edth-operator}]. This means that (\ref{E01})
corresponds to the operator $-i\bar{\textrm\dh}$ instead of \dh\;
discussed in the appendix. Observing this and (\ref{spin-2-T}) in
(\ref{NPQ-semifinal-definition}) we obtain the formula
\begin{equation}
\begin{array}{lll}
G_m=\displaystyle{ 
i^{2-m}(5\pi)^{\frac{1}{2}}\oint\bar{T}_{4\;\;\;4}^{2-m}E^{\bullet}_{00'}(\phi_0)\mu}
& {\textrm{for}} &  m=-2,\dots,2,
\label{NPQ-for-evaluation}
\end{array}
\end{equation}
where $\mu=\frac{1}{4\pi^2}d{\cal{S}}d\alpha$ is the Haar-measure on
$SU(2)$.

To calculate (\ref{NPQ-for-evaluation}) we expand the integrand
in terms of $\rho$ and take the limit as $\rho\rightarrow{0}$. 
For this we have to determine for $E^{\bullet}_{00'}(\phi_0)$ only the
terms of order O(1). In the limit only these terms give a
contribution while the terms of order $\rho^{-1}$ cancel each other.   
Using the explicit results of the previous chapters we arrive after
some lengthy but straightforward calculations at the expression 
\begin{equation}
\begin{array}{@{}l}
\multicolumn{1}{c}{G_m\bigl|_{I^+}=
\displaystyle{\lim_{\rho\rightarrow{0}}}\,G_m=
i^{2-m}(10\pi)^{\frac{1}{2}}} \\[6pt]
\displaystyle{\oint} 
\bar{T}_{4\;\;\;4}^{2-m}\Bigl(-\frac{5}{32}X_-X_-r^2
+\frac{635}{8}\,m\,X_-X_-W_2 
-\frac{1905}{2}\,(X_-W_1)^2+\frac{16}{3}X_-X_-W_3\Bigr)\mu.
\end{array}
\end{equation}
Expanding the functions in the brackets in terms of the functions
$T_{m\;j}^{\;\;k}$ and using the orthogonality relations satisfied by these
functions we can perform the integration. All terms expect the last one
give some contributions. Using the formulae (\ref{W:T-expansion-2}),
(\ref{SR:T-expansion}) and (\ref{quadratic-W-expansions}) we get the
final expression
\begin{equation}
G_m\bigl|_{I^+}= \frac{i^{2-m}}{2}(15\pi)^{\frac{1}{2}}\left\{
127\left(m\,W_{2;4,2-m}-6\,a_{2-m}\right)
-\frac{1}{2\sqrt{6}}\,R^*_{2-m}\right\},
\end{equation}
where the coefficients $a_{2-m}$, which are quadratic in $W_{1;2,k}$, 
are given by (\ref{def:a}). We note that the structure of this more general
expression is essentially the same as that of the expression obtained by
Newman and Penrose in the case of static and stationary solutions.

\sect{Concluding remarks}

We have seen that, under the assumptions explained above, certain fields which
are given near space-like infinity in terms of Bondi-type systems can be
expressed in a straightforward way in terms of the gauge conditions used in
\cite{Fr:space-like-infinity} and can thus be related directly to the
structure of the Cauchy data which give rise to the space-times by
Einstein evolution. The calculations involved are quite lengthy but taking
into account that we relate quantities which are obtained by a non-linear
evolution over an infinite domain of space-time to the data from which they
arise, the overall structure of the argument is surprisingly simple.  

\vspace{.5cm}

ACKNOWLEDGMENTS: One of us (H.F.) would like to
thank the ITP in Santa Barbara and the KFKI RMKI in Budapest for
hospitality, the work of H.F. was supported in part by the National
Science Foundation and Grant No.  PHY94-07194. J.K. would like to
thank the AEI in Potsdam for hospitality where part of this work was
completed, his research was supported by the grant OTKA-D25135.

\appendix

\sect{Appendix}

\subsection{$X_+$ and the \dh-operator} 
\label{edth-operator}

In this section we describe the relation between the
operators ${\textrm\dh}$, $\bar{\textrm\dh}$ introduced in
\cite{NP:BMS} and the operators $X_+$, $X_-$, $X$ used in 
\cite{Fr:space-like-infinity}.

Consider on the group $SU(2)$, which is diffeomorphic to $S^3$,
coordinates $\{x$, $y$, $\alpha\}$ such that outside a set of measure
zero the general group element $t^a_{\phantom{a}b}\in{SU(2)}$ is given
by
\begin{equation}
t^a_{\;b}= \frac{1}{\sqrt{1+\zeta\bar{\zeta}}}
\begin{pmatrix}
e^{i\alpha} & ie^{-i\alpha}\zeta \\
ie^{i\alpha}\bar{\zeta} & e^{-i\alpha}
\end{pmatrix},
\end{equation}
with $\zeta =x+iy$. Then $\alpha$ is a parameter and $x$ and $y$ are
constant on the orbits of the the subgroup $U(1)$. The tangent vectors
$\partial_x$, $\partial_y$, respectively $\partial_{\alpha}$ at the
unit element coincide with the generators $u_1$, $u_2$, and $u_3$ of
the Lie algebra of $SU(2)$. Writing $P =\frac{1}{2}(1+\zeta\bar\zeta)$, we
get for the corresponding left invariant vector fields the expressions
\begin{equation}
\begin{array}{l}
Z_{u_1}= P\cos(2\alpha)\partial_x +P\sin(2\alpha)\partial_y
+\frac{1}{2}[x\sin(2\alpha)-y\cos(2\alpha)]\partial_{\alpha}, \\[6pt]
Z_{u_2}= -P\sin(2\alpha)\partial_x +P\cos(2\alpha)\partial_y
+\frac{1}{2}[y\sin(2\alpha)+x\cos(2\alpha)]\partial_{\alpha}, \\[6pt]
Z_{u_3}= \frac{1}{2}\partial_{\alpha},
\end{array}
\end{equation}
whence
\begin{equation}
\begin{array}{lr}
X_+ =-Z_{u_2}-iZ_{u_1}= 
e^{2i\alpha}\{-i\sqrt{2}(m-\frac{i}{2\sqrt{2}}\bar\zeta\partial_{\alpha}\}, &
X =-2iZ_{u_3}=-i\partial_{\alpha}, \\[6pt]
X_- =-Z_{u_2}+iZ_{u_1}= 
e^{-2i\alpha}\{i\sqrt{2}(\bar{m}+\frac{i}{2\sqrt{2}}\zeta\partial_{\alpha}\},
&
\label{X-op}
\end{array}
\end{equation}
where the vectors $m =\sqrt{2}P\partial_{\zeta}$ and $\bar{m}=
\sqrt{2}P\partial_{\bar\zeta}$ define a complex dyad 
tangent to the surfaces $\{\alpha=const.\}$ which is null
with respect to the standard $S^2$-metric
$ds^2=P^{-2}d\zeta\,d\bar\zeta$ on these surfaces.

We may identify $SU(2)$ with the spin frame bundle over the base
manifold $S^2$ with structure group $U(1)$. The section $\{\alpha=0\}$
can be identified with the base manifold (with a point omitted). Here
we take the complex null frame $\{m,\bar{m}\}$ defined above, where a
group element $u^a_{\;b}=diag(e^{i\alpha},e^{-i\alpha}) \in U(1)$ acts
as $u(\{m,\bar{m}\})=\{e^{2i\alpha}m,e^{-2i\alpha}\bar{m}\}$.  A
function $\eta$ on $S^3$ is said to have spin weight N, if it can be
decomposed as $\eta|_{\zeta,\alpha}=e^{2Ni\alpha}\eta_0$, where the
function $\eta_0$ is independent of the parameter $\alpha$ along the
fibers. The \dh-operator is defined by the complex null vector $m$ and
acts on a spin-N function as
\begin{equation}
{\textrm\dh}\eta|_{\zeta,\alpha}= \sqrt{2}\bigl\{m(\eta_0) 
+N{\eta_0}\,\bar{m}^{\gamma}m^{\beta}\delta_{\beta}m_{\gamma}\bigr\} 
e^{2(N+1)i\alpha}= 
\sqrt{2}\bigl\{m(\eta_0)+\frac{1}{\sqrt{2}}N\bar\zeta\eta_0\bigr\}e^{2(N+1)i\alpha},
\end{equation}
where $\delta$ denotes the Levi-Civita differential operator induced
by the standard $S^2$-metric. This means that ${\textrm\dh}\eta$ has
spin weight $N+1$. (This treatment of the functions with spin weight
and the \dh\; operator is a bit different from the one which can be
found in the literature (cf. \cite{NP:BMS,G:edth-operator,Ko-Newman}),
where the expressions are evaluated on some cross-section of $S^3$.)

The horizontal lift of the vector $m$ defined with respect to the
Levi-Civita connection $\delta$ is given by
\begin{equation}
\left.m_H\right|_{\zeta,\alpha}= 
m-\frac{i}{2\sqrt{2}}\hat\zeta\partial_{\alpha}.
\label{m-hor-lift}
\end{equation}
This means that the \dh-operator on $S^3$ is given by
\begin{equation}
\left.{\textrm\dh}\right|_{\zeta,\alpha}=\sqrt{2}\,e^{2i\alpha}m_H.
\label{edth-op}
\end{equation}
Comparing the formulae (\ref{X-op}), (\ref{m-hor-lift})
and (\ref{edth-op}) we get the relations
\begin{equation}
\begin{array}{lll}
X_+=-i{\textrm\dh}, & X_-=i\bar{\textrm\dh}, &
X=-[{\textrm\dh},\bar{\textrm\dh}]. 
\label{X-edth}
\end{array}
\end{equation}

The spherical harmonics $Y_{l,m}$ are defined as an orthogonal
function system on the sphere $S^2$. They can be extended to $S^3$ as
functions with zero spin weight, i.e. they became independent on the
parameter along the fibers. This means that they can be expanded as
$Y_{l,m}=\displaystyle{\sum_{k,j}c_{kj}T_{2k\;\;k}^{\;\;\;j}}$ in
terms of the functions $T_{m\;\;k}^{\;\;j}$. The spherical harmonics
satisfy the equation ${\textrm\dh}\bar{\textrm\dh}Y_{l,m}=
-l(l+1)Y_{l,m}$, so using the relations (\ref{X-edth}) and
(\ref{Xact}) we arrive at the relation
\begin{equation}
Y_{l,m}=\displaystyle{\sum_{j}c_jT_{2l\;\;l}^{\;\;\;j}}.
\label{Y-T}
\end{equation}
Taking into account the explicit coordinate expressions of the group
elements one could determine the expansion coefficients $c_j$. Using
the definition of the spin harmonics $_sY_{l,m}$ (cf.
\cite{G:edth-operator}) and equations (\ref{Xact}), (\ref{X-edth}) and
(\ref{Y-T}) one can also derive the relation between the functions
$_sY_{l,m}$ and the functions $T_{m\;\;k}^{\;\;j}$. We shall only need
the transformation formulae
\begin{equation}
\begin{array}{r@{}ll}
Y_{2,m}&= 
(-i)^{4-m}\bigl(\frac{5}{4\pi}\bigr)^{\frac{1}{2}}T_{4\;\;2}^{2-m}, & \\[6pt]
_2Y_{2,m}&= 
(-i)^{2-m}\bigl(\frac{5}{4\pi}\bigr)^{\frac{1}{2}}T_{4\;\;0}^{2-m}, &
_{-2}Y_{2,m}=
(-i)^{2-m}\bigl(\frac{5}{4\pi}\bigr)^{\frac{1}{2}}T_{4\;\;4}^{2-m}. 
\label{spin-2-T}
\end{array}
\end{equation}

\subsection{Some useful spinor identities}

\label{spinorial-identities}

Here we describe irreducible decompositions of spinors with four
unprimed indices in terms of the ``primary spinors'' $\varepsilon^i_{abcd}$,
$h_{abcd}$, $x_{ab}$, $y_{ab}$, $z_{ab}$ and $\epsilon_{ab}$, where
\begin{equation}
\begin{array}{r@{}lr@{}lr@{}l}
x_{ab}= &\; \sqrt{2}\epsilon_{(a}^{\phantom{(a}0}
\epsilon_{b)}^{\phantom{b)}1}, & 
y_{ab}= &\; -\frac{1}{\sqrt{2}}\epsilon_{a}^{\phantom{a}1}
\epsilon_{b}^{\phantom{b}1}, & 
z_{ab}= &\; \frac{1}{\sqrt{2}}\epsilon_{a}^{\phantom{a}0}
\epsilon_{b}^{\phantom{b}0}, \\[6pt] 
\varepsilon^i_{abcd}= &\; \epsilon_{(a}^{\phantom{(a}(e}
\epsilon_{b}^{\phantom{b}f} \epsilon_{c}^{\phantom{c}g}
\epsilon_{d)}^{\phantom{d)}h)_i}, & h_{abcd}= &\;
-\epsilon_{a(c}\epsilon_{d)b}. & & \label{primary-spinors}
\end{array}
\end{equation}
It is well known that a spinor $A_{abcd}$ satisfying $A_{abcd} =
A_{(ab)(cd)} = -A_{cdab}$ can be decomposed in the form $A_{abcd} =
\epsilon_{ac}A_{bd}+\epsilon_{bd}A_{ac}$ with $A_{ab} =
\frac{1}{2}\,A_{afb}^{\phantom{afb}f} = A_{(ab)}$ and that a spinor
$S_{abcd}$ satisfying $S_{abcd} = S_{(ab)(cd)} = S_{cdab}$ can be
written in the form $S_{abcd}=S_{(abcd)}+\frac{1}{3}h_{abcd}S$ with
$S:=S_{ef}^{\phantom{ef}ef}$. It follows from this that an arbitrary
four index spinor with symmetries $X_{abcd}= X_{(ab)(cd)}$ can be
expanded in terms of $\varepsilon^i_{abcd}$,
$\epsilon_{ac}x_{bd}+\epsilon_{bd}x_{ac}$,
$\epsilon_{ac}y_{bd}+\epsilon_{bd}y_{ac}$,
$\epsilon_{ac}z_{bd}+\epsilon_{bd}z_{ac}$ and $h_{abcd}$.

The following relations were frequently used in the calculations:
\begin{equation*}
y_{ab}x_{cd} =-\varepsilon^3_{abcd}
-{\frac{1}{2\sqrt2}}(\epsilon_{ac}y_{bd} +\epsilon_{bd}y_{ac}), 
\;\;
z_{ab}x_{cd} =\varepsilon^1_{abcd}
+{\frac{1}{2\sqrt2}}(\epsilon_{ac}z_{bd} +\epsilon_{bd}z_{ac});
\end{equation*}
\begin{equation*}
\begin{array}{llllll}
x_{ab}x^{ab}= -1, & x_{ab}y^{ab}= 0, & x_{ab}z^{ab}= 0, &
y_{ab}y^{ab}= 0, & y_{ab}z^{ab}= -\frac{1}{2}, & z_{ab}z^{ab}= 0; 
\end{array}
\end{equation*}
\begin{equation*}
\begin{array}{r@{}lr@{}lr@{}l}
x_a^{\;\;f}x_{bf}= &\;\frac{1}{2}\epsilon_{ab}, &
y_a^{\;\;f}x_{bf}= &\;\frac{1}{\sqrt2}y_{ab}, &
z_a^{\;\;f}x_{bf}= &\;-\frac{1}{\sqrt2}z_{ab}, \\[6pt]
y_a^{\;\;f}y_{bf}= &\;0, & y_a^{\;\;f}z_{bf}=
&\;-\frac{1}{2}\epsilon_a^{\;1}\epsilon_b^{\;0}, & 
z_a^{\;\;f}z_{bf}= &\;0; 
\end{array}
\end{equation*}
\begin{equation*}
\begin{array}{r@{}lr@{}lr@{}lr@{}l}
\varepsilon^0_{abcd}x^{cd}= &\;0, & 
\varepsilon^0_{abcd}y^{cd}= &\;-z_{ab}, &
\varepsilon^0_{abcd}z^{cd}= &\;0, &
\varepsilon^1_{abcd}x^{cd}= &\;-\frac{1}{2}z_{ab}, \\[6pt]
\varepsilon^1_{abcd}y^{cd}= &\;-\frac{1}{4}x_{ab}, & 
\varepsilon^1_{abcd}z^{cd}= &\;0, &
\varepsilon^2_{abcd}x^{cd}= &\;-\frac{1}{3}x_{ab}, &
\varepsilon^2_{abcd}y^{cd}= &\;\frac{1}{6}y_{ab}, \\[6pt]
\varepsilon^2_{abcd}z^{cd}= &\;\frac{1}{6}z_{ab}, &
\varepsilon^3_{abcd}x^{cd}= &\;\frac{1}{2}y_{ab}, &
\varepsilon^3_{abcd}y^{cd}= &\;0, &
\varepsilon^3_{abcd}z^{cd}= &\;\frac{1}{4}x_{ab}, \\[6pt]
\varepsilon^4_{abcd}x^{cd}= &\;0, &
\varepsilon^4_{abcd}y^{cd}= &\;0, &
\varepsilon^4_{abcd}z^{cd}= &\;-y_{ab}; & &
\end{array}
\end{equation*}
\begin{equation*}
\begin{array}{r@{}lr@{}lr@{}l}
x_{(ab}x_{cd)}= &\;2\,\varepsilon^2_{abcd}, & 
x_{(ab}y_{cd)}= &\;-\varepsilon^3_{abcd}, & 
x_{(ab}z_{cd)}= &\;\varepsilon^1_{abcd}, \\[6pt]
y_{(ab}y_{cd)}= &\;\frac{1}{2}\varepsilon^4_{abcd}, & 
y_{(ab}z_{cd)}= &\;-\frac{1}{2}\varepsilon^2_{abcd}, &
z_{(ab}z_{cd)}= &\;\frac{1}{2}\varepsilon^0_{abcd}; 
\end{array}
\end{equation*}
\begin{equation*}
\begin{array}{l@{}c@{}lr@{}c@{}ll@{}c@{}l}
x_{(a}^{\phantom{(a}f}\varepsilon^0_{b)cdf} \, &=
&\;\frac{1}{\sqrt{2}}\varepsilon^0_{abcd}, &
x_{(a}^{\phantom{(a}f}\varepsilon^1_{b)cdf} \, &=
&\;\frac{1}{2\sqrt{2}}z_{ab}x_{cd}, &
x_{(a}^{\phantom{(a}f}\varepsilon^2_{b)cdf} \, &=
&\;\frac{1}{12}(\epsilon_{ac}x_{bd}+\epsilon_{bd}x_{ac}), \\[6pt]
x_{(a}^{\phantom{(a}f}\varepsilon^3_{b)cdf} \, &=
&\;\frac{1}{2\sqrt{2}}y_{ab}x_{cd}, &
x_{(a}^{\phantom{(a}f}\varepsilon^4_{b)cdf} \, &=
&\;-\frac{1}{\sqrt{2}}\varepsilon^4_{abcd}, &
h_{ab(c}^{\phantom{ab(c}f}x_{d)f} \, &=
&\;\frac{1}{2}(\epsilon_{ac}x_{bd}+\epsilon_{bd}x_{ac}); 
\end{array}
\end{equation*}
\begin{equation*}
\begin{array}{l@{}c@{}lr@{}c@{}l}
y_{(d}^{\phantom{(a}f}\varepsilon^2_{c)abf}\, &=& \, 
-\frac{1}{2\sqrt{2}}\varepsilon^3_{abcd} 
+\frac{1}{24}(\epsilon_{ac}y_{bd}+\epsilon_{bd}y_{ac}), &
z_{(d}^{\phantom{(a}f}\varepsilon^2_{c)abf}\, &=& \, 
-\frac{1}{2\sqrt{2}}\varepsilon^1_{abcd} 
+\frac{1}{24}(\epsilon_{ac}z_{bd}+\epsilon_{bd}z_{ac}); \\[12pt]
\varepsilon_{\;ab}^{2\;\;\;ef}\varepsilon^1_{cdef}\, &=& \,
-\frac{1}{12}\varepsilon^1_{abcd} 
+\frac{1}{8\sqrt{2}}(\epsilon_{ac}z_{bd}+\epsilon_{bd}z_{ac}), &
\varepsilon_{\;ab}^{2\;\;\;ef}\varepsilon^3_{cdef}\, &=& \,
-\frac{1}{12}\varepsilon^3_{abcd} 
+\frac{1}{8\sqrt{2}}(\epsilon_{ac}y_{bd}+\epsilon_{bd}y_{ac});
\end{array}
\end{equation*}
\begin{equation*}
\begin{array}{ccc}
\varepsilon^2_{abcd}\varepsilon^{2\;abcd}= \frac{1}{6}, & &
\varepsilon_{ab}^{2\;\;ef}\varepsilon^2_{cdef}=
-\frac{1}{6}\varepsilon^2_{abcd} +\frac{1}{18}h_{abcd}.
\end{array}
\end{equation*}

\subsection{The detailed expressions for $u^p$, $p = 0, \ldots, 3$}
\label{tau-polynomials} 

\noindent
The $\tau$-dependent functions occuring in (\ref{1-order-solution}).
\begin{equation*}
\begin{array}{l@{}c@{}lr@{}c@{}ll@{}c@{}l}
c^{01}(\tau) \, &=& \, m\,({\frac {4}{3}} \,\tau ^{3} -
{\frac {1}{3}} \,\tau ^{5}), &
c^{\pm1}(\tau) \, &=& \, m\,(\tau ^{2} - {\frac {1}{6}}  
\,\tau ^{4}), &
S^1(\tau) \, &=& \, \sqrt{2}\,m\,({\frac {1}{2}} \,\tau ^{2}
 - { \frac {1}{4}} \,\tau ^{4}), \\[6pt]
K^1(\tau) \, &=& \, m\,( - 12\,\tau  + 4\,\tau ^{3}), &
F^1(\tau) \, &=& \, {\frac {1}{3}} \,m\,\tau ^{4}, &
t^1(\tau) \, &=& \, \sqrt{2}\,4\,\tau \,m, \\[6pt]
T^1(\tau) \, &=& \, 6\,m\,(1 - \tau ^{2}), &
\phi^1_1(\tau) \, &=& \, - 12\,(1 - \tau )^{2}, &
\phi^1_2(\tau) \, &=& \, - m^{2}\,(18\,\tau ^{2} - 3\,\tau ^{4}), \\[6pt]
\phi^1_3(\tau) \, &=& \, - 36 + 36\,\tau ^{2}. & &
\end{array}
\end{equation*}
The $\tau$-dependent functions occuring in 
(\ref{2-order-solution}). 
\begin{equation*}
\begin{array}{r@{}lr@{}l}
c^{02}_1(\tau)= &\; m^{2}\,( - 2\,\tau ^{3} - 3\,\tau ^{5} +
{ \frac {8}{7}} \,\tau ^{7} - { \frac {1}{7}} \,\tau ^{9}), &
c^{02}_2(\tau)= &\; 16\,\tau ^{3} - { \frac {26}{5}} \,\tau ^{5} 
+ { \frac {6}{5}} \,\tau ^{7}, \\[6pt]
c^{02}_3(\tau)= &\; 8\,\tau ^{3} - { \frac {7}{5}} \,\tau ^{5} - 
{ \frac {3}{5}} \,\tau ^{7}, &
c^{12}(\tau)= &\; m\,(-4\,\tau^2+\frac{2}{3}\tau^4), \\[6pt]
c^{\pm2}_1(\tau)= &\; m^{2}\,( - 2\,\tau ^{2} + 3\,\tau ^{4} -
{ \frac {8}{9}} \,\tau ^{6} + { \frac {1}{14}} \,\tau ^{8}), &
c^{\pm2}_2(\tau)= &\; 12\,\tau ^{2} - 3\,\tau ^{4} 
+ {\frac {3}{5}} \,\tau ^{6}, \\[6pt]
c^{\pm2}_3(\tau)= &\; - 6\,\tau ^{2} - { \frac {1}{2}} \,
\tau ^{4} + { \frac {3}{10}} \,\tau ^{6}, &
S^2_1(\tau)= &\; \sqrt{2}\,m^{2}\,({ \frac {4}{3}} \,\tau
 ^{4} - { \frac {2}{9}} \,\tau ^{6} -
{ \frac {1}{28}} \,\tau ^{8}), \\[6pt]
S^2_2(\tau)= &\; \sqrt{2}\,(6\,\tau ^{2} - { \frac {5}{2}
} \,\tau ^{4} + { \frac {9}{10}} \,\tau ^{6}), &
S^2_3(\tau)= &\; \sqrt{2}\,( - \frac{5}{4}\,\tau^{4} +3\,\tau^2
-\frac{9}{20} \,\tau ^{6}), \\[6pt]
S^2_4(\tau)= &\; -36\,\tau^{2} + 11\tau^{4} +\frac{3}{5}\,\tau^{6}, &
K^2_1(\tau)= &\; m^{2}\,(24\,\tau  - 8\,\tau ^{3} + 4\,\tau ^{5} -
{ \frac {4}{21}} \,\tau ^{7}), \\[6pt]
K^2_2(\tau)= &\;  - 144\,\tau  + 72\,\tau ^{3} - {
\frac {108}{5}} \,\tau ^{5}, &
K^2_3(\tau)= &\; m^{2}\,( - { \frac {20}{3}} \,\tau ^{3}
 + { \frac {8}{3}} \,\tau ^{5} - {
\frac {20}{63}} \,\tau ^{7}), \\[6pt]
K^2_4(\tau)= &\; -\sqrt{2}\,2\,\tau^3, &
K^2_5(\tau)= &\; -48\,\tau +\frac{36}{5}\,\tau^5, \\[6pt]
F^2_1(\tau)= &\; m^{2}\,( - 2\,\tau ^{2} + { \frac {1}{3}
} \,\tau ^{4} - { \frac {4}{9}} \,\tau ^{6} +
{ \frac {1}{7}} \,\tau ^{8}), &
F^2_2(\tau)= &\; 2\,\tau ^{4} - { \frac {6}{5}} \,\tau ^{
6}, \\[6pt]
F^2_3(\tau)= &\; 3\,\tau ^{4} + { \frac {3}{5}} \,\tau ^{
6}, &
t^2_1(\tau)= &\; \sqrt{2}\,m^{2}\,( - 12\,\tau  - {
\frac {8}{3}} \,\tau ^{3} + { \frac {4}{3}} \,\tau
^{5}), \\[6pt]
t^2_2(\tau)= &\; \sqrt{2}\,(48\,\tau  - 16\,\tau ^{3}), &
t^2_3(\tau)= &\; \sqrt{2}\,(24\,\tau  + 8\,\tau ^{3}), \\[6pt]
T^2_1(\tau)= &\; m^{2}\,( - 12 + 12\,\tau ^{2} - 10\,\tau ^{4} +
{ \frac {2}{3}} \,\tau ^{6}), &
T^2_2(\tau)= &\; 72 - 72\,\tau ^{2} + 36\,\tau ^{4}, \\[6pt]
T^2_3(\tau)= &\; m^{2}\,(4\,\tau ^{2} - { \frac {8}{3}}
\,\tau ^{4} + { \frac {4}{9}} \,\tau ^{6}), &
T^2_4(\tau)= &\; -\sqrt{2}\,6\,\tau^{2}, \\[6pt]
T^2_5(\tau)= &\; 24 -12\,\tau^{4}, &
\phi^2_1(\tau)= &\;  - ( - 1 + \tau )^{4}, \\[6pt]
\phi^2_2(\tau)= &\; 4\,m\,({ \frac {37}{10}} \,\tau ^{6} -
{ \frac {41}{5}} \,\tau ^{5} - {
\frac {41}{2}} \,\tau ^{4} + 46\,\tau ^{3} - 18\,\tau ^{2}), &
\phi^2_3(\tau)= &\; 16\,(1 + \tau )\,( - 1 + \tau )^{3}, \\[6pt]
\phi^2_4(\tau)= &\; 6\,( - { \frac {8}{21}} \,\tau ^{8} +
{ \frac {14}{3}} \,\tau ^{6} - 15\,\tau ^{4} + 6\,
\tau ^{2})\,m^{3}, &
\phi^2_5(\tau)= &\; 6\,m\,( - { \frac {46}{5}} \,\tau ^{6} +
62\,\tau ^{4} - 72\,\tau ^{2}), \\[6pt]
\phi^2_6(\tau)= &\;  - 72\,(1 + \tau )^{2}\,( - 1 + \tau )^{2}. &
\end{array}
\end{equation*}
The $\tau$-dependent functions occuring in 
(\ref{3-order-solution-1}).
\begin{equation*}
\begin{array}{r@{}l}
c^{03}_1(\tau)= &\; (3\,\tau ^{3} + 18\,\tau ^{5} + {
\frac {283}{21}} \,\tau ^{7} - { \frac {1510}{189}}
\,\tau ^{9} + { \frac {2972}{2079}} \,\tau ^{11} -
{ \frac {74}{693}} \,\tau ^{13})\,m^{3}, \\[6pt]
c^{03}_2(\tau)= &\; ( - 44\,\tau ^{3} - { \frac {588}{5}}
\,\tau ^{5} + { \frac {268}{7}} \,\tau ^{7} -
{ \frac {58}{7}} \,\tau ^{9} + {
\frac {6}{5}} \,\tau ^{11})\,m, \\[6pt]
c^{03}_3(\tau)= &\; 48\,\tau ^{3} - { \frac {96}{5}} \,\tau
 ^{5} + { \frac {312}{35}} \,\tau ^{7} -
{ \frac {12}{7}} \,\tau ^{9}, \\[6pt]
c^{03}_4(\tau)= &\; ( - 20\,\tau ^{3} - 6\,\tau ^{5} + {
\frac {439}{70}} \,\tau ^{7} - { \frac {573}{280}}
\,\tau ^{9} - { \frac {1}{40}} \,\tau ^{11})\,m, \\[6pt]
c^{03}_5(\tau)= &\; 16\,\tau ^{3} - 4\,\tau ^{5} - {
\frac {4}{7}} \,\tau ^{7} + { \frac {4}{7}} \,\tau
^{9}, \\[6pt]
c^{13}_1(\tau)= &\; (12\,\tau ^{2} + 15\,\tau ^{4} - {
\frac {14}{3}} \,\tau ^{6} + { \frac {3}{7}} \,\tau
^{8})\,m^{2}, \\[6pt]
c^{13}_2(\tau)= &\;  - 72\,\tau ^{2} + 18\,\tau ^{4} - {
\frac {18}{5}} \,\tau ^{6}, \\[6pt]
c^{13}_3(\tau)= &\;  - 36\,\tau ^{2} + 3\,\tau ^{4} + {
\frac {9}{5}} \,\tau ^{6}, \\[6pt]
c^{\pm3}_1(\tau)= &\; (18\,\tau ^{2} + 12\,\tau ^{4} - {
\frac {31}{5}} \,\tau ^{6} + { \frac {3}{2}} \,\tau
^{8} - { \frac {3}{40}} \,\tau ^{10})\,m, \\[6pt]
c^{\pm3}_2(\tau)= &\;  - 12\,\tau ^{2} + { \frac {4}{5}}
\,\tau ^{6} - { \frac {2}{7}} \,\tau ^{8}, \\[6pt]
c^{\pm3}_3(\tau)= &\; ({ \frac {9}{2}} \,\tau ^{2} -
{ \frac {33}{2}} \,\tau ^{4} + {
\frac {50}{3}} \,\tau ^{6} - { \frac {515}{84}} \,
\tau ^{8} + { \frac {25}{27}} \,\tau ^{10} -
{ \frac {34}{693}} \,\tau ^{12})\,m^{3}, \\[6pt]
c^{\pm3}_4(\tau)= &\; ( - 48\,\tau ^{2} + 105\,\tau ^{4} -
{ \frac {453}{10}} \,\tau ^{6} + {
\frac {2847}{280}} \,\tau ^{8} - { \frac {7}{8}} \,
\tau ^{10})\,m, \\[6pt]
c^{\pm3}_5(\tau)= &\; 36\,\tau ^{2} - 12\,\tau ^{4} + {
\frac {24}{5}} \,\tau ^{6} - { \frac {6}{7}} \,\tau
^{8}, \\[6pt]
c^{\pm3}_6(\tau)= &\;  - 3\,\tau ^{2} - 2\,\tau ^{4} + {
\frac {3}{5}} \,\tau ^{6} + { \frac {1}{14}} \,\tau
^{8}, \\[6pt]
S^3_1(\tau)= &\;  - 9\,\tau ^{2} - 2\,\tau ^{4} + {
\frac {13}{5}} \,\tau ^{6} + { \frac {1}{14}} \,\tau
 ^{8}, \\[6pt]
S^3_2(\tau)= &\; (108\,\tau ^{2} - 168\,\tau ^{4} + 86\,\tau ^{6} 
- { \frac {39}{5}}\,\tau ^{8} -\frac{3}{20}\,\tau^{10})\,m, \\[6pt]
S^3_3(\tau)= &\;  - 72\,\tau ^{2} + 48\,\tau ^{4} 
- \frac{72}{5}\,\tau ^{6} -\frac{4}{7}\,\tau^8, \\[6pt]
S^3_4(\tau)= &\; ( - { \frac {9}{4}} \,\tau ^{2} -
{ \frac {37}{4}} \,\tau ^{4} + {\frac {19}{2}} \,\tau ^{6} 
- { \frac {827}{168}} \,\tau ^{8} + { \frac {355}{378}} \,\tau ^{10} 
-{ \frac {6}{77}} \,\tau ^{12})\,\sqrt{2}\,m^{3}, \\[6pt]
S^3_5(\tau)= &\; (6\,\tau ^{2} + { \frac {69}{2}} \,\tau^{4} 
- { \frac {333}{20}} \,\tau ^{6} +{ \frac {1999}{560}} \,\tau ^{8} 
+ {\frac {13}{80}} \,\tau ^{10})\,\sqrt{2}\,m, \\[6pt]
S^3_6(\tau)= &\; ( 18\,\tau ^{2} - 6\,\tau ^{4} 
+{\frac {24}{5}} \,\tau ^{6} - { \frac {9}{7}} \,\tau^{8})\,\sqrt{2}, \\[6pt]
S^3_7(\tau)= &\; ( -3\,\tau ^{2} - \frac{33}{2}\,\tau ^{4} 
+ { \frac {177}{20}} \,\tau ^{6} -{ \frac {379}{112}} \,\tau ^{8} 
+ { \frac {1}{40}} \,\tau ^{10} )\sqrt{2}\,m^{3}, \\[6pt]
S^3_8(\tau)= &\; (6\,\tau ^{2} -2\,\tau ^{4} 
+{ \frac {3}{7} \,\tau ^{8}})\,\sqrt{2}\,m, \\[6pt]
K^3_1(\tau)= &\;  - 6\,\tau  - 8\,\tau ^{3} + { \frac {18
}{5}} \,\tau ^{5} + { \frac {4}{7}} \,\tau ^{7}, \\[6pt]
K^3_2(\tau)= &\; (144\,\tau  +12\,\tau ^{3} 
- { \frac {351}{5}} \,\tau ^{5} + { \frac {237}{5}} \,\tau ^{7}
- { \frac {17}{4}} \,\tau ^{9})\,m, \\[6pt]
K^3_3(\tau)= &\;  - 96\,\tau  + 16\,\tau ^{3} 
+ { \frac {72}{5}} \,\tau ^{5} - { \frac {64}{7}} \,\tau ^{7}, \\[6pt]
K^3_4(\tau)= &\; ( - 54\,\tau  + 12\,\tau ^{3} - 216\,\tau ^{5} +
{ \frac {796}{7}} \,\tau ^{7} - {
\frac {440}{21}} \,\tau ^{9} + { \frac {16}{11}} \,
\tau ^{11})\,m^{3}, \\[6pt]
K^3_5(\tau)= &\; (576\,\tau  - 216\,\tau ^{3} + { \frac {
1962}{5}} \,\tau ^{5} - { \frac {714}{5}} \,\tau ^{7
} + { \frac {23}{2}} \,\tau ^{9})\,m, \\[6pt]
K^3_6(\tau)= &\;  - 432\,\tau  + 288\,\tau ^{3} - {
\frac {864}{5}} \,\tau ^{5} + { \frac {288}{7}} \,\tau ^{7}, \\[6pt]
K^3_7(\tau)= &\; (40\,\tau ^{3} - 16\,\tau ^{5} + {
\frac {100}{21}} \,\tau ^{7} - { \frac {160}{189}}
\,\tau ^{9} + { \frac {20}{693}} \,\tau ^{11})\,m^{3}, \\[6pt]
K^3_8(\tau)= &\; ( - 240\,\tau ^{3} + { \frac {582}{5}}
\,\tau ^{5} - { \frac {218}{7}} \,\tau ^{7} +
{ \frac {23}{6}} \,\tau ^{9})\,m, \\[6pt]
K^3_9(\tau)= &\; (9\,\tau ^{3} - { \frac {33}{20}}\,\tau ^{5} 
- { \frac {13}{20}} \,\tau ^{7} +
{ \frac {1}{80}} \,\tau ^{9})\sqrt{2}\,m, \\[6pt]
K^3_{10}(\tau)= &\;  (- 4\,\tau ^{3} 
+ { \frac {6}{5}} \,\tau ^{5})\,\sqrt{2}, 
\end{array}
\end{equation*}

\noindent
\begin{equation*}
\begin{array}{r@{}l}
F^3_1(\tau)= &\; (9\,\tau ^{2} + 2\,\tau ^{4} - { \frac {
7}{3}} \,\tau ^{6} + { \frac {26}{7}} \,\tau ^{8} -
{ \frac {20}{21}} \,\tau ^{10} + {
\frac {74}{693}} \,\tau ^{12})\,m^{3}, \\[6pt]
F^3_2(\tau)= &\; ( - 60\,\tau ^{2} + 36\,\tau ^{4} - 12\,\tau ^{6} +
{ \frac {106}{35}} \,\tau ^{8} - {
\frac {6}{5}} \,\tau ^{10})\,m, \\[6pt]
F^3_3(\tau)= &\;  - { \frac {24}{5}} \,\tau ^{6} +
{ \frac {12}{7}} \,\tau ^{8}, \\[6pt]
F^3_4(\tau)= &\; ( - 12\,\tau ^{2} - 6\,\tau ^{4} + {
\frac {7}{2}} \,\tau ^{6} + { \frac {169}{56}} \,
\tau ^{8} + { \frac {1}{40}} \,\tau ^{10})\,m, \\[6pt]
F^3_5(\tau)= &\; 4\,\tau ^{4} - { \frac {4}{5}} \,\tau ^{
6} - { \frac {4}{7}} \,\tau ^{8}, \\[6pt]
t^3_1(\tau)= &\; (36\,\tau  + 20\,\tau ^{3} + 46\,\tau ^{5} -
{ \frac {296}{21}} \,\tau ^{7} + {
\frac {272}{189}} \,\tau ^{9})\,\sqrt{2}\,m^{3}, \\[6pt]
t^3_2(\tau)= &\; ( - 312\,\tau  - 24\,\tau ^{3} - {
\frac {72}{5}} \,\tau ^{5} - { \frac {40}{7}} \,\tau
 ^{7})\,\sqrt{2}\,m, \\[6pt]
t^3_3(\tau)= &\; (144\,\tau  - 96\,\tau ^{3} + { \frac {
144}{5}} \,\tau ^{5})\,\sqrt{2}, \\[6pt]
t^3_4(\tau)= &\; ( - 96\,\tau  - 12\,\tau ^{3} + {
\frac {294}{5}} \,\tau ^{5} - { \frac {86}{35}} \,
\tau ^{7})\,\sqrt{2}\,m, \\[6pt]
t^3_5(\tau)= &\; (48\,\tau  - { \frac {48}{5}} \,\tau ^{
5})\,\sqrt{2}, \\[6pt]
T^3_1(\tau)= &\; 3 + 9\,\tau ^{2} - 3\,\tau ^{4} - \tau ^{6}, \\[6pt]
T^3_2(\tau)= &\;  (- 72 -36\,\tau ^{2} +81\,\tau ^{4} 
- \frac{423}{5}\,\tau^{6} + { \frac {33}{4}} \,\tau ^{8})\,m, \\[6pt]
T^3_3(\tau)= &\; 48 - 24\,\tau ^{2} + 16\,\tau ^{6}, \\[6pt]
T^3_4(\tau)= &\; (27 - 18\,\tau ^{2} + 180\,\tau ^{4} - 134
\,\tau ^{6} + { \frac {204}{7}} \,\tau ^{8} -
{ \frac {16}{7}} \,\tau ^{10})\,m^{3}, \\[6pt]
T^3_5(\tau)= &\;  (- 288 + 216\,\tau ^{2} - 558\,\tau ^{4} +
{ \frac {1326}{5}} \,\tau ^{6} - {\frac {243}{10}} \,\tau ^{8})\,m, \\[6pt]
T^3_6(\tau)= &\; 216 - 216\,\tau ^{2} 
+ 216\,\tau ^{4} - 72\,\tau^{6}, \\[6pt]
T^3_7(\tau)= &\; ( - 24\,\tau ^{2} + 16\,\tau ^{4} - {
\frac {20}{3}} \,\tau ^{6} + { \frac {32}{21}} \,
\tau ^{8} - { \frac {4}{63}} \,\tau ^{10})\,m^{3}, \\[6pt]
T^3_8(\tau)= &\; (144\,\tau ^{2} - 102\,\tau ^{4} + {
\frac {178}{5}} \,\tau ^{6} - { \frac {57}{10}} \,
\tau ^{8})\,m, \\[6pt]
T^3_9(\tau)= &\; ( 27\,\tau ^{2} -\frac{81}{4}\,\tau ^{4} 
- {\frac {11}{20}} \,\tau ^{6} 
+ { \frac {9}{80}} \,\tau^{8})\sqrt{2}\,m, \\[6pt]
T^3_{10}(\tau)= &\; (-12\,\tau ^{2} +6\,\tau ^{4})\sqrt{2}.
\end{array}
\end{equation*}
The $\tau$-dependent functions occuring in 
(\ref{phi-equations-right-side}).
\begin{equation*}
\begin{array}{r@{}l}
A_1(\tau)= &\;\left (36\,\tau-78\,{\tau}^{2}+82\,{\tau}^{3}
-{\frac {97}{2}}\,{\tau}^{4}+\frac{6}{5}\,{\tau}^{5}
+{\frac{169}{5}}\,{\tau}^{6}-{\frac {208}{7}}\,{\tau}^{7}
+{\frac {54}{7}}\,{\tau}^{8}\right )m, \\[6pt]
A_2(\tau)= &\; - 648\,\tau  + 1728\,\tau ^{2} - 1692\,\tau ^{3} +
432\,\tau ^{4} + { \frac {2592}{5}} \,\tau ^{5} -
{ \frac {2286}{5}} \,\tau ^{6} + {
\frac {756}{5}} \,\tau ^{7} - { \frac {162}{5}} \,
\tau ^{8}, \\[6pt]
B_1(\tau)= &\;(108\,\tau  - 234\,\tau ^{2} - 396\,\tau ^{3}
 + 1503\,\tau ^{4} - 579\,\tau ^{5} - { \frac {14939
}{20}} \,\tau ^{6} \\[6pt]
&\; + { \frac {11682}{35}} \,\tau ^{7
} + { \frac {40413}{560}} \,\tau ^{8} -{ \frac {2591}{70}} \,\tau ^{9} +
{ \frac {177}{80}} \,\tau ^{10})m^{2}, \\[6pt]
B_2(\tau)= &\; - 648\,\tau  + 1404\,\tau ^{2} - 540\,\tau ^{3} - 810
\,\tau ^{4} + { \frac {1404}{5}} \,\tau ^{5} +
{ \frac {1458}{5}} \,\tau ^{6} - 108\,\tau ^{7} +
{ \frac {108}{5}} \,\tau ^{8}, \\[6pt]
B_3(\tau)= &\;( - 72\,\tau  + 168\,\tau ^{2} + 24\,\tau ^{3} - 274\,
\tau ^{4} + 120\,\tau ^{5} + { \frac {306}{5}} \,
\tau ^{6} - 32\,\tau ^{7} + { \frac {6}{7}} \,\tau
^{8})\,m, \\[6pt]
C_1(\tau)= &\;( - 27\,\tau  + 342\,\tau ^{3} - 696\,\tau ^{5} +
{ \frac {2598}{7}} \,\tau ^{7} - {
\frac {4555}{63}} \,\tau ^{9} + { \frac {1079}{231}
} \,\tau ^{11})\,m^{4}, \\[6pt]
C_2(\tau)= &\;(504\,\tau  - 3492\,\tau ^{3} + { \frac {
17607}{5}} \,\tau ^{5} - { \frac {41289}{35}} \,\tau
 ^{7} + { \frac {16559}{140}} \,\tau ^{9})\,m^{2}, \\[6pt]
C_3(\tau)= &\; - 1296\,\tau  + 2376\,\tau ^{3} - {
\frac {4752}{5}} \,\tau ^{5} + 216\,\tau ^{7}, \\[6pt]
C_4(\tau)= &\;( - 432\,\tau  + 792\,\tau ^{3} - {
\frac {3072}{5}} \,\tau ^{5} + { \frac {816}{7}} \,
\tau ^{7})\,m, \\[6pt]
C_5(\tau)= &\; - 216\,\tau  + 108\,\tau ^{3} + {
\frac {648}{5}} \,\tau ^{5} + { \frac {324}{5}} \,\tau^{7}. 
\end{array}
\end{equation*}

\noindent
The $\tau$-dependent functions occuring in 
(\ref{3-order-solution-2}). 
\begin{equation*}
\begin{array}{l@{}c@{}l}
f_0(\tau) \, &=& \,  - 18 + 216\,\tau ^{2} - 240\,\tau ^{3} + 18\,\tau ^{4
} - 48\,\tau ^{5} + 204\,\tau ^{6} - 144\,\tau ^{7} + 30\,\tau ^{
8}, \\[6pt]
f_1(\tau) \, &=& \,  - 9 - 216\,\tau ^{2} + 696\,\tau ^{3} - 198\,\tau ^{4
} - { \frac {2544}{5}} \,\tau ^{5} + {
\frac {984}{5}} \,\tau ^{6} + { \frac {936}{7}} \,
\tau ^{7} - { \frac {411}{7}} \,\tau ^{8}, \\[6pt]
f_2(\tau) \, &=& \,  - 3 - 216\,\tau ^{2} + 372\,\tau ^{4} -
{ \frac {936}{5}} \,\tau ^{6} + {
\frac {219}{7}} \,\tau ^{8}, \\[6pt]
g_0(\tau) \, &=& \, 108 - 1944\,\tau ^{2} + 4752\,\tau ^{3} - 5724\,\tau
^{4} + { \frac {19008}{5}} \,\tau ^{5} -
{ \frac {6264}{5}} \,\tau ^{6} + {
\frac {864}{5}} \,\tau ^{7} - { \frac {108}{5}} \,
\tau ^{8}, \\[6pt]
g_1(\tau) \, &=& \, 54 - 972\,\tau ^{2} + 1620\,\tau ^{3} + 378\,\tau ^{4}
 - { \frac {11448}{5}} \,\tau ^{5} + {
\frac {5778}{5}} \,\tau ^{6} + { \frac {108}{5}} \,
\tau ^{7} - { \frac {108}{5}} \,\tau ^{8}, \\[6pt]
g_2(\tau) \, &=& \, 18 - 540\,\tau ^{2} + 972\,\tau ^{4} - {
\frac {2808}{5}} \,\tau ^{6} + { \frac {108}{5}} \,
\tau ^{8}, \\[6pt]
\multicolumn{3}{l}{h_0(\tau)= \;{ \frac {3}{2}}, \;\;\;
h_1(\tau)= \;{ \frac {3}{4}}, \;\;\; h_2(\tau)= \;{ \frac {1}{4}},} \\[6pt]
k_1(\tau) \, &=& \,  108\,\tau ^{2} - 276\,\tau ^{3} - 129\,\tau ^{4} + 
{ \frac {4077}{5}} \,\tau ^{5} - { 
\frac {3289}{10}} \,\tau ^{6} - { \frac {9439}{35}} 
\,\tau ^{7} + { \frac {32803}{280}} \,\tau ^{8} + 
{ \frac {463}{20}} \,\tau ^{9} - { 
\frac {2721}{280}} \,\tau ^{10}, \\[6pt]
k_2(\tau) \, &=& \,  252\,\tau ^{2} - 942\,\tau ^{4} + { \frac {3614}{5}
} \,\tau ^{6} - { \frac {6341}{35}} \,\tau ^{8} + 
{ \frac {99}{7}} \,\tau ^{10}, \\[6pt]
p(\tau) \, &=& \,  - { \frac {27}{2}} \,\tau ^{2} + { 
\frac {171}{2}} \,\tau ^{4} - 116\,\tau ^{6} + { 
\frac {1299}{28}} \,\tau ^{8} - { \frac {911}{126}} 
\,\tau ^{10} + { \frac {1079}{2772}} \,\tau ^{12}, \\[6pt]
q(\tau) \, &=& \,  {\frac {216}{5}} \,\tau ^{8} 
- {\frac {576}{5}} \,\tau ^{6} + 648\,\tau ^{4} - 864\,\tau^{2}. 
\end{array}
\end{equation*}
The $\tau$-dependent functions occuring in 
(\ref{rhs phi-constraints}).
\begin{equation*}
\begin{array}{l@{}l}
F_1(\tau) \, =& \, (72\,\tau ^{2} - { \frac {1071}{2}} \,\tau ^{4}
+ { \frac {4077}{5}} \,\tau ^{5} - {
\frac {2639}{20}} \,\tau ^{6} - { \frac {18878}{35}
} \,\tau ^{7} + { \frac {113287}{560}} \,\tau ^{8}
+ { \frac {1389}{20}} \,\tau ^{9} - {
\frac {15087}{560}} \,\tau ^{10})\,m^{2}, \\[6pt]
F_2(\tau) \, =& \,  - 864\,\tau ^{2} + 1584\,\tau ^{3} - 810\,\tau ^{4}
 - { \frac {1296}{5}} \,\tau ^{5} + {
\frac {882}{5}} \,\tau ^{6} + { \frac {432}{5}} \,
\tau ^{7} - { \frac {108}{5}} \,\tau ^{8}, \\[6pt]
F_3(\tau) \, =& \, ( - 36\,\tau ^{2} - 40\,\tau ^{3} + 156\,\tau ^{4} -
{ \frac {888}{5}} \,\tau ^{5} + {
\frac {194}{5}} \,\tau ^{6} + { \frac {456}{7}} \,
\tau ^{7} - { \frac {198}{7}} \,\tau ^{8})\,m, \\[6pt]
G_1(\tau) \, =& \, ({ \frac {27}{2}} \,\tau ^{2} +
{ \frac {171}{2}} \,\tau ^{4} - 348\,\tau ^{6} +
{ \frac {6495}{28}} \,\tau ^{8} - {
\frac {911}{18}} \,\tau ^{10} + { \frac {1079}{308}
} \,\tau ^{12})\,m^{4}, \\[6pt]
G_2(\tau) \, =& \, ( - 144\,\tau ^{2} - 1071\,\tau ^{4} + {
\frac {3679}{2}} \,\tau ^{6} - { \frac {220837}{280}
} \,\tau ^{8} + { \frac {24999}{280}} \,\tau ^{10})
\,m^{2}, \\[6pt]
G_3(\tau) \, =& \, 1116\,\tau ^{4} - 468\,\tau ^{6} + {
\frac {648}{5}} \,\tau ^{8}, \\[6pt]
G_4(\tau) \, =& \, 174\,m\,\tau ^{4} - { \frac {1824}{5}} \,
m\,\tau ^{6} + { \frac {684}{7}} \,m\,\tau ^{8}, \\[6pt]
G_5(\tau) \, =& \, 432\,\tau ^{2} - 234\,\tau ^{4} + {
\frac {306}{5}} \,\tau ^{6} + { \frac {216}{5}} \,\tau ^{8}.
\end{array}
\end{equation*}
The $\tau$-dependent functions occuring in 
(\ref{reduced-equations}).
\begin{equation*}
\begin{array}{l@{}l}
T_{1}(\tau) \, =& \, -46656\,\tau  + 124416\,\tau ^{2} - 138240\,\tau ^{3} 
+ 51840\,\tau ^{4} + 31104\,\tau ^{5}-{ \frac  {133056}{5}}\,\tau^{6}  \\[6pt]
&\;+12096 \tau ^{7}-{ \frac {50112}{5}} \,\tau ^{8} 
+{ \frac {10368}{5}} \,\tau ^{9}, \\[6pt]
T_{2}(\tau) \, =& \, (5184\,\tau  - 3456\,\tau ^{2} - 5088\,
m\,\tau ^{3} - 6048\,\tau ^{4} + 12288\,\tau ^{5} 
+384\,\tau ^{6}-4128\,\tau ^{7} \\[6pt]
&\;+1824\,\tau ^{8}-1344\,\tau ^{9} + 384\,\tau^{10})\,m. 
\end{array}
\end{equation*}

\end{document}